\documentclass[twocolumn]{aastex631}

\def\lesssim{\mathrel{\hbox{\rlap{\hbox{\lower4pt\hbox{$\sim$}}}\hbox{$<$}}}}
\def\gtrsim{\mathrel{\hbox{\rlap{\hbox{\lower4pt\hbox{$\sim$}}}\hbox{$>$}}}}
\newcommand{\bea}{\begin{eqnarray}}
\newcommand{\eea}{\end{eqnarray}}

\newcommand{\dr}{\mathrm{d}r}

\newcommand{\md}{\mathrm{d}}

\newcommand{\beq}[1]{\begin{equation} #1 \end{equation}}

\newcommand{\nnn}{\nonumber \\}

\newcommand{\xx    }[1]{x^{\left(#1\right)}}

\newcommand{\xone  }{\xx{1}}
\newcommand{\xtwo  }{\xx{2}}
\newcommand{\xthree}{\xx{3}}

\usepackage{amsmath}
\usepackage{ulem}
\usepackage[T1]{fontenc} 

\graphicspath{{./}{./figures/}}

\begin{document}


\title{Magnetized Accretion onto Neutron Stars: from Photon-trapped to Neutrino-cooled Flows}
\author[0000-0002-5427-1207]{Luciano Combi}
\affiliation{Perimeter Institute for Theoretical Physics, 31 Caroline St. North, Waterloo, ON N2L 2Y5, Canada }
\affiliation{Department of Physics, University of Guelph, Guelph, Ontario N1G 2W1, Canada}

\author[0000-0003-4305-5653]{Christopher Thompson}
\affiliation{Canadian Institute for Theoretical Astrophysics, 60 St. George Street, Toronto, ON M5S 3H8, Canada}

\author[0000-0001-6374-6465]{Daniel M.~Siegel}
\affiliation{Institute of Physics, University of Greifswald, D-17489 Greifswald, Germany}
\affiliation{Department of Physics, University of Guelph, Guelph, Ontario N1G 2W1, Canada}

\author[0000-0001-7801-0362]{Alexander Philippov}
\affiliation{Department of Physics, University of Maryland, College Park, MD 20742, USA}

\author[0000-0002-7301-3908]{Bart Ripperda}
\affiliation{Canadian Institute for Theoretical Astrophysics, 60 St. George Street, Toronto, ON M5S 3H8, Canada}
\affiliation{Department of Physics, University of Toronto, 60 St. George Street, Toronto, ON M5S 1A7, Canada}
\affiliation{David A. Dunlap Department of Astronomy, University of Toronto, 50 St. George Street, Toronto, ON M5S 3H4, Canada}
\affiliation{Perimeter Institute for Theoretical Physics, 31 Caroline St. North, Waterloo, ON N2L 2Y5, Canada }

\begin{abstract}

When a neutron star (NS) intercepts gas from a non-degenerate star, e.g., in a tidal disruption event, a common-envelope phase, or the collapsing core of a massive star, photons become trapped in the hot flow around the NS.  This gas forms a radiatively inefficient accretion flow (RIAF) until the density and temperature close to the NS surface grow large enough for binding energy to be converted to neutrinos. 
Here we present three-dimensional, general-relativistic, magnetohydrodynamic simulations of accretion onto a non-rotating, unmagnetized NS. 
These connect,  for the first time, an extended accretion disk with a self-consistent hydrostatic atmosphere around the star.  The impact of different seed magnetic fields and accretion rates is studied by approximating the radiation-pressure dominated flow as an ideal gas with an adiabatic index of $4/3$, coupled to a variable neutrino emissivity.
At low accretion rates, the hydrostatic atmosphere shows slow rotation and weak magnetization, transitioning to an outer RIAF structure.  A toroidal magnetic field mediates the inward flow of energy and angular momentum through the atmosphere, which reaches a steady state when neutrino emission balances the accretion power.  
We develop a one-dimensional analytical model connecting these results with more general initial conditions and describing the main features of the flow.  Our results have implications for the spin and mass evolution of hypercritically accreting NSs.

\end{abstract}



\section{Introduction}\label{sec:intro}

A very compact star, such as a neutron star (NS) or white dwarf, may sometimes collide with a less
evolved star that still retains a hydrogen-rich envelope (see \citealt{Hirai2022} and references therein).
There is a longstanding interest, going
back at least to \cite{thorne1975red, thorne1977stars}, in the consequences of this interaction.   How much
mass and angular momentum does a NS accrete, and how is the gravitational energy of the accreted material
transported away?  
At such high accretion rates, radiation is at first trapped in the material circulating around the NS. Eventually, the rate of accretion may become limited
by photon pressure and nuclear shell burning \citep{cannon1992structure, cannon1993massive}.

The suggestion that runaway accretion near the Bondi rate could add enough mass
to the NS to transform it to a black hole (BH; \citealt{bb1998,blb2000}) predated the
realization that non-radiative accretion onto compact stars can be very
inefficient.  Circulating flows
transfer increasingly small proportions of the bound gas at closer distances to	the central
mass, even as they transport energy and angular momentum outward \citep{narayan_advection-dominated_1995, quataert2000convection}.

In the absence of an event horizon, such a radiatively inefficient flow (RIAF) must
release energy to neutrinos to settle onto the NS.  Neutrino radiation is effective
when the inflow onto the NS is nearly spherical \citep{zel1972nonstationary, colgate_luminosity_1980, houck1991steady, klein1980supercritical, bernal2013}, but may not be
if the flow has the milder density profile characteristic of a RIAF.	
Existing simulations that resolve the Bondi sphere in a realistic accretion geometry
only extend the flow to intermediate distances from the accretor, in the approximation
where it is treated as a point particle and cooling is neglected
\citep{armitage2000, macleod2014accretion, de_common_2020}.
A hydrostatic core is expected to form around a NS	that is	embedded in a more
extended RIAF,
in the situation where neutrino cooling is initially weak.  The spherical models
of \cite{houck1991steady} and  \cite{chevalier1993neutron} suggest that this atmosphere may reach a large
radius at low accretion rates and even become dynamically unstable.  
The circulation must slow near the NS, due to
friction with its surface or its magnetic field.  

A consequence of this effect is that the
angular	momentum added to the star by accretion, and the circulatory stress that is available
for amplifying a magnetic field, may both be substantially smaller than	a naive	application of
a RIAF disk model would	suggest.  A self-consistent treatment of the accretion flow is
also needed to evaluate when a magnetic field anchored in the NS could emerge
to form a magnetosphere, allowing the star eventually to form a relativistic pulsar outflow \citep{piro2011supernova}.

There is a complementary body of work on numerical simulations of extreme accretion onto NSs in the context of core-collapse supernovae and binary neutron-star mergers where weak and nuclear interactions become highly relevant (e.g. \citealt{janka2007theory,fernandez2009stability, ott2013general, fahlman2023secular, combi_jets_2023-2, camilletti2024geometric}). Matter here is assembled close to the hot proto-NS (remnant NS) after core collapse (merger) and is accreted at enormous rates of $\dot{M} \sim 1 \, M_{\odot}\,s^{-1}$, approaching $\sim 10^{16}$ times the
Eddington rate.  Densities ($\rho \gtrsim 10^{12}$\,g\,cm$^{-3}$) and temperatures ($\sim 10$ MeV) are so high in the disk that matter becomes dissociated into nucleons and cools significantly by neutrino emission, deleptonizing the disk in the process \citep{chen2007Neutrinocooled, metzger2008Conditions, siegel2019collapsars}. The transition from radiatively inefficient accretion to a significantly neutrino-cooled, self-neutronized (electron or proton fraction $Y_e\ll 0.5$) accretion flow occurs above a critical ``ignition'' accretion rate $\dot{M}_{\rm ign}\propto \alpha^{5/3} M^{4/3}$, which, depending on the effective viscosity parameter $\alpha$ of the accretion flow ($\alpha \sim 10^{-2}-1$), ranges between $\dot{M}_{\rm ign} \sim 10^{-3}-1 M_{\odot}$ s$^{-1}$ for BHs with mass $M\gtrsim 3\,M_\odot$ \citep{chen2007Neutrinocooled,de_igniting_2021, siegel2019collapsars}.

{Although there are several three-dimensional (3D) Newtonian and general-relativistic magnetohydrodynamic (GRMHD) simulations of accreting NS with magnetospheres (e.g. \citealt{das2024three, parfrey2024accreting, romanova2012mri}), there are few multi-dimensional calculations of accretion through a surface boundary layer.  In the super/hypercritical regime, notable exceptions are \cite{takahashi2018supercritical} and \cite{abarca2018radiative}, who performed two-dimensional (2D) GRMHD simulations of supercritical accretion onto a non-magnetized NS, including a two-momentum closure scheme for photon transport and reflective boundary conditions to model the NS. These simulations target ultra-luminous X-ray binaries, where the photon trapping radius is not too far from the NS surface and neutrino emission can be neglected.  To our knowledge, full 3D MHD simulations of accretion onto an NS with a physical surface have not been performed before.}

We present GRMHD simulations of a RIAF around a NS with a resolved outer 
hydrostatic structure. The accretion rate is much less than $\dot M_{\rm ign}$ but still high enough that one can assume photons to be tightly bound to the flow over the duration of the simulation, and thus we assume an ideal gas with adiabatic index of $\Gamma=4/3$. We add an optically thin cooling term in the equations representing neutrino energy release. The radiative losses are chosen to have the same strong dependence on density and temperature as expected for neutrino emission; these are concentrated in a thin layer around the star in the simulations.  The flow is magnetized and transports both mass and magnetic flux toward the initially unmagnetized star.  We observe the development of an inner hydrostatic flow with weak rotation gradient;  but, in contrast with a spherical inflow of near vanishing energy \citep{houck1991steady}, no shock appears near the transition to a hydrostatic flow.

The paper is organized as follows. We start in Section~\ref{sec:analytic} by presenting an analytical model for the formation and accretion properties of the hydrostatic atmosphere formed around the NS.  The
simulation setup is described in Section \ref{sec:setup}, followed by the simulation results in Section~\ref{sec:results}. We comment on some of the long-term implications of the accretion flow in Section~\ref{sec:implications} and summarize our findings in Section~\ref{sec:conclusion}. 
The Appendices give analytical details governing neutrino emission, atmosphere structure, and the flow of angular and radial momentum. We also present a convergence study of the numerical evolution.  

\textbf{Conventions:} Throughout this paper, 
we write dimensionful quantities as $X = X_n\times 10^n$ in cgs units. We use a $+2$ signature for the spacetime metric, $a,b,c,...=0,1,2,3$ for spacetime index, and $i,j,k,..=1,2,3$ for spatial index. The spacetime coordinates are denoted by $x^a=\lbrace ct,x,y,z\rbrace$, where $c$ is the speed of light.

\begin{figure}
        \centering
        \includegraphics[width=1\linewidth]{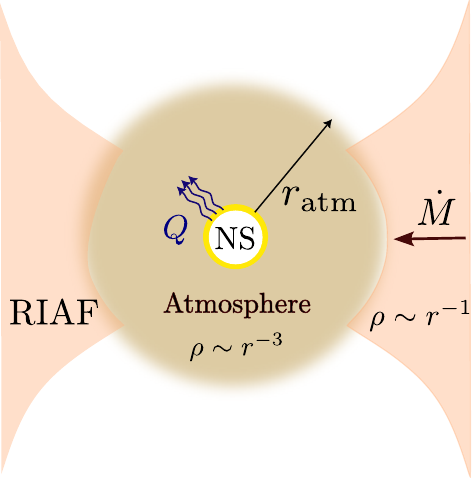}
        \caption{Schematic of the accretion structure considered in this paper: a dense and geometrically thick flow around and onto a NS.  Photons are trapped in the flow over the timescales considered; neutrino radiative losses ($Q$) are concentrated in a thin layer above the star.}
        \label{fig:schematic}
\end{figure}

\section{Accretion of High-Entropy Gas Mediated by Neutrino Emission}
\label{sec:analytic}

A NS may intercept gas rapidly enough that radiation is trapped where the gas angular momentum approaches the Keplerian angular momentum, and some material becomes gravitationally bound.   Radiative diffusion in this bound gas can be negligible
over the accretion timescale $\sim \alpha^{-1} \Omega^{-1}$, where $\alpha$ measures the ratio of
viscous and thermal stresses within the gas.\footnote{This conclusion follows straightforwardly by noting that the Kelvin time of the Sun is some 11 orders of magnitude
longer than its dynamical time, meaning that radiation
must be trapped over many dynamical times
when the flow is 
more compact than the Sun but carries even a small fraction
of the Solar mass.}

The flow closer to the NS must have a very different structure from a RIAF around a BH, because accretion onto the NS depends on the loss of energy to neutrinos (Figure \ref{fig:schematic}) and the advected energy cannot be absorbed through a horizon. As the gas starts accumulating near the NS surface, thermal pressure increases and forms an atmosphere that backreacts against the inflow. Neutrino cooling is greatly enhanced by the formation of this nearly hydrostatic density cusp around the NS. As we now explain, such a slowly rotating atmosphere
must form below a critical accretion rate $\dot M \sim 10^{-3}\,M_\odot\,{\rm s}^{-1}$; at higher accretion rates, cooling is fast enough to allow a RIAF structure very close to the surface.

Neutrino emission is a very strong function of temperature, $T$:  the neutrino energy, $E_\nu$, released per unit time, $t$, over a volume, $V$, is
\begin{equation}
 Q=\frac{\md^2E_\nu}{\md V \md t} = K_\nu^{\rm cap} \rho T^6 + K_\nu^{\rm ann}T^9,   
\end{equation}
representing losses by $e^\pm$ capture and $e^\pm$ annihilation, respectively; here, $K$ are constants and $\rho$ is the fluid rest-mass density.

The flow entropy is high enough at low $\dot M$ such that pressure $p$ is dominated by photons (and also by $e^\pm$ pairs close enough to the NS). In a RIAF with density profile $\rho(r) \propto r^{-1}$ and mass accretion $\dot M(r) \sim 2\pi \alpha \Omega r^3\rho(r) \propto r^{1/2}$, one has $p = (f_\pm/3)aT^4 \simeq GM\rho/4r$, where $G$ is the gravitational constant, $M = 1.4\,M_{1.4}\,M_\odot$ is the mass of the star, $r$ is the radius from the center, and the coefficient $f_\pm = 1$ or $11/4$ in the absence or presence of pairs. Then $T \propto r^{-1/2}$ in the RIAF, as compared with $\sim r^{-1}$ in an adiabatic atmosphere with density cusp $\rho(r) \propto r^{-3}$. The neutrino luminosity at radius $r$ therefore scales as $r^3 \md^2 E_\nu/ \md V \md t \sim r^{-1}$ in the RIAF (for electron-capture cooling), as compared with $\sim r^{-6}$ in the atmosphere.

The structure of the atmosphere is largely determined by its specific entropy $s_{\rm atm}$, which equals $h/T$ in a relativistic gas with enthalpy $h$.
The mass flow rate $\dot M$ toward the NS is approximately independent of radius within the atmosphere,
$\dot M = \dot M_{\rm atm}$;
the transition to a RIAF structure (at radius $r_{\rm atm}$) occurs where the inflow speed
$|v_r|$ reaches $\sim \alpha (GM/r_{\rm atm})^{1/2}$.  The RIAF entropy $s_d$ outside the central core
decreases outward approximately as $T \md s/\md r = \Omega r^3 \md \Omega/ \md r$ \citep{blandford1999fate}.
This relation suggests that, as the flow approaches the hydrostatic, isentropic atmosphere, $\Omega(r)$ approaches a constant, $\Omega_{\rm atm} \sim (GM/r_{\rm atm}^3)^{1/2}$, a result that our simulations largely confirm (Section \ref{sec:results}).
As a result, the material reaching the NS surface will carry
much less angular momentum than would be expected in a quasi-Keplerian disk.

The atmosphere entropy $s_{\rm atm}$ takes on a characteristic value for the range of accretion rates considered here. Outside a thin surface neutrino cooling layer that merges with the outer layers of the NS, the equation of hydrostatic equilibrium gives
a specific enthalpy $h \simeq h(r_{\rm atm}) + GM(r^{-1} -r_{\rm atm}^{-1}) \simeq GM/r$,
a temperature $T = h/s_{\rm atm} \simeq s_{\rm atm}^{-1}GM/r$, and a density $\rho = 4p/h =                                  
(4f_\pm a/3)(GM/r)^3s_{\rm atm}^{-4}$.
Balancing the released gravitational energy and neutrino energy, $GM\dot M_{\rm atm}/r_\star = \int_{\rm r_\star} \md r 4\pi r^2  \md^2 E_\nu/ \md V \md t$, the
integral over radius is dominated by a shell of thickness $\sim r/7$.  A full integration of the coupled
energy and hydrostatic equations gives (when electron capture cooling dominates; Appendix \ref{sec:cooling_layer})
\begin{equation}\label{eq:svsmdot}
  m_p s_{\rm atm} = 46\,\left({\dot M_{\rm atm}\over 10^{-3}\,M_\odot~{\rm s^{-1}}}\right)^{-1/10} M_{1.4}^{4/5}
  r_{\star,6}^{-1/2}
\end{equation}
and an inflow rate
\bea\label{eq:vratm}
  |v_r| &=& {\dot M_{\rm atm}\over 4\pi r^2\rho} \nnn
  &=& 6.3\times 10^{-3}c\,\left({r\over r_\star}\right)
  \left({\dot M_{\rm atm}\over 10^{-3}\,M_\odot~{\rm s^{-1}}}\right)^{3/5}{M_{1.4}\over r_{\star,6}}. \nnn
\eea
The atmosphere is present when the mass transfer rate in the inner parts of the RIAF drops below
$\sim 10^{-2}\,M_\odot~{\rm s}^{-1}$ with $\alpha \sim 0.1$. 
Electron-positron cooling dominates when the accretion rate
drops below $\sim 10^{-5}\,M_\odot$ s$^{-1}$.

The rate $\dot M_{\rm atm}$ at which the NS receives mass is connected in a nonlinear way with the structure of the outer RIAF and its accretion rate, $\dot M_{\rm d}(r)$.  Representing the outer flow by a characteristic (ingoing) accretion rate $\dot M_0$
at a radius $r_0 \gg r_\star$, decreasing inward as $\dot M_{\rm d}(r) = \dot M_0(r/r_0)^{1/2}$ (a power-law dependence suggested in inflow-outflow models for RIAFs, see \cite{blandford1999fate}),
the quantities $\dot M_{\rm atm}$ and $r_{\rm atm}$ are readily determined
by matching $\dot M_{\rm atm} = \dot M_{\rm d}(r_{\rm atm})$ and
$|v_r(r_{\rm atm})| = \alpha\Omega(r_{\rm atm})r_{\rm atm}$.  In this situation, $e^\pm$ annihilation
cooling tends to dominate near the NS and Eq.~\eqref{eq:svsmdot} changes slightly to 
$m_p s_{\rm atm} = 83\,(\dot M_{\rm atm}/10^{-6}\,M_\odot~{\rm s}^{-1})^{-1/9} M_{1.4}^{8/9}
r_{\star,6}^{-5/9}$.  We obtain
\begin{equation}\label{eq:Ratm}
  {r_{\rm atm}\over r_\star} = 94\,\alpha_{-1}^{1/4}r_{0,10}^{5/8}
  \left({M_0\over 10^{-2}\,M_\odot}\right)^{-5/16};
\end{equation}
\beq{
  \dot M_{\rm atm} = 7\times 10^{-7}\,{\alpha_{-1}^{9/8}\over r_{0,10}^{27/16}}
  \left({M_0\over 10^{-2}\,M_\odot}\right)^{27/32}\,  M_\odot~{\rm s}^{-1}.
}

{The accretion model and simulations we present here are applicable in astrophysical scenarios where (i) the photons remain trapped in the flow and (ii) neutrinos
carry away a small proportion of the internal energy from the inner parts of
a RIAF.
The trapped mass $M_0$ must lie below a critical value $M_0^{\rm ad}$ for neutrino cooling
to be weak.  This depends on the radius $r_0$ where the flow becomes
centrifugally supported;  $M_0^{\rm ad}$ is pushed below $0.1\,M_\odot$ when
$r_0$ is smaller than $10^8$ cm (see Equation (\ref{eq:nueff}) below).
The combination of weak neutrino cooling and strong photon
trapping will hold for a wide range of $M_0$ below $M_0^{\rm ad}$, with the caveat
that magnetocentrifugal friction from a rotating, magnetized NS 
will modify the flow when $M_0$ is too small (Section \ref{sec:implications}).}

The condition that photons are strongly tied to the accreted material is
readily checked.  Consider a geometrically thick torus that has accreted
close to the NS with small net binding energy.  The torus
has partial hydrostatic support and a radial pressure gradient
$dp/dr \sim -\rho GM/2r^2$.   Given a density $\rho(r) \sim
GM_0/2\pi r_0^2r$ and pressure $p(r) \sim GM\rho/4r$
at radii $r < r_0$, the photon pressure
exceeds the gas pressure $\rho k_{\rm B}T/\mu_g$ 
when the trapped mass is lighter than
$M_0 \lesssim 0.017\,M_{1.4}^3 (\mu_g/m_p)^4\,M_\odot$.  This is
independent of $r_0$ but depends strongly on the stellar mass 
and the gas mean molecular weight $\mu_g$.
The temperature is high enough for H and He to be fully ionized,
$T = 5\times 10^7\,M_{1.4}^{3/4}r_{0,10}^{-1}(M_0/10^{-2}M_\odot)^{1/4}
(r/r_0)^{-1/2}$ K.  The radiative diffusion time as limited by
electron scattering (opacity $\kappa_{\rm es} \sim \sigma_{\rm T}/m_p$)
is $t_{\rm rad} \sim \kappa_{\rm es}\rho r^2/c$;  compared with
the radial flow time $t_{\rm flow} \sim (\alpha\Omega)^{-1}$, this is
\begin{eqnarray}\label{eq:trad}
{t_{\rm rad}\over t_{\rm flow}} &\sim&
\alpha {\kappa_{\rm es}M_0\over 2\pi r_0^2}\left({r\over GM/c^2}\right)^{-1/2}\nnn
&=&6\times 10^6\,\alpha_{-1}{M_{1.4}^{1/2}\over r_{0,10}^{5/2}}
\left({M_0\over 10^{-2}\,M_\odot}\right)\,\left({r\over r_0}\right)^{-1/2}.\nnn
\end{eqnarray}
The corresponding entropy per baryon is $m_p s = 4.6\,M_{1.4}^{3/4}(M_0/10^{-2}M_\odot)^{-1/4}
(r/r_0)^{-1/2}\,k_{\rm B}$.

The weakness of neutrino cooling is readily checked when the flow
maintains this RIAF structure all the way down to the NS.
The energy lost to neutrinos, normalized
to the internal heat of the flow, is
\begin{equation}
{Q\, t_{\rm flow}\over h \rho} =
{K_\nu T^6(1+m_ps/60k_{\rm B})\over Ts\cdot\alpha\Omega}.
\end{equation}
This expression remains small when extrapolated close to the NS,
because $s$  and $T = GM/r s$ both diverge slowly 
inward as $\sim r^{-1/2}$.  One finds, substituting Equation (\ref{eq:qnu}),
\begin{equation}\label{eq:nueff}
{Q\,t_{\rm flow}\over h\rho} = 5\times 10^{-10}
\left(1+{m_ps\over 60k_{\rm B}}\right)
\,{(M_0/10^{-2}M_\odot)^{3/2}\over \alpha_{-1}r_{10}^{1/2} r^3_{0,10}}.
\end{equation}
The main assumption underlying this estimate is that the electrons in
the flow are non-degenerate. 

{Once the hydrostatic atmosphere settles, the neutrino emission is concentrated in a thin layer near the NS.  Neutrino losses have a stronger
feedback on the flow dynamics in binary NS mergers and the collapsing cores of
massive stars, when gas is assembled near the NS and accretes initially at $\dot{M} \gtrsim  10^{-2} M_{\odot} \, s^{-1}$.}

\section{Simulation set-up}\label{sec:setup}

\subsection{Evolution Code and Grid Setup}

We perform numerical simulations of accretion from a magnetized fluid torus in orbit around a non-rotating and unmagnetized star in General Relativity using the code {\sc HARM3D} \citep{noble2009DIRECT, noble2011RADIATIVE}.
This code is a version of {\sc HARM} \citep{gammie2003harm}; it solves the flux-conservative ideal GRMHD equations in arbitrary coordinates and metric. We evolve a magnetized fluid assuming an ideal gas equation of state, $p=(\Gamma-1)u$
for pressure $p$ and internal energy density $u$;  the adiabatic index $\Gamma=4/3$ corresponds to pressure dominated by photons and relativistic pairs.  We use piecewise parabolic interpolation for reconstruction and a Harten-Lax-van Leer-Einfeldt (HLLE) Riemann solver \citep{harten1983upstream}. The magnetic field is evolved maintaining the solenoidal constraint using the constrained transport method \citep{toth2000Constraint}, with a piecewise parabolic interpolation of the electromotive forces (EMFs). We enforce radially-dependent minimum values for the energy density $u$ and proper mass density $\rho$
in combination with the 2D conservative-to-primitive variable conversion developed in \cite{noble2006primitive}. 

The uniform numerical grid is mapped to spherical Boyer-Lindquist coordinates, $(r,\theta, \phi)$, using an exponential function for the radial coordinate, a slightly stretched $\theta$ coordinate using the transformation in \cite{gammie2003harm} with parameter $h=0.7$, and uniform $\phi$ coordinate (Figure \ref{fig:id}).  At the resolutions considered here, we find that a spherical grid is necessary to avoid excitation of strong azimuthal $m=4$ density modes, which are observed in simulations of stars in Cartesian coordinates \citep{ott_dynamics_2011, takasao2025connecting}. In our experience, this problem is exacerbated when the density contrast between the stellar surface and the accretion flow becomes large, leading to numerical spiral shocks that can dominate angular momentum transport.

The fiducial resolution is $n_{r}\times n_{\theta} \times n_{\phi} = 400 \times 256 \times 160$, {which allows us to resolve the magneto-rotational instability (MRI) in the disk; see Appendix \ref{sec:convergence}. We also perform an additional, shorter run at a higher azimuthal resolution using $n_{r}\times n_{\theta} \times n_{\phi} = 400 \times 256 \times 320$ (suffix \texttt{HR} in Table~\ref{tab:simulations}), with a goal
of testing numerical convergence in the hydrostatic atmosphere around the NS (see Appendix \ref{sec:convergence}).

\subsection{Boundary Conditions and Neutron Star}

The NS is represented explicitly in the center of the simulation domain. We first solve the Tolman-Oppenheimer-Volkoff (TOV) equations assuming (an adiabatic) piecewise polytropic equation of state, $p(\rho) = K_i \rho^{\Gamma_i}$, where $\lbrace \Gamma_1, K_1 \rbrace = \lbrace 4/3, 0.08951\,[K_{1}] \rbrace$ if $\rho < \rho_0 = 9.13\times10^{13}$ g cm$^{-3}$ and $\lbrace \Gamma_2, K_2 \rbrace = \lbrace 3,170427\, [K_{2}] \rbrace$ if $\rho >\rho_0$; the adiabatic coefficients are $[K_{i}] = \textrm{g}^{1-\Gamma_i} \textrm{cm}^{2+3\,(\Gamma_i-1)}\textrm{s}^{-2}$ in cgs units. We choose a central density of $\rho_{\rm ns}=5.2 \times 10^{14}$ g cm$^{-3}$, corresponding to a gravitational mass of $M=1.4\:M_{\odot}$ for the NS; in terms of gravitational radius, the star is at $r_\star=6.4 \,GM/c^2\approx 13.4$\,km. 

The same EOS is adopted in the accretion flow and a thin boundary layer connecting it with the NS;  as a result,
we place only a thin outer stellar layer of subnuclear density (the ``crust'') in the computational domain. 
This layer is is restricted to $\rho<\rho_0$, corresponding to a radius larger than $r_{\rm in}=6.2\,GM/c^2$ in the TOV solution.  We use fixed inner boundary conditions at $r_{\rm in}$ using the hydrostatic TOV solution values at that radius and outflow boundary conditions at $r_{\rm out} = 800\,GM/c^2$.  The portion of the star covering $r \in (r_{\rm in}, r_\star)$ is placed and evolved explicitly in the domain, and is initially resolved by 10 radial points. We found that it is important to set these boundary values appropriately with the TOV solution to obtain a stable, hydrostatic inner boundary.

The spacetime is described by the Schwarzschild metric, which is a good approximation because most of the mass sits inside the boundary radius $r_{\rm in}$. We use periodic boundary conditions in coordinate $\phi$ and transmissive boundary conditions at the spherical polar axes. 

\subsection{Torus Initial Data}
\label{sec:id}

We initialize an axisymmetric hydrostatic gas torus with constant specific angular momentum as described in \cite{fishbone1976relativistic}. The torus has an inner edge at $r_{\rm in} =20\: GM/c^2 = 2.857 \: r_{*}$ and maximum pressure is attained at $r_{p,\rm max} = 40 \:GM/c^2 = 5.714 \:r_{*}$, which sets the outer edge of the torus at $r_{\rm out}=45.7\:r_{*}$ (Figure
\ref{fig:id}).  The torus density is normalized so that the maximum density is much lower than the crustal boundary,  $\rho_{\rm max} = 10^{-6} \rho_0$. 

The torus is threaded by a seed poloidal loop of magnetic field, which is set by the azimuthal vector potential 
\begin{equation}\label{eq:Aphi}
    A_{\phi} = A_{\rm b} \: \textrm{max}(q, 0).
\end{equation}
Following previous GRMHD simulations modeling RIAFs onto BHs (e.g. \citealt{tchekhovskoy_efficient_2011}), we choose two different expressions for $q$.  One represents a small loop,
\begin{equation}
    q_{\rm small} = \frac{\rho}{\rho_{\rm max}} -0.2,
\end{equation}
the other a large loop,
\begin{equation}
    q_{\rm large} = \frac{\rho}{\rho_{\rm max}} \sin^3\theta \Big(\frac{r}{r_{\rm in}}\Big)^3 \exp(-r/400) -0.2,
\end{equation}
with $r$ in units of $GM/c^2$.  The choice of $q_{\rm large}$ corresponds to an increased magnetic flux threading the inner torus,
in comparison with the small-loop configuration.  If the central object is a BH, this leads to a so-called magnetically arrested disk (MAD) state around the BH.
The coefficient $A_{\rm b}$ in Equation (\ref{eq:Aphi}) is fixed 
by the minimum value of plasma $\beta$ in the torus, where $\beta=2p/b^2$, and $b=\sqrt{b^\mu b_\mu}$ denotes the co-moving magnetic field strength. We choose $\beta = 100$ in all simulations, except for $\beta = 50$ in the \texttt{LargeLoop}\_$\beta50$ setup (see Table~\ref{tab:simulations}). Furthermore, we introduce small random pressure perturbations in the initial torus configuration to trigger the MRI, which mediates angular momentum transport and accretion onto the NS.

\begin{table*}[ht]
\caption{Summary of the setup of all simulation runs presented in this paper. From left to right: configuration name, number of grid points $n_r \times n_\theta \times n_\phi$, cooling prescription (see Section~\ref{sec:cooling}), physical evolution time in units of the Keplerian period at the star's surface, and initial seed magnetic field setup with maximum plasma $\beta$ (see Section~\ref{sec:id}).} 
\begin{center}
\begin{tabular}{lllll}
\hline\hline
Configuration & Resolution       & Cooling                            & Duration             & Initial $\mathbf{B}$-field                     \\ \hline
\texttt{Fiducial}            & $400\times 256 \times 160$ & No                                           & 750 $P_{\rm Kep\star} (8\times10^4\:GM/c^3)$ & $\beta=100,\: A_{\phi} \propto q_{\rm small}$ \\ 
\texttt{Cooling}           & $400\times 256 \times 160$ & $Q(r_\star) \propto 10^{-1}\times \dot{M}_{\rm Ad}(5 r_\star)$                    & 750 $P_{\rm Kep\star} (8\times 10^4 \:GM/c^3)$ & $\beta=100,\: A_{\phi} \propto q_{\rm small}$ \\ 
\texttt{StrongCooling}       & $400\times 256 \times 160$ & $Q(r_\star) \propto 10^{2}\times \dot{M}_{\rm Ad}(5 r_\star)$ & 450 $P_{\rm Kep\star} (6 \times 10^4\:GM/c^3)$ & $\beta=100,\: A_{\phi} \propto q_{\rm small}$ \\ 
\texttt{LargeLoop\_$\beta100$}           & $400\times 256 \times 160$ & No                                           & 750 $P_{\rm Kep\star} (8\times10^4\:GM/c^3)$ & $\beta=100,\: A_{\phi} \propto q_{\rm large}$ \\ 
\texttt{LargeLoop\_$\beta50$}           & $400\times 256 \times 160$ & No                                           & 750 $P_{\rm Kep\star} (8\times10^4\:GM/c^3)$ & $\beta=50,\: A_{\phi} \propto q_{\rm large}$ \\ 
\texttt{LargeLoop\_$\beta100$\_HR}        & $400\times 256 \times 320$ & No                                           & 200 $P_{\rm Kep\star} (8\times10^4\:GM/c^3)$ & $\beta=100,\: A_{\phi} \propto q_{\rm large}$ \\
\hline
\end{tabular}
\end{center}
\label{tab:simulations}
\end{table*}

\begin{figure}
    \centering
    \includegraphics[width=1\linewidth]{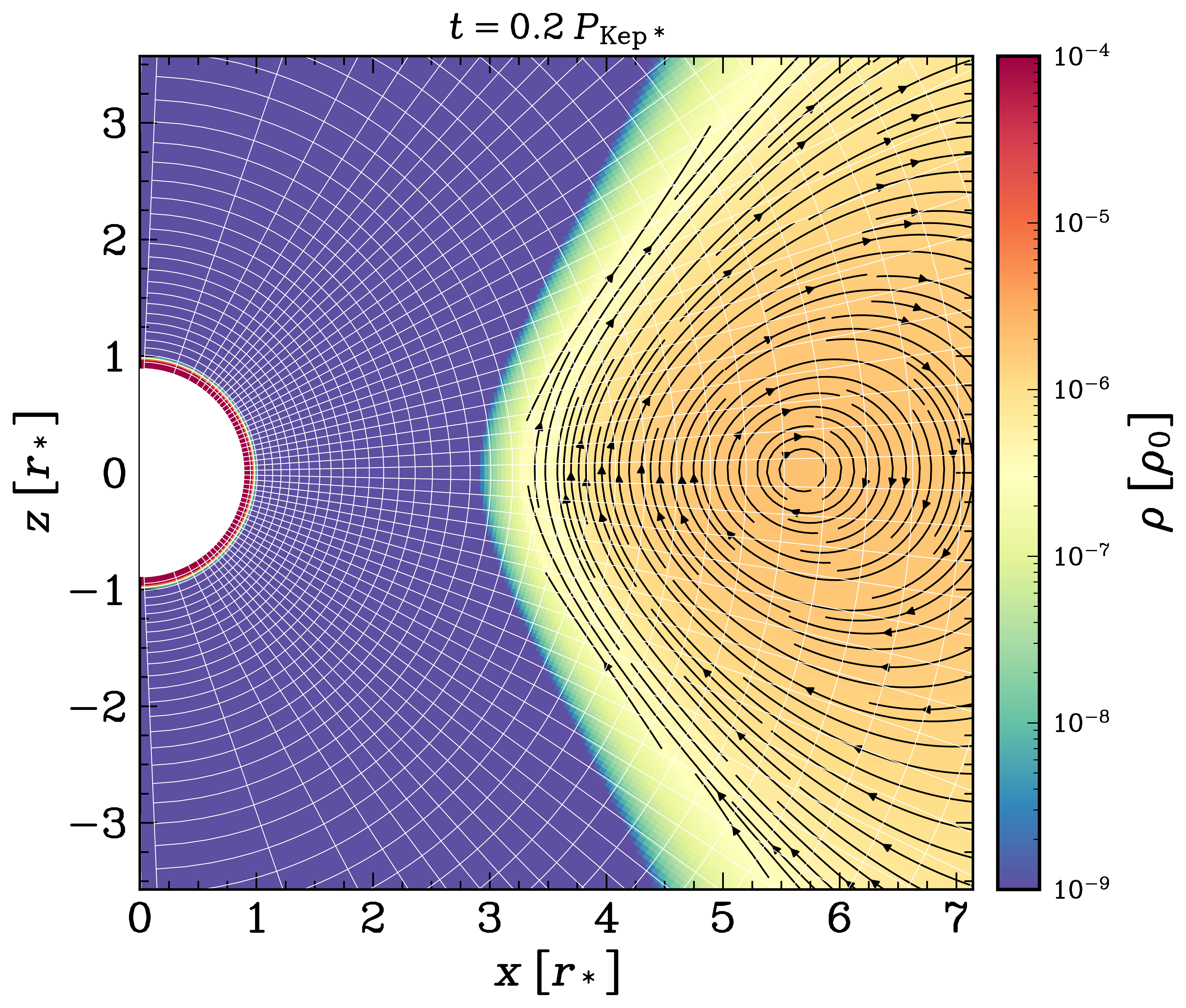}
    \caption{Meridional slice through the initial data of the \texttt{Fiducial} simulation, showing the grid (white lines) with the initial proper
    mass density as well as magnetic field lines superimposed (black lines). The white circle marks the inner radial boundary. 
    }
    \label{fig:id}
\end{figure}

\subsection{Optically-thin Neutrino Cooling}
\label{sec:cooling}

Near the surface of the star, the settling flow can reach a high enough temperature to effectively 
release energy in neutrinos through 
relativistic $e^{\pm}$ capture on nucleons or pair annihilation. We include 
neutrino cooling as an optically thin, isotropic sink term in the energy-momentum conservation equation,
\begin{equation}
    \nabla_{a} T^{ab} = - Q u^b,
\end{equation}
where $u^a$ denotes the four-velocity.  The energy radiated away per unit volume and time is
\begin{equation}\label{eq:Qfunc}
    Q = {\mathrm{d}^2 E_\nu\over \mathrm{d}V\mathrm{d}t} = A_Q \rho p^{3/2},
\end{equation}
with $A_Q$ a normalization factor.  The dependence on $\rho$ and $p \propto T^4$
(in a relativistic $e^\pm-$photon gas) arises from the weak cross section 
($\propto T^2$).  In an approximately adiabatic atmosphere, this scaling 
is consistent with both neutrino emission channels;  see Equation (\ref{eq:qnu}).

{Before introducing a cooling function, the global flow density scale in our simulations is free. The 
flow acquires a density scale in our simulations through this loss of entropy (see Section 
\ref{sec:analytic} and Appendix \ref{app:density_scale} for details). 
Where the gas is hydrostatic and nearly isentropic, $Q$ has effectively the same
scaling with $\rho$ and $p$ whether cooling is dominated by $e^\pm$ captures or by $e^\pm$
annihilation;  in the latter case, $A_Q$ contains an additional factor dependent on entropy, $s$.
From Equations (\ref{eq:svsmdot}) and (\ref{eq:qnu}), 
$e^\pm$ capture dominates if $\dot{M} \gtrsim 10^{-4}\,M_{1.4}^8\,r_{\star,6}^{-5} M_{\odot}$ s$^{-1}$, whereas $e^{\pm}$-annihilation dominates 
at lower accretion rates.} 

The strong dependence of $Q$ on flow parameters allows for a large density contrast between
the non-radiative disk flow and a thin, outer cooling layer of the NS.   
In hydrostatic equilibrium, we expect $Q \propto \rho^{5/2}r^{-3/2} \propto r^{-9}$.  
This shows that the appearance of an extended hydrostatic envelope is insensitive to the 
precise value of $A_Q$.  Our choice of $A_Q$ is guided by taking the cooling time at the radius $r_\star$ 
to be a moderate fraction $\varepsilon_{\rm cool} \sim 0.1-0.2$ of the time $t_{\rm acc}$ for a significant fraction of the initial torus mass $M_{\rm tor}$ to be transferred to the envelope.
More precisely, neglecting general relativistic effects, we may combine the condition for quasi-steady mass accretion through the envelope,
$\dot M = -v_r 4\pi r^2\rho(r) \sim \text{const.}$, with the equation of hydrostatic equilibrium, $\partial_rp = 
-\rho GM/r^2$, and the equation of energy conservation,
\begin{equation}
Q + {\dot M\over 4\pi r^2}\partial_r\left({4p\over\rho} - {GM\over r}\right) = 0,
\end{equation}
to give $\md\ln\rho/\md\ln p = 3/4 + Q t_{\rm adv}/(\rho GM/r)$, where $t_{\rm adv} = r/4|v_r|$
is the local advection time.  Normalizing $Q$ at density $\rho_\star$ and pressure $p_\star
\sim \rho_\star GM/4r_\star$ at the base of the envelope, we estimate $\dot M
\sim M_c/t_{\rm acc}$ and set
\begin{equation}
Q(\rho_\star,p_\star) = {3\over 4\varepsilon_{\rm cool}}{GM\dot M\over \pi r_\star^4};
\qquad \varepsilon_{\rm cool} \sim 0.1-0.2.
\label{eq:coolingest}
\end{equation}

Given a fixed rate of mass transfer $\dot M$ at the outer boundary (radius $r_{\rm atm}$)
of the hydrostatic atmosphere, choosing $\varepsilon_{\rm cool} < 1$ will
force an increase in $\rho_\star$ by a factor $\sim \varepsilon_{\rm cool}^{-1/2}$.  The boundary
radius decreases as $r_{\rm atm} \sim \varepsilon_{\rm cool}^{1/4}$ when the exterior RIAF
maintains a density profile $\rho(r) \propto r^{-1}$.

To normalize $Q$, we first evolve an adiabatic simulation without neutrino cooling
(denoted as $\texttt{Fiducial}$, see Table~\ref{tab:simulations}) for a duration $\approx 80000\, GM/c^3$, when the nearly hydrostatic atmosphere has expanded to $\sim 6\,r_{*}$.  At this time, we measure $\rho_\star$ and $p_\star$ at the NS surface and the accretion rate at $5 r_\star$ (approximately the size of the hydrostatic core). We then derive $Q(\rho_\star,p_\star)$ from Equation~\eqref{eq:coolingest} for $\varepsilon_{\rm cool} = 0.1$ and $r_\star= 7\,GM/c^2$. Subsequently, we perform a separate simulation, referred to below as \texttt{Cooling}, setting $A_Q = Q(\rho_\star,p_\star)/\rho_\star p_\star^{3/2}$. We also perform an additional run with shorter evolution time and enhanced cooling (\texttt{StrongCooling}), using a cooling function with the larger normalization $A_Q = 10^3 \times Q(\rho_\star,p_\star)/\rho_\star p_\star^{3/2}$. Our choices for the cooling normalization are meant to represent different accretion rate regimes found in various astrophysical scenarios (see the end of Section 2 and Section 5).

In addition to neutrino cooling, dissociation of nuclei (if present) acts as another source of cooling for the accretion flow.  As the binding energy per nucleon of $\simeq 160\,\text{MeV} (M_{1.4}/r_{\star,6})$ is much larger than the dissociation energy of $\approx \!8$\,MeV per nucleon for heavier nuclei, we may safely ignore this cooling source.

\subsection{Simulation Diagnostics}

Quantities integrated over a spherical surface of radius $r$ are denoted by
\begin{equation}
    \lbrace \mathcal{Q} \rbrace \equiv \int \mathcal{Q}\: \sqrt{-g}\, \mathrm{d}\theta \mathrm{d}\phi,
\end{equation}
where $g(r,\theta)$ is the determinant of the metric. We define the spherical average of a quantity $\mathcal{Q}$ as
\begin{equation}
    \langle \mathcal{Q} \rangle \equiv \frac{\lbrace \mathcal{Q} \rbrace}{A},
\end{equation}
where the area $A\equiv  \int \sqrt{-g}\, \mathrm{d}\theta \mathrm{d}\phi$ and the integrals are
performed by summing over grid cells.
In Schwarzschild coordinates, $\sqrt{-g} = r^2 \sin^2\theta$ and $A = 4\pi r^2$.  
Averages weighted by
fluid energy are similarly computed via $\langle \mathcal{Q} \rangle_{\rho h} \equiv \langle \mathcal{Q} \rho h \rangle/\langle \rho h \rangle$.  An angular average $\langle Q\rangle$ may be further averaged 
over a finite time interval according to
\begin{equation}
\langle \mathcal{Q} \rangle_t \equiv (\Delta t)^{-1}\int_t^{t+\Delta t} \langle \mathcal{Q} \rangle (t') \mathrm{d}t' .
\end{equation}

We display flow diagnostics with time normalized by the Keplerian orbital period extrapolated to the NS surface, 
$P_{\rm Kep\star} = 2\pi/\Omega_{\rm Kep}(r_\star)$, where $\Omega_{\rm Kep}(r) = (GM)^{1/2} r^{-3/2}$.\\
\\
\section{Simulation Results}\label{sec:results}
 
\subsection{Flow Structure and Formation of a Nearly Hydrostatic Atmosphere}
\label{sec:atmosphere}

\begin{figure*}[t]
    \centering
    \includegraphics[width=0.49\linewidth]{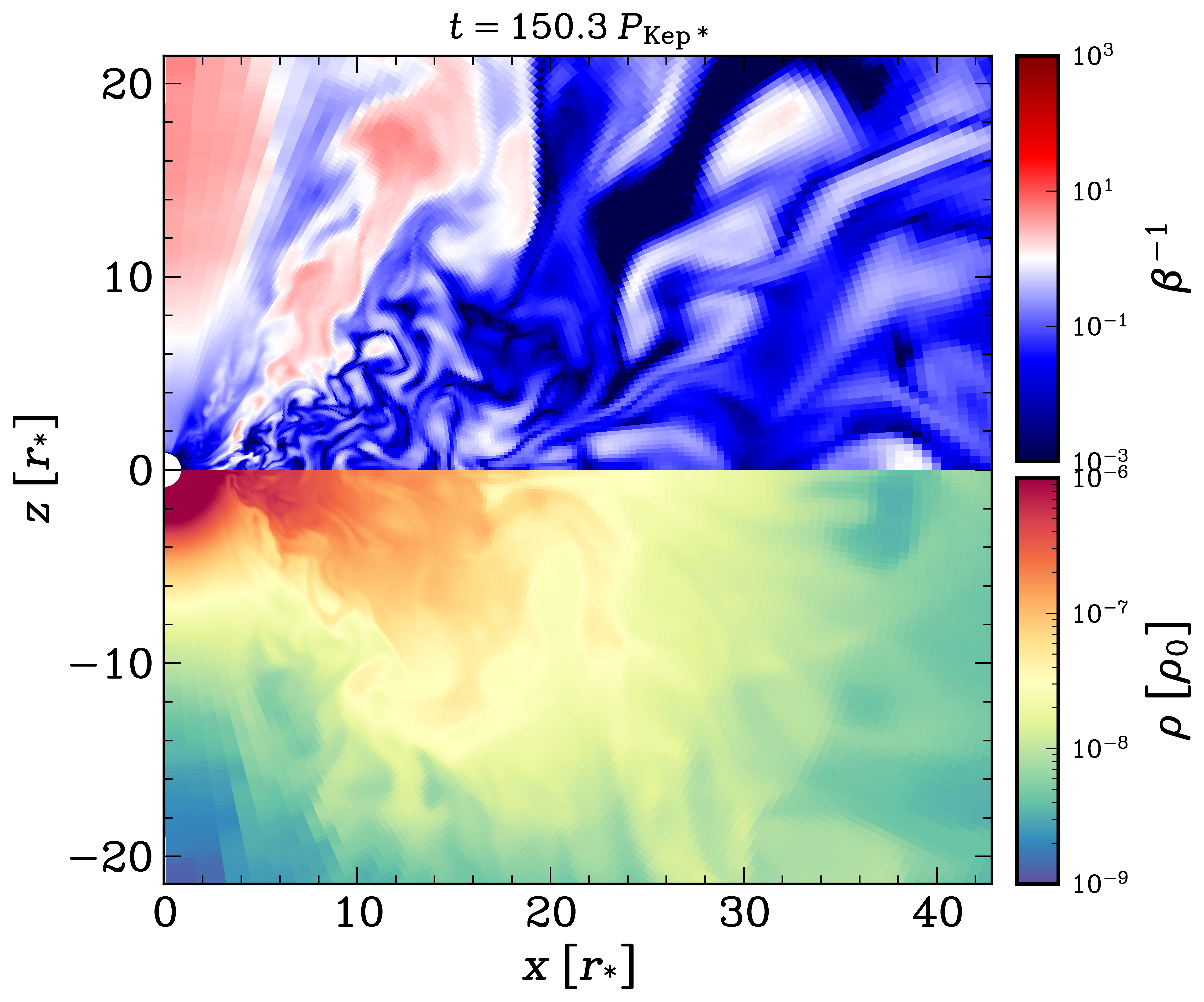}
    \includegraphics[width=0.49\linewidth]{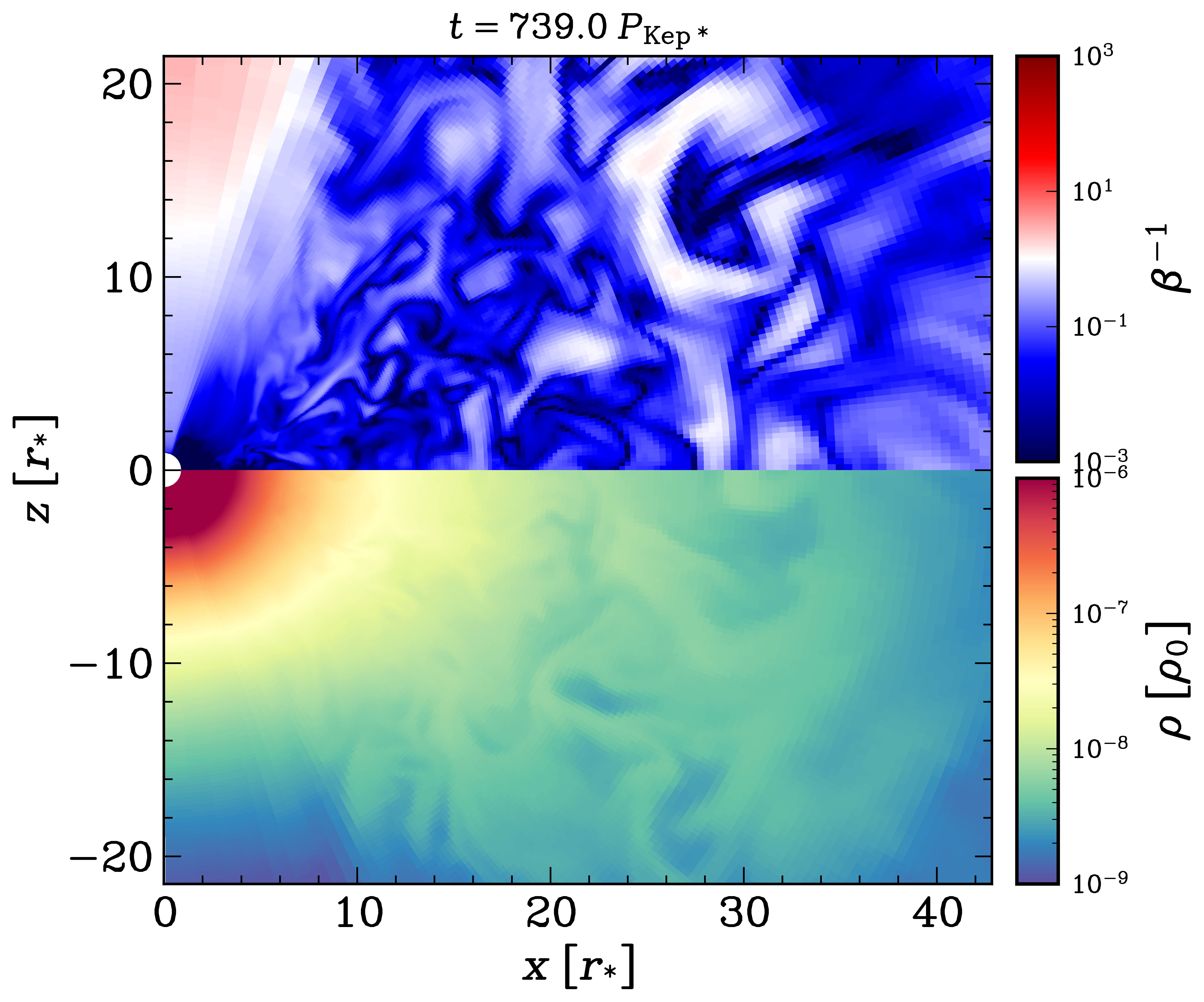}
    \caption{Vertical slices through the meridional plane showing rest-mass density (lower panel) and the ratio of magnetic to fluid pressure $\beta^{-1}$ (upper panel) for the \texttt{Fiducial} simulation at an early time (left) and late time (right). The inner part of the flow is in approximate hydrostatic balance, growing in mass as time evolves. 
    MRI-generated turbulence is prevalent outside the atmosphere, 
    with $\beta^{-1} \sim 0.1$ in the saturated state, while 
    $\beta^{-1}<10^{-2}$ in the inner atmosphere.
    }
    \label{fig:rhoinvb_xz}
\end{figure*}

\begin{figure}[t]
    \centering
    \includegraphics[width=1\linewidth]{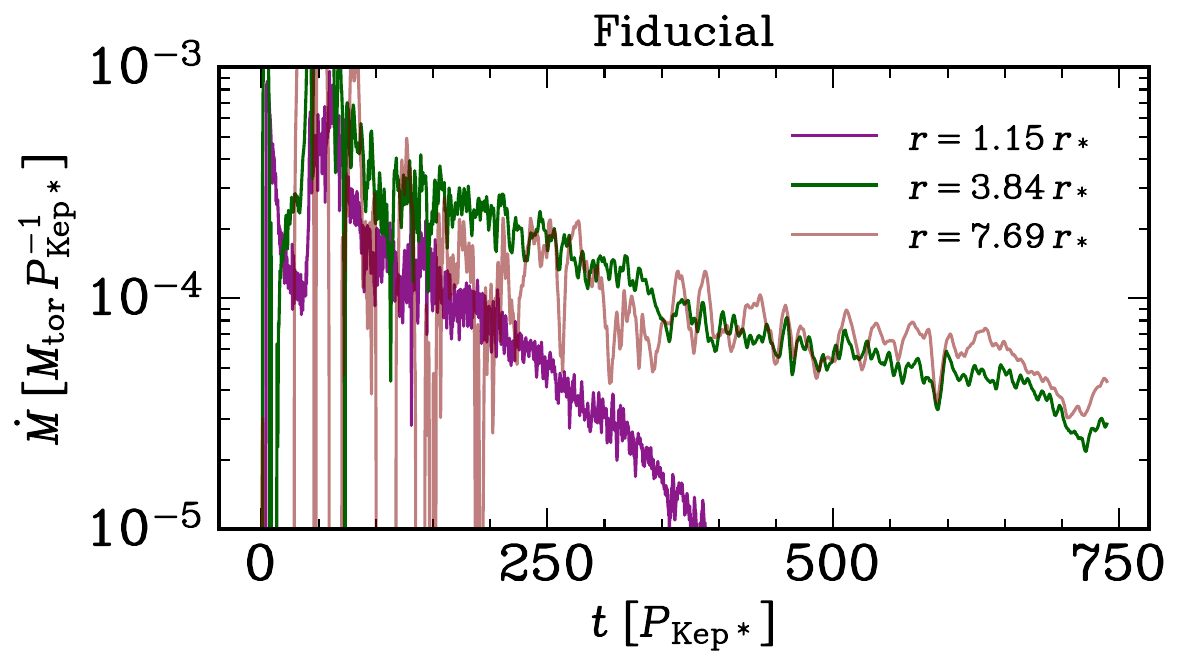}
    \caption{Accretion rate $\dot{M}$ evaluated at different radii in the \texttt{Fiducial} simulation (without neutrino cooling).
    The plotted value is normalized by $M_{\rm tor}/P_{\rm Kep\star}$;
    it increases by a factor 15.6 when $P_{\rm Kep}$ is evaluated at the 
    pressure maximum of the initial torus ($r_{p,\rm max} = 40 \: GM/c^2$).  Closer to the NS, the accretion rate drops rapidly with time (purple curve), as the atmosphere approaches hydrostatic equilibrium.}
    \label{fig:mdot_t_sane}
\end{figure}

The shearing motion of the gas ($\mathrm{d}\Omega / \mathrm{d}r < 0$), aided by the seed pressure perturbation, triggers the onset of the MRI in the first few tens of rotational periods. The bulk of the accretion flow becomes turbulent, and magnetic field amplification quickly saturates in a state with $\langle \beta^{-1} \rangle \sim 0.1$ (see Figure~\ref{fig:rhoinvb_xz}, upper panel). As a result, MHD stresses start driving sustained mass accretion towards the NS from $t=50\,P_{\rm Kep\star}$ onward. 
The accretion rate
\begin{equation}\label{eq:mdot}
    \dot{M} = -\int \rho u^r \sqrt{-g}\, \mathrm{d}\theta \mathrm{d}\phi,
\end{equation}
{is extracted at a range of distances from the NS as a function of time} (see Figure~\ref{fig:mdot_t_sane} for the \texttt{Fiducial} simulation). 

\begin{figure}[t]
    \centering
    \includegraphics[width=1\linewidth]{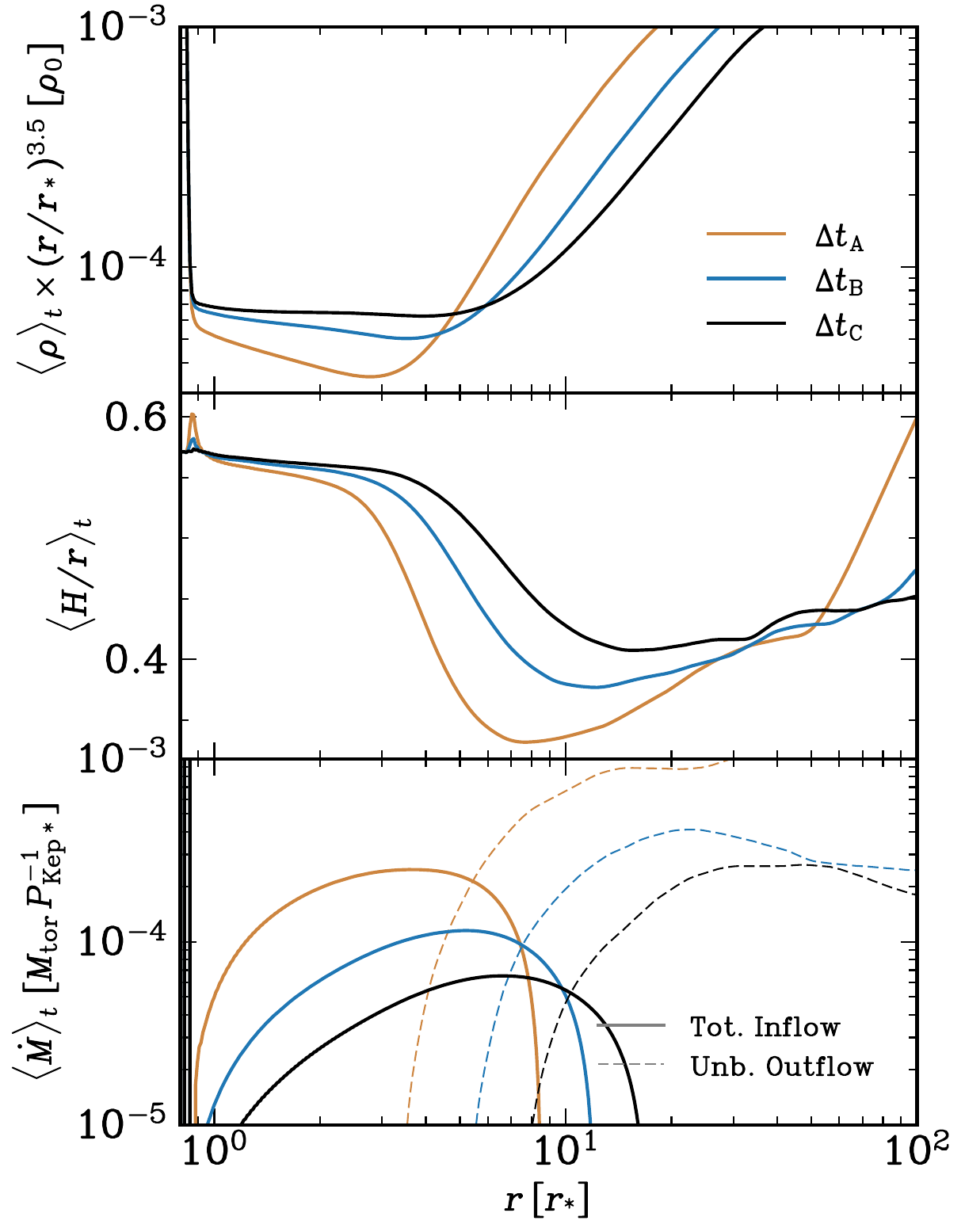}
    \caption{Various quantities versus radius $r$ in the \texttt{Fiducial} 
    simulation, as recorded in three successive time intervals. 
    Top to bottom: spherically averaged proper mass density, scale height, and accretion rate (here, thick lines represent total inflow accretion rate, and dashed lines represent unbound outflows). Quantities are further averaged over time in the intervals 
    $\Delta t_{\rm A} = [90, 270]\,P_{\rm Kep\star}$, $\Delta t_{\rm B} =
    [270, 449]\,P_{\rm Kep\star}$, and $\Delta t_{\rm C} = [449, 750]\,P_{\rm Kep\star}$.
    The inner hydrostatic core is characterized here by a strong drop in $\dot{M}$, an almost spherical and constant scale-height, and a density profile $\rho\propto r^{-3.5}$.
    }
    \label{fig:qvsr}
\end{figure}

In the boundary layer between the accretion flow and the NS surface, the hot flow quickly spreads in latitude (see Figure~\ref{fig:rhoinvb_xz}, left bottom panel). 
As matter accumulates around the NS surface, the increasing thermal pressure slows down the infalling gas and kinetic energy is dissipated into heat.  As a result, the rate of mass accretion decreases close to the star 
(see the purple curve in Figure~\ref{fig:mdot_t_sane}).  Gas above the NS surface settles
into a slowly rotating, nearly spherical, and mostly hydrostatic \textit{atmosphere}. 

Figure \ref{fig:qvsr} illustrates the atmosphere structure as characterized by the sphere-averaged
density $\langle \rho \rangle_t(r)$, which is further averaged over a sequence of narrow
time intervals.  This quantity follows a power-law $\langle \rho \rangle_t \propto r^{-3.5}$ out
to $r_{\rm atm} \sim 3-5$ stellar radii.  The atmosphere grows as more mass is accreted from the disk.  This density profile is compatible with an isentropic hydrostatic structure, 
$\rho_{\rm eq} \sim r^{1/(\Gamma-1)} = r^{-3}$ with $\Gamma=4/3$, slightly modified by the 
presence of rotation.  Outside the atmosphere, the density profile transitions to a softer 
$\langle\rho\rangle_t \sim r^{-1}$ scaling typical of a RIAF disk.

The part of the 
accretion flow interacting with the NS surface only briefly sustains a near-Keplerian angular momentum. 
Once the nearly hydrostatic atmosphere emerges and expands, its rotation slows to a nearly uniform angular 
velocity  $\Omega_{\rm atm}$ that is comparable to $\Omega_{\rm Kep}$ at the transition radius $r_{\rm atm}$.
Figure \ref{fig:omegar} shows 
$\langle \Omega \rangle_t \equiv \langle \rho h u^\phi \rangle_t /\langle \rho h u^t \rangle_t$ 
in the same time intervals
as in Figure \ref{fig:qvsr}.  A small positive $\md \langle\Omega\rangle_t/\md r$ develops inside the atmosphere, where $\langle \Omega\rangle_t \propto r^{1/2}$,
beyond which there is a transition to a nearly Keplerian profile, $\langle\Omega\rangle_t \propto r^{-3/2}$.
As the atmosphere expands, $\Omega_{\rm atm}$ decreases, reaching 
$\sim 1 \%$ of the surface Keplerian angular velocity $\Omega_{\rm Kep\star}$ toward the end of the simulation.

To track the extent of the atmosphere, we define $r_{\rm atm}(t)$ as the outer radius where $\rho \times r^{3.5}$ is roughly constant (Figure~\ref{fig:qvsr}, top panel).  In the \texttt{Fiducial} simulation, 
this evolves as $r_{\rm atm} \sim t^{0.37}$ (Figure~\ref{fig:r_t}, upper panel, black line). By the end of the simulation, the atmosphere has expanded to $r_{\rm atm} \sim 5.5 r_\star$.  The atmosphere can be independently measured by tracking the maximum of 
$\Omega/\Omega_{\rm Kep\star}$.   Both this radius and $r_{\rm atm}$ have the same time scaling (Figure~\ref{fig:r_t}).
Although both azimuthal and radial velocities decrease toward the star, we find 
$\langle h \rho u^{\phi} \rangle \gg \langle h \rho u^r \rangle $ everywhere and at all times.

The final panel in Figure~\ref{fig:qvsr} shows the accretion rate (\ref{eq:mdot}) in the same three snapshots of
the \texttt{Fiducial} simulation,
both in the inner part (thick lines, where the flow is mostly bound, $-hu_t < 1$) and in the exterior
(dashed lines, where $-hu_t > 1$ and the flow is mostly unbound). 
The mass inflow rate $\langle\dot M\rangle_t$ reaches a maximum near the outer atmosphere, but in the absence of 
neutrino radiation drops rapidly toward the NS surface as the hydrostatic structure emerges.  
This is in striking contrast to BH accretion, where an inner absorbing boundary (the horizon) allows for an inflow equilibrium to be established from the horizon to increasingly larger radii as time evolves, provided there is sufficient mass in the initial disk configuration (e.g. \citealt{white2020, chatterjee2022flux}). In the presence of cooling, however, we show in Section~\ref{sec:coolingresults} that inflow equilibrium starting at the NS surface can indeed be reached.

The accretion rate generally decreases over time and its peak moves outward (see the lower and upper panels of Figure~\ref{fig:qvsr}). Given that the torus spreads outward due to the transport of angular momentum by the MRI, 
net inflow ($\dot{M}>0$) is established outside $\sim 10\,r_\star$ only at later times.  
We quantify this by tracking the radius $r_{\rm in-out}(t)$ marking the boundary between net inflow and outflow
(Figure~\ref{fig:r_t}, bottom panel). This inflow-outflow radius closely follows the expected viscous 
evolution of a thick disk with scale height $H \sim r$, sound speed $c_s \sim v_{\rm Kep}$,
and viscosity $\nu = \alpha c_s H$, namely $r \sim t \nu /r = (\alpha c_{\rm s}) (H/r) \, t 
\propto t r^{-1/2}$, and thus $r \propto t^{2/3}$ when $\alpha=\text{const.}$

\begin{figure}
    \centering
    \includegraphics[width=1\linewidth]{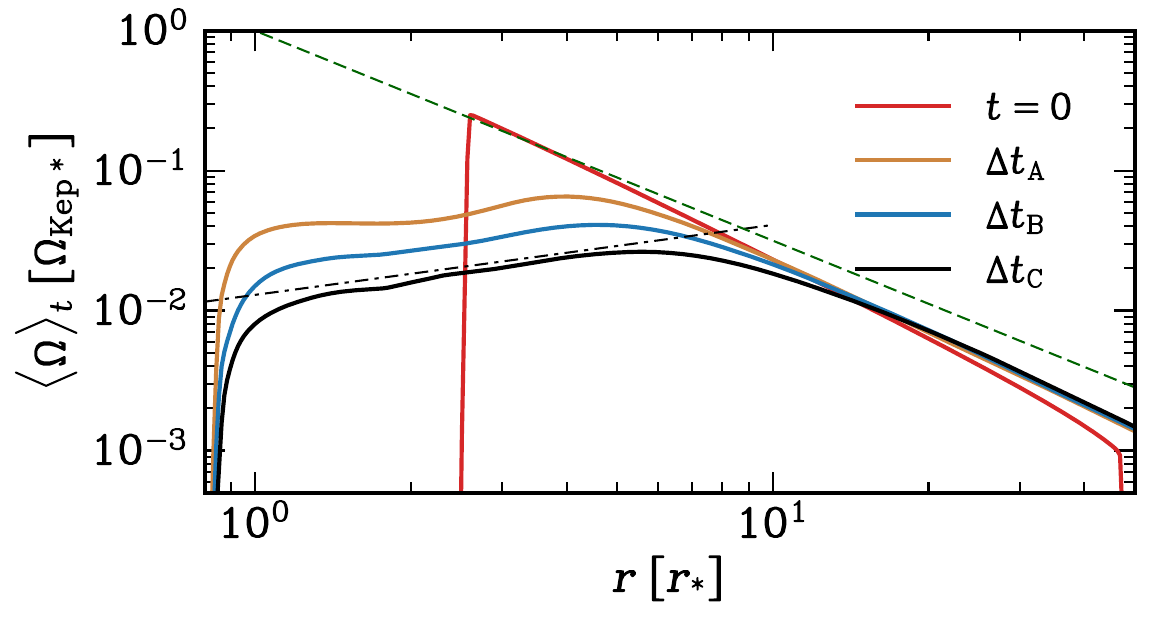}
    \caption{Angular velocity profile of the \texttt{Fiducial} simulation,
    at the same three epochs shown in Figure~\ref{fig:qvsr}.
    Added for reference: initial $\Omega$ profile corresponding to uniform specific angular momentum ($\sim r^{-2}$; red solid line), Keplerian profile ($\sim r^{-3/2}$; dashed green line) and profile inside the atmosphere ($\sim r^{1/2}$; dot-dashed black lines). As the atmosphere grows, the rotation of the atmosphere decreases as $\Omega_{\rm atm} \sim \Omega(r_{\rm atm})$.
    }
    \label{fig:omegar}
\end{figure}

Figure \ref{fig:qvst} reports the total mass contained in the region $r<3\,r_\star$. We observe a rapid increase in the first interval $100\,P_{\rm Kep\star}$, followed by a phase of slow growth.  By the end of the simulation, the total mass in this region plateaus at an accumulated $\sim\!10\%$ of the initial torus mass $M_{\rm tor}$. Modest deviations from spherical symmetry are present in the inner atmosphere.  The scale height is nearly
constant inside the atmospheric boundary, $H/r \equiv \langle |\theta-\pi/2| \rho \rangle_\rho \approx 0.6$,
(see the middle panel of Figure~\ref{fig:qvsr} and the density structure in Figure~\ref{fig:rhoinvb_xz}). 
This compares with $H/r\approx \pi/4\approx 0.79$ for an exactly spherical distribution. The inner hydrostatic core smoothly connects to the outer thick disk, which has a smaller average scale height, $H/r\approx 0.4$.  

The low Mach number atmosphere found here, in which $\mathcal{M} \sim |v^\phi|/c_{\rm s} < 1$ near the NS, is qualitatively different from the supersonic boundary/spreading layer regime expected in other accretion regimes \citep{belyaev2012angular,  philippov_spreading_2016, belyaev2018inefficient, coleman2022boundarya, coleman2022boundaryb}. When $\mathcal{M}\gtrsim1$ at the NS surface, acoustic waves, excited via hydrodynamical shearing instabilities, can efficiently transport angular momentum onto the star \citep{coleman2022boundaryb}. We do not observe evidence of such modes in our simulations, as expected. Furthermore, we do not observe gas mixing due to the Kelvin-Helmholtz instability, which could operate in hotter boundary layers ($\mathcal{M} \lesssim 1$; \citealt{miles1958disturbed}) in the presence of sharp rotational gradients e.g. at $r\sim r_{*}$. However, future simulations at much higher resolution than employed here, e.g., better resolving the scale height near the surface of the star, will be needed to establish whether or not such modes are present definitively.

\begin{figure}
    \centering
    \includegraphics[width=1\linewidth]{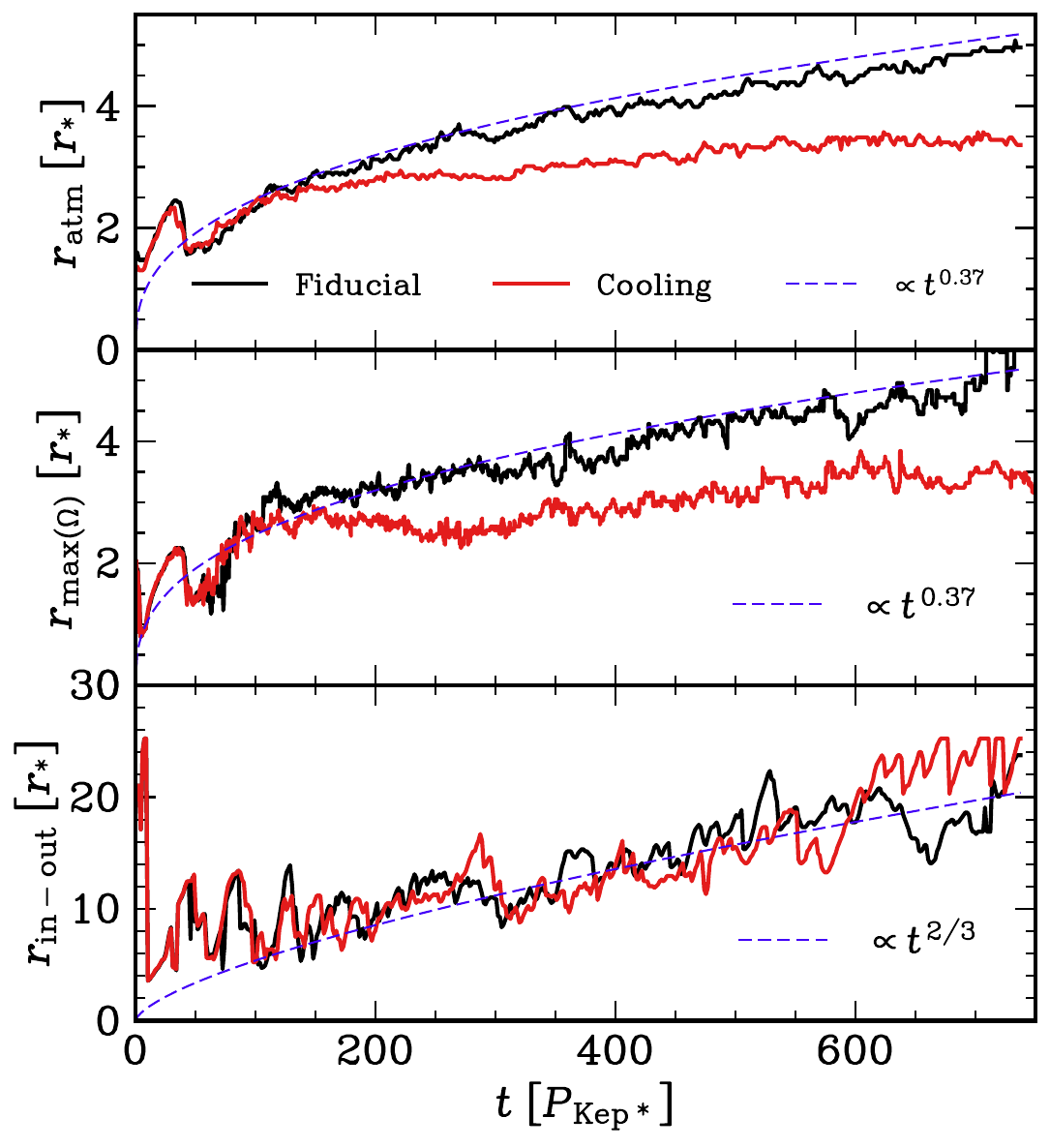}
    \caption{Growth of the nearly hydrostatic atmosphere as measured by the
    radius $r_{\rm atm}$ where $\md \ln\rho/ \md\ln r = -3.5$ (upper panel), the radius of maximum angular velocity $\Omega$ (middle panel), and the radius where the flow transitions from outflow to inflow (lower panel).  Whereas the atmospheric boundary as defined in any of these ways grows continuously in the adiabatic simulation \texttt{Fiducial} (black curves), growth slows down considerably once neutrino cooling becomes efficient in the \texttt{Cooling} simulation (red curves).
    }
    \label{fig:r_t}
\end{figure}

\begin{figure}
    \centering
    \includegraphics[width=1\linewidth]{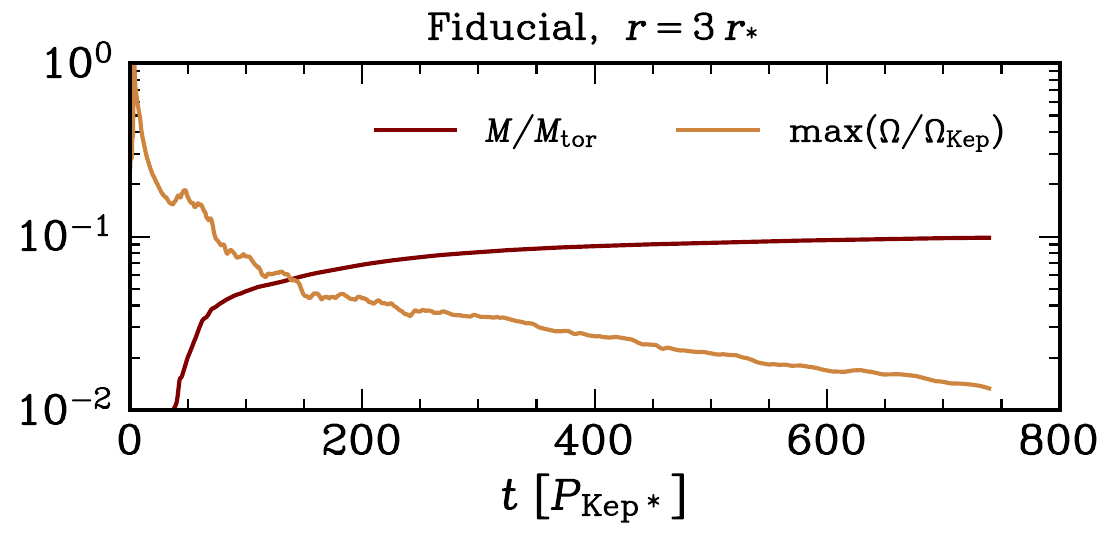}
    \caption{Net mass accreted (red solid line), and maximum value of $\Omega/\Omega_{\rm Kep}$ (orange line) within $r=3\,r_\star$. The inner part of the atmosphere accumulates the accreted mass until a nearly hydrostatic state is reached; the angular velocity keeps decreasing as the atmosphere expands.
    }
    \label{fig:qvst}
\end{figure}

The atmosphere is largely supported by thermal pressure, with a small contribution from magnetic stresses and 
negligible centrifugal support.  Figure \ref{fig:forces} shows the different components of the force density, as applied
on the sphere and normalized by the gravitational force.  (See Appendix \ref{sec:force} for a definition of these
components.)  At the outer edge of the atmosphere ($r\sim 5\,r_\star$) the centrifugal force becomes comparable to the thermal pressure gradient, as the flow transitions to the partially rotationally-supported RIAF. Turbulent (convective) pressure gradients are subdominant in the outer torus, and are negligible at smaller radii.

\begin{figure}
    \centering
    \includegraphics[width=1\linewidth]{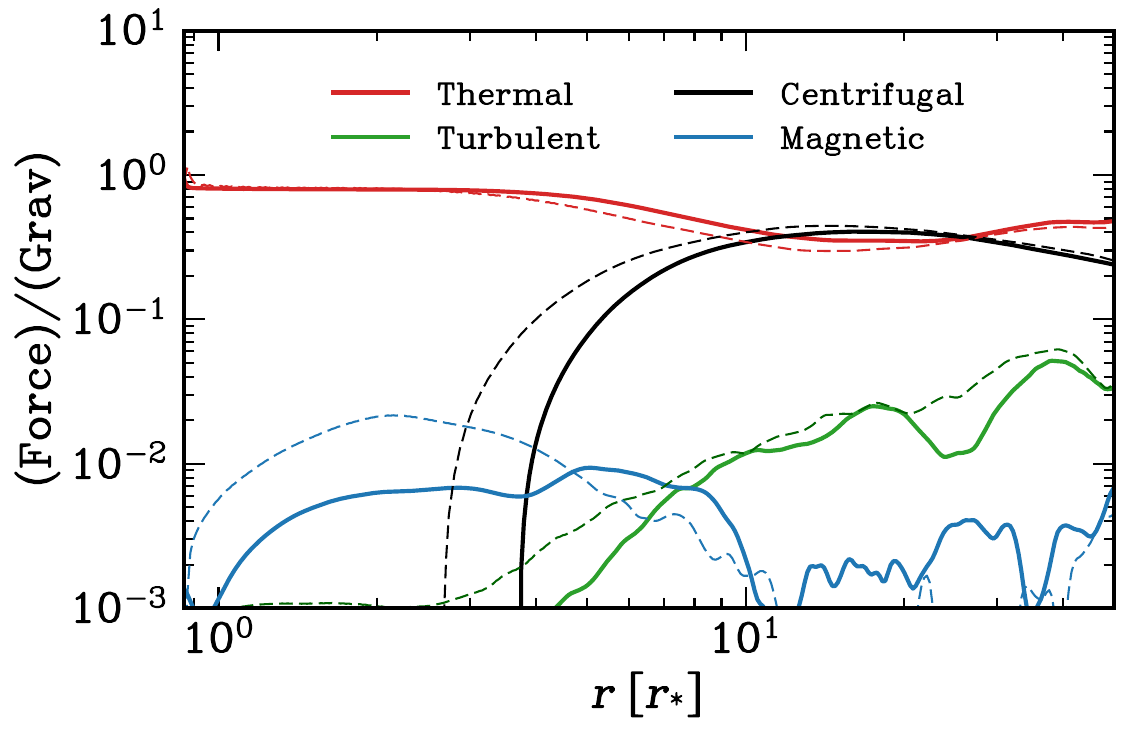}
    \caption{Spherically averaged pressure gradients acting on the accreting plasma, time-averaged over $\Delta t_{\rm C}= [449, 750]\,P_{\rm Kep\star}$ as in Figure~\ref{fig:qvsr} and normalized by the gravitational force; these include turbulent (Reynolds) pressure, isotropic thermal pressure (including magnetic field pressure), centrifugal motion, and magnetic stresses (see Appendix \ref{sec:force}). Dashed lines show results for the \texttt{Fiducial} and \texttt{Cooling} simulations, respectively.}
    \label{fig:forces}
\end{figure}

\subsection{Magnetized Accretion: Weakly vs.~Strongly Magnetized Flows}
\label{sec:magnetized_accretion}

Magnetic fields are essential to sustain mass accretion onto compact objects, with MRI-driven turbulence 
being a dominant mechanism to transport angular momentum in ionized accretion flows \citep{balbus1991Powerful}. Moreover, when sufficient poloidal magnetic flux is advected onto the compact object, the field can become dynamically important and backreact onto the gas, thereby suppressing and regulating accretion \citep{igumenshchev2003three, tchekhovskoy2011Efficient}. In this so-called magnetically arrested disk (MAD) state, the magnetic flux in the inner part of the accretion disk saturates and the rotation locally becomes sub-Keplerian. Frequent flux eruptions and powerful outflows may occur near the compact object (see e.g. \citealt{ressler2021magnetically, lalakos2022bridging, begelman2022really, galishnikova2025strongly}). In the case of BH accretion, numerical simulations have shown that vertical magnetic flux accumulates near the horizon until plasmoid-mediated reconnection ejects flux bundles that mix with the accretion flow \citep{takasao2019giant, ripperda2020, ripperda2022black}. In the following, we investigate whether related configurations develop during accretion onto a NS.

\begin{figure}
    \centering
    \includegraphics[width=1\linewidth]{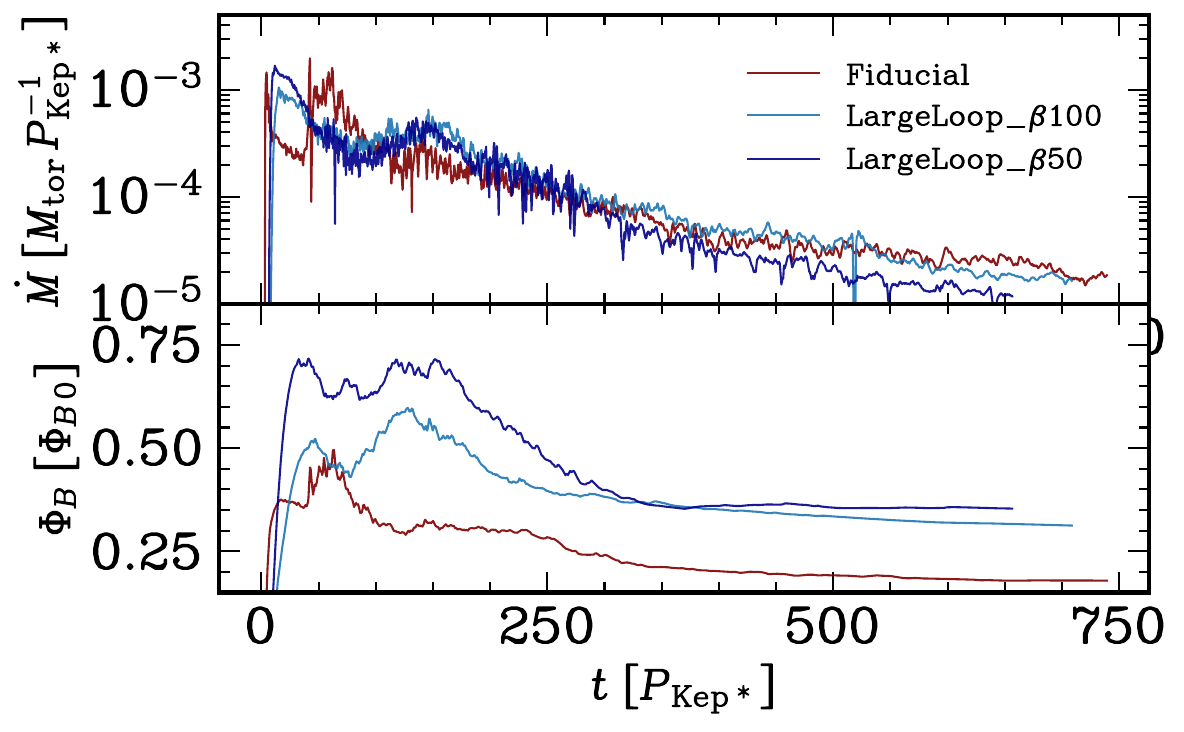}
    \caption{Accretion rate (top) and net enclosed vertical magnetic flux (bottom),
    as measured at $r = 2\,r_\star$, in the \texttt{Fiducial} and 
    \texttt{LargeLoop} simulations.
    \label{fig:qmvst}
    }
\end{figure}

\begin{figure}
    \centering
    \includegraphics[width=1\linewidth]{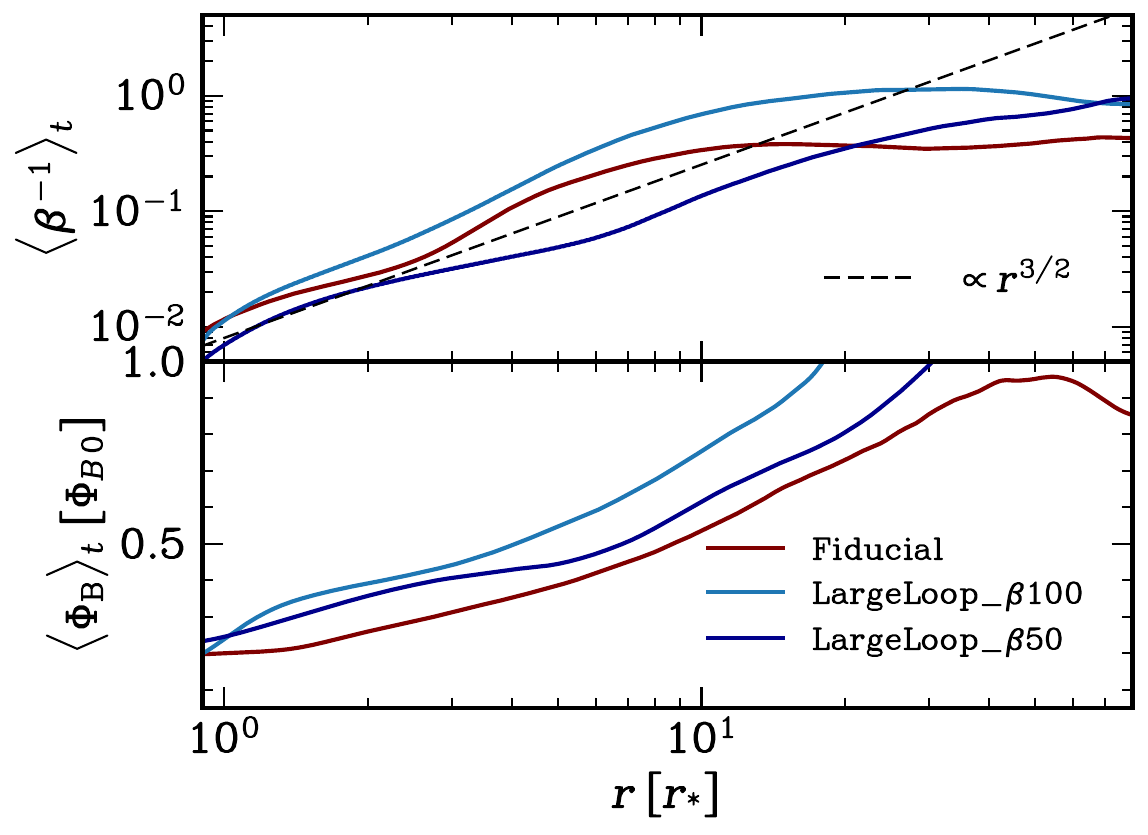}
    \caption{Magnetic properties of the flow in the final state of the 
    \texttt{Fiducial} and \texttt{LargeLoop} simulations, time averaged over $\Delta t_{\rm C}= [449, 750]\,P_{\rm Kep\star}$. Top: spherically averaged ratio $\beta^{-1}$ of magnetic to fluid pressure. 
    Bottom: enclosed vertical magnetic flux in units of ${\Phi_{B}}_0= 1/\sqrt{(M_{\rm tor}/P_{\rm Kep\star})\,(G^2M^2)/c^2}$. 
    }
    \label{fig:qmvsr}
\end{figure}

\begin{figure}
    \centering
    \includegraphics[width=1\linewidth]{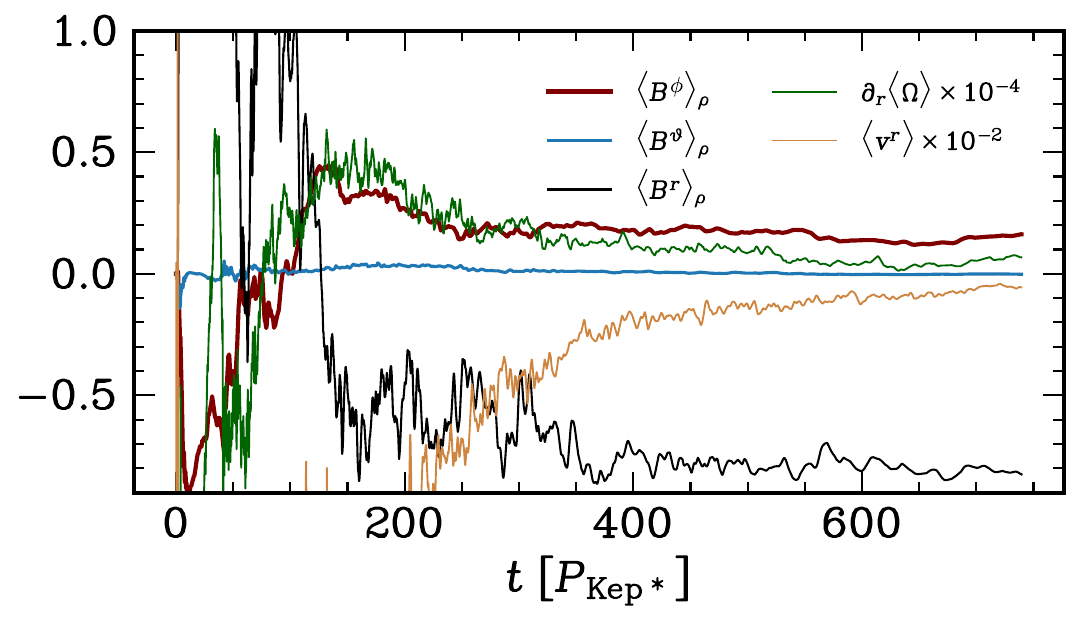}
    \caption{Components of the magnetic field (in units of $c^2/\sqrt{G}\,(GM/c^2)^{-1}$) in the \texttt{Fiducial} run weighted by rest-mass density within the atmosphere zone $3\,r_\star<r<3.5\,r_\star$ in the upper hemisphere $z>0$. The values are normalized to $2\times10^{-8}$.
    }
    \label{fig:bavst}
\end{figure}

\begin{figure*}
    \centering
    \includegraphics[width=.50\linewidth]{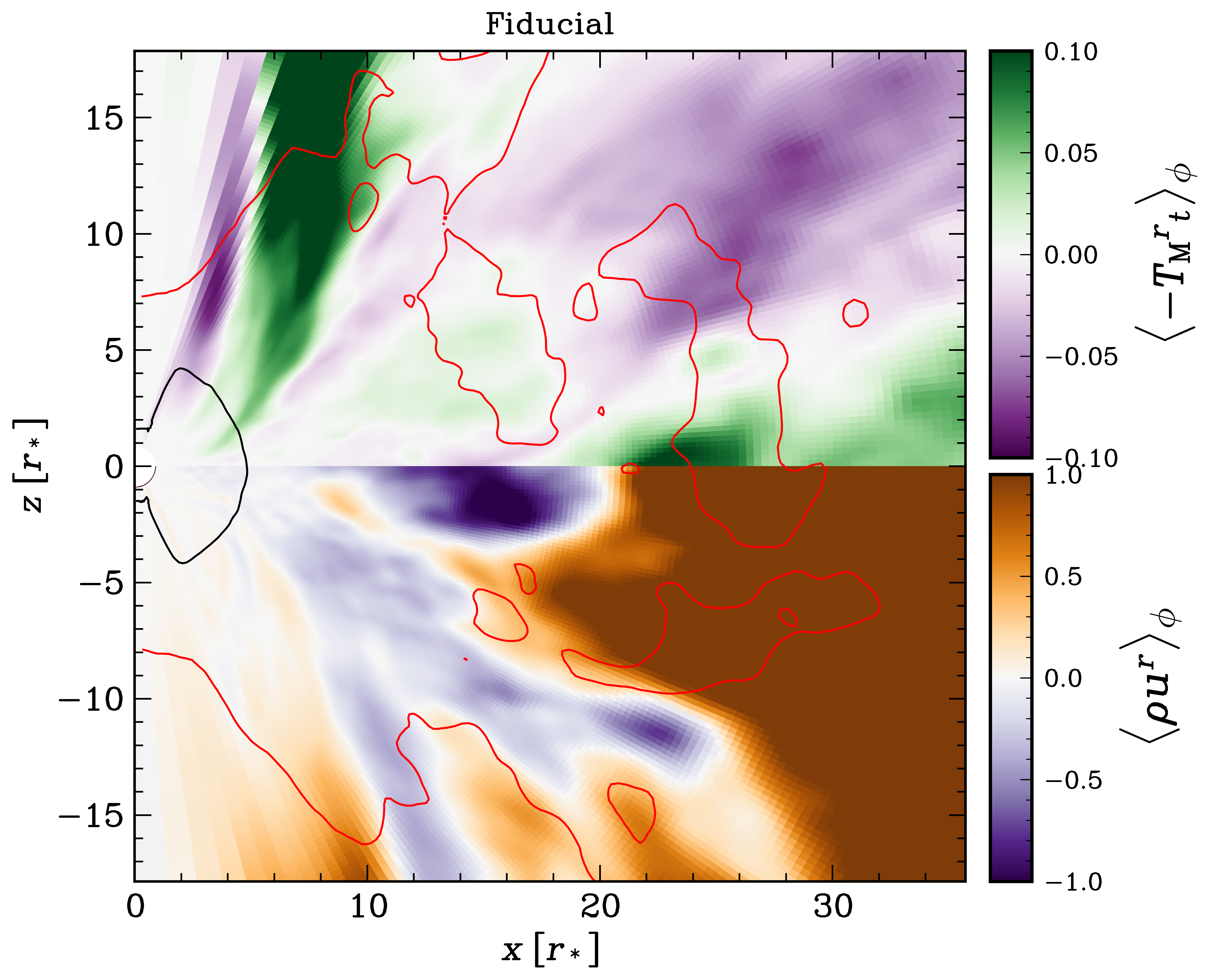}
    \includegraphics[width=0.49\linewidth]{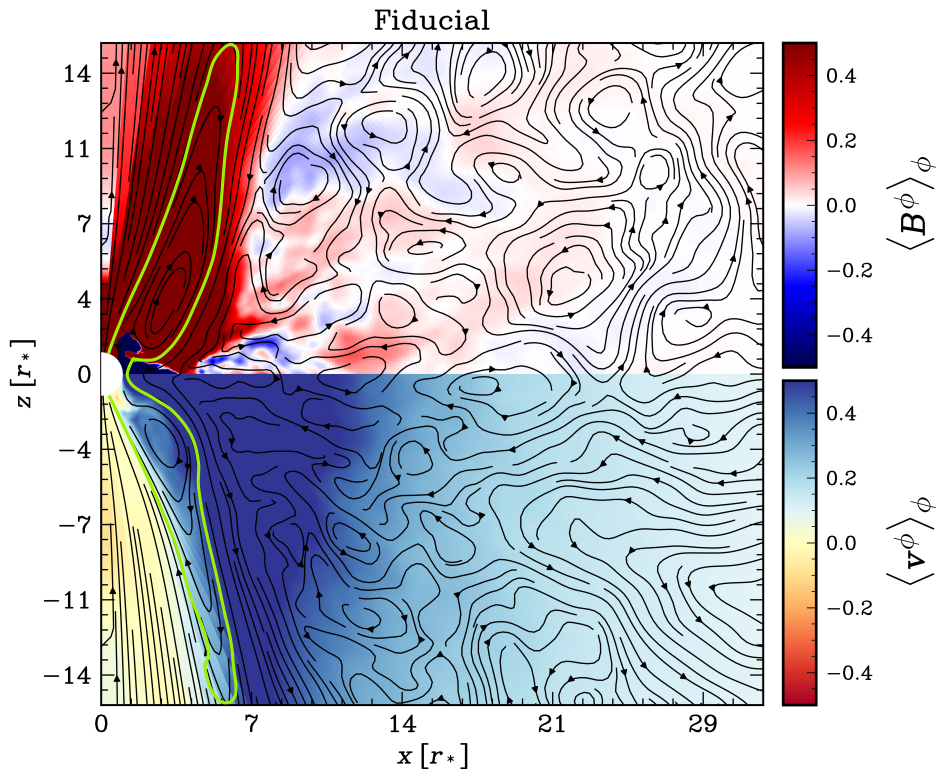}
    \includegraphics[width=0.49\linewidth]{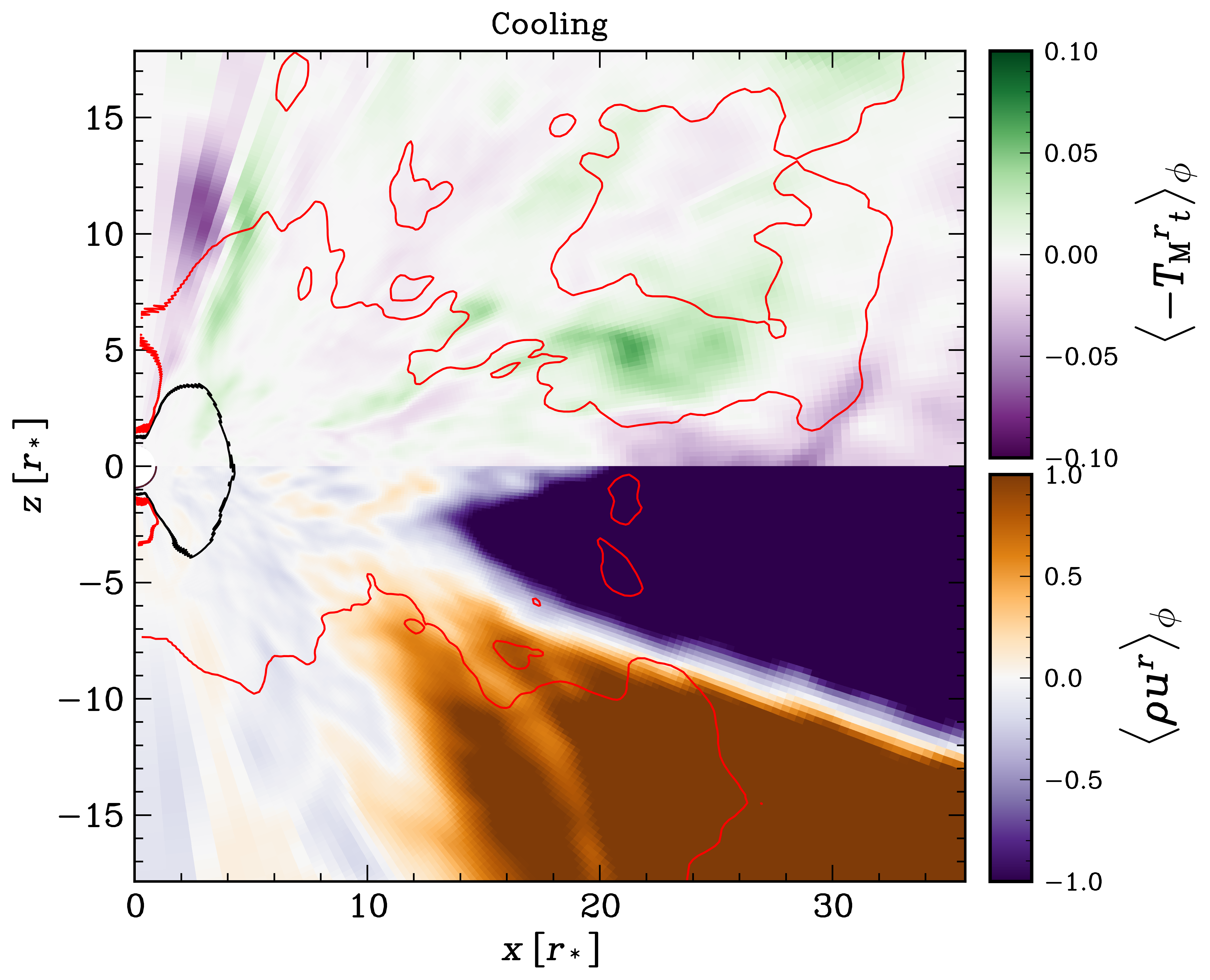}
    \includegraphics[width=0.49\linewidth]{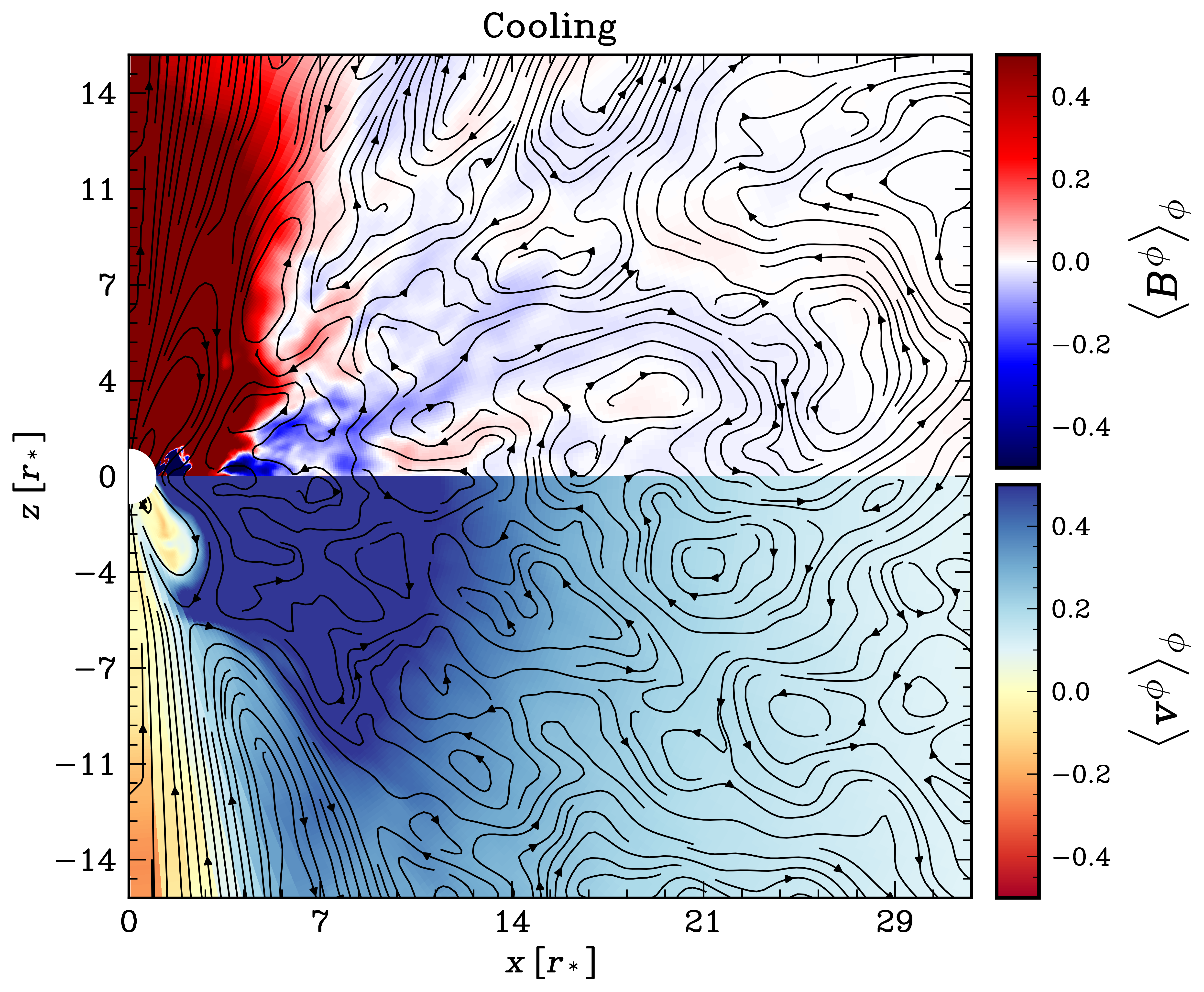}
    \includegraphics[width=.49\linewidth]{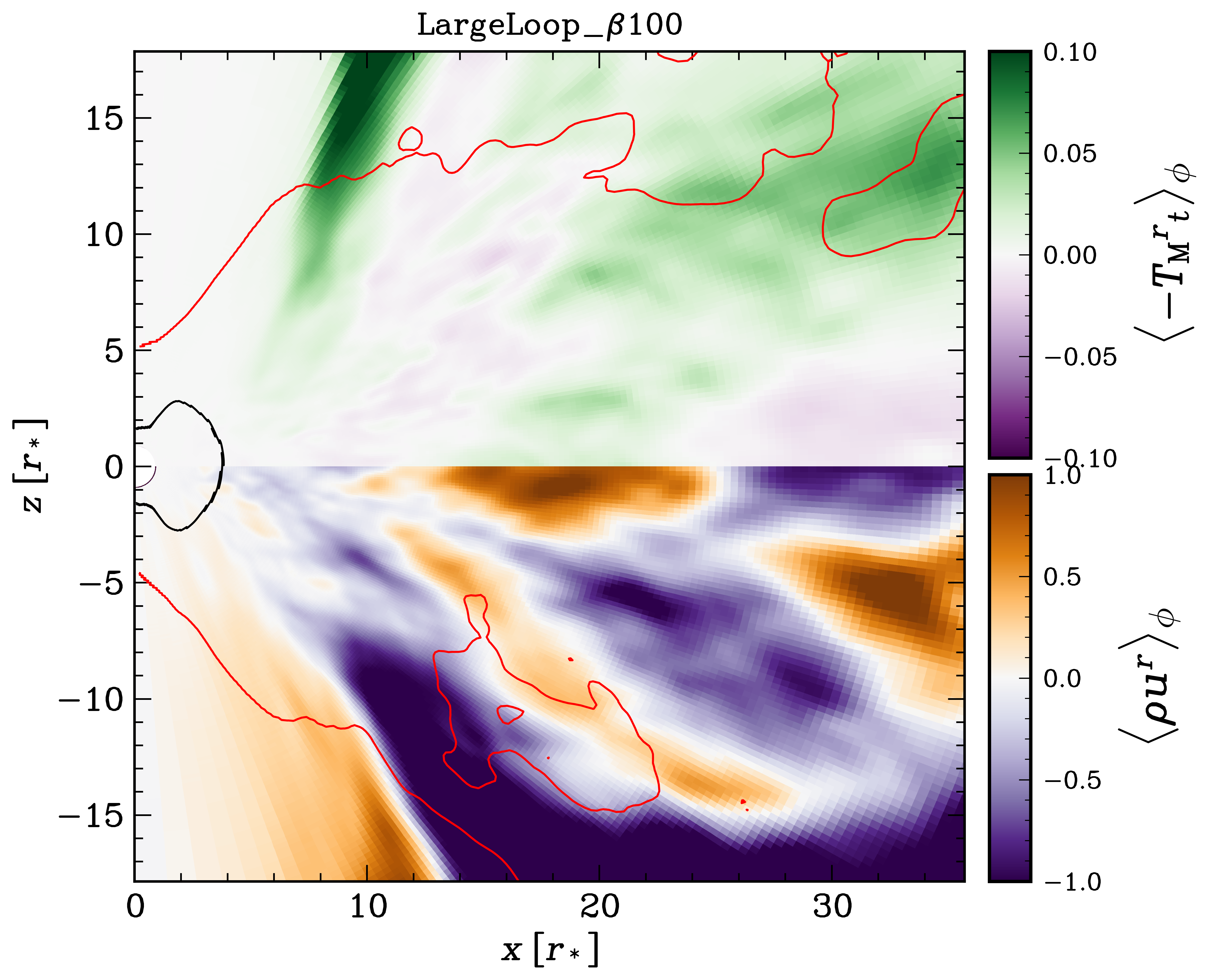}
    \includegraphics[width=0.49\linewidth]{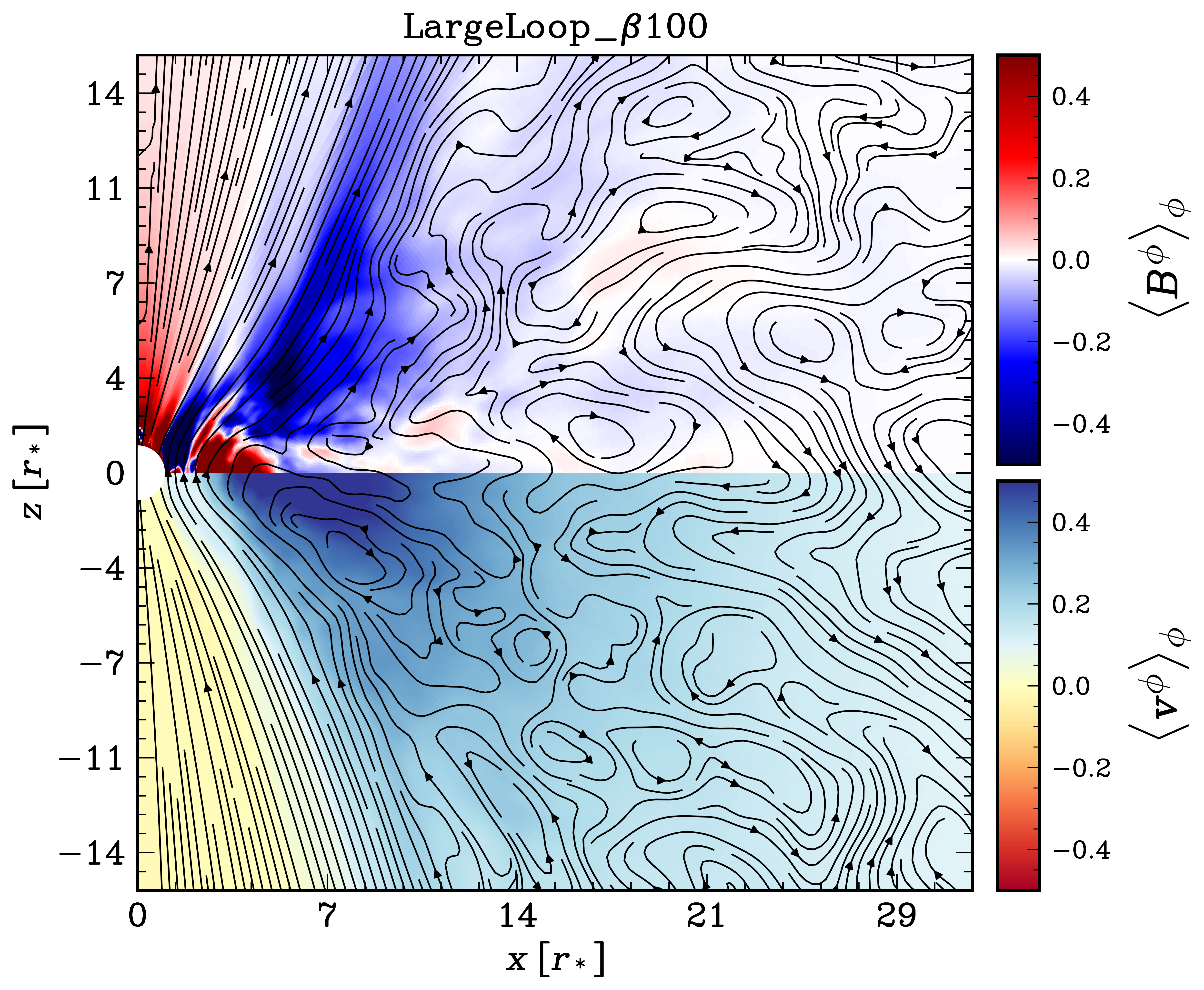}
    \caption{
    Azimuthally averaged meridional profiles of the \texttt{Fiducial} (top), \texttt{Cooling} (middle), and \texttt{LargeLoop\_}$\beta100$ (bottom) simulations, averaged over the time interval $720-730 \,P_{\rm Kep\star}$. Left panels: radial Poynting flux normalized to $6\times 10^{-11}$ (top half-planes) and isotropic equivalent radial mass flux normalized to $6\times10^{-9}$ (lower half-planes). Red contours encapsulate bound matter with averaged specific energy $-hu_t<1.0$ and black contours represent bound matter with $-hu_t<0.995$. Right panels: toroidal magnetic field {normalized to $1\times10^{-9}$ in Gaussian units} (top half-planes) and azimuthal velocity in units of $c$ (lower half-planes). Field lines represent the in-plane magnetic field; a selected field line connecting the polar cap of the star with the rotating belt in the equatorial plane is highlighted in green.}
    \label{fig:outflowsfield}
\end{figure*}

To measure the impact of magnetic fields in NS RIAFs, we compare simulations with different initial seed magnetic fields representing small (\texttt{Fiducial}) and large (\texttt{LargeLoop}\_{$\beta100$}, \texttt{LargeLoop}\_{$\beta50$}) flux reservoirs with different magnetizations. Although the first stages of the two \texttt{LargeLoop} simulations show some features of BH MAD states, e.g.~flux eruptions, we find that their time-averaged properties converge toward the \texttt{Fiducial} simulation
in the latter part of the run.
This is because the growing atmosphere efficiently traps magnetic flux. Since there is a finite amount of flux in the disk, the trapping of magnetic flux in the atmosphere ultimately quenches further eruptions and interchange instabilities typical of MADs, in striking contrast to an otherwise equivalent accretion scenario onto a BH.

In the \texttt{LargeLoop} simulations, large amounts of poloidal flux are advected inward as matter is accreted onto the NS. Field lines are pushed against the star surface and migrate to the polar cap, forming a wide open magnetically dominated ($\beta^{-1}\gg 1$) funnel region along the $z$-axis. This funnel evacuates the polar regions and blocks lateral plasma motion into the funnel. Accretion proceeds mostly through the equatorial zone, as gas accumulates onto the NS and a matter-dominated ($\beta \gg 1$) atmosphere starts forming. Magnetic fields are frozen in, becoming `buried' within this growing, nearly hydrostatic atmosphere, and newly advected flux from the disk anchors to the atmosphere instead of piling up against the star's surface. 

We monitor the vertical magnetic flux threading the atmosphere, defined as
\begin{equation}
    \Phi_{B}  =\frac{1}{2} \int |B^r| \sqrt{-g} \,\mathrm{d}\theta \mathrm{d}\phi,
\end{equation}
where the integration is over a sphere of radius $r=2\,r_\star$. This flux first peaks around a time $50\,P_{\rm Kep\star}$ and then decreases rapidly (see Figure~\ref{fig:qmvst}, bottom panel), signaling an eruption of a flux bundle back towards the disk. A magnetized ($\beta^{-1}\sim 1$) bubble is ejected, similar to those seen in simulations of MADs around BHs. In the BH case, flux eruptions are triggered as a result of the lateral accretion flow interacting with the low-$\beta$ magnetic funnel regions, while in the NS RIAF case, flux eruptions originate at the disk-atmosphere boundary (around $3\,r_\star$ radius)
and thus carry away significant mass from the extended NS atmosphere. 

After this first ejection event, the bubble migrates outward, dissipates, and no further such ejections are detected in the \texttt{LargeLoop}\_$\beta100$ run. The \texttt{LargeLoop}\_$\beta50$ simulation shows a few additional eruption events, before such activity also ceases. After the eruptions, the magnetic flux within the atmosphere increases again and reaches a global maximum around $t \sim 130\,P_{\rm Kep\star}$ and $t \sim 160\, P_{\rm Kep\star}$ in the \texttt{LargeLoop}\_$\beta100$ and \texttt{LargeLoop}\_$\beta50$ runs, respectively. At this point, the atmosphere radius freezes, and the accretion rate starts to decrease in time. This monotonic decrease of $\dot M$ commences much earlier in the \texttt{Fiducial} case (at $\sim 50\,P_{\rm Kep\star}$). In all runs, the vertical magnetic flux in the NS atmosphere decreases and eventually approaches a nearly constant value, with $\Phi_{\rm B}$ in both \texttt{LargeLoop} simulations being $\sim 2$ times larger than in the \texttt{Fiducial} setup. The nearly saturated nature of the magnetic flux inside the atmosphere is reflected by the behavior of the time-averaged $\Phi_{\rm B}$ as a function of radius (Figure~\ref{fig:qmvsr}). Even though the \texttt{LargeLoop}\_$\beta50$ setup is initially more magnetized, the final state of $\Phi_{\rm B}$ is very similar to that in the \texttt{LargeLoop}\_$\beta100$ case. The total magnetic energy inside the atmosphere closely follows the behavior of the vertical flux: it increases in the first $t \sim 50\,P_{\rm Kep\star}$, decays, and then saturates at $t \gtrsim 200\,P_{\rm Kep\star}$.

\subsection{Rotation Profile}

As the atmosphere emerges and spreads radially, the MRI ceases to be active at $r< r_{\rm atm}$:
in this zone, the angular velocity develops a positive gradient, which then
saturates at a small positive value ($\mathrm{d}\Omega/\dr \gtrsim 0$). The toroidal field inside the atmosphere is wound up linearly,
with $\langle B^\phi\rangle_\rho$ and $\langle B^r\rangle_\rho$ sharing the same sign out to a time
$\sim 100\, P_{\rm Kep*}$, 
as illustrated for the \texttt{Fiducial} run in Figure~\ref{fig:bavst}.
Then the relative sign of these two field components becomes negative, and their ratio saturates
at $\langle B^\phi \rangle_\rho \sim -0.25 \langle B^{r}\rangle_\rho$. 
In a perfectly static atmosphere, positive $\mathrm{d}\Omega/\mathrm{d}r$ winds up positive $B_r$ to positive $B_\phi$. In contrast, negative $B_r$ (negative flux $\Phi_r$) can generate positive $B_\phi$ with negative $v_r$ (positive $\dot M$), as we observe here. The large-scale toroidal field at the boundary of the atmosphere launches an outward Poynting flux in all simulations (see Section~\ref{sec:coolingresults}).

Rotation near the poles is slowed by the accumulation of vertical magnetic flux, as shown in Figure~\ref{fig:outflowsfield} (azimuthal velocity $\langle v^\phi\rangle_\phi\approx 0$). In the \texttt{Fiducial} and \texttt{Cooling} runs, some slowly-rotating vertical field lines anchored close to the polar axis (e.g. with $B^z>0$) bend and connect back to the inner atmosphere ($B^z<0$) near the NS surface
(see Figure~\ref{fig:outflowsfield}, right panels; a sample field 
line is shown in green). The twisting of these half-open magnetic arcades
causes them to open out and reconnect, forming closed poloidal loops that are eventually ejected along the funnel sheath (Figure~\ref{fig:outflowsfield}, right panels).  

The geometry of the accreted magnetic field influences details of angular momentum transport, as we explore in more detail in Section \ref{sec:amt}.
In contrast with the simulations that are initialized with a smaller-scale
field, the \texttt{LargeLoop}\_$\beta100$ run maintains the same field polarity across the funnel and atmosphere.  This is because the large poloidal seed loop 
in the initial state is only partially accreted into the atmosphere;
mostly vertical field of positive polarity is advected inward 
until the end of the evolution. 


Although the gas distribution is roughly spherically symmetric within the atmosphere, a magnetized funnel forms near the poles.  In this polar zone, 
$\beta^{-1}\lesssim 1$ within and $\beta^{-1} \gtrsim 1$ above the atmosphere in all simulations (see Figure~\ref{fig:rhoinvb_xz} for the \texttt{Fiducial} case). Because the magnetic field is only weakly amplified inside the slowly rotating and nearly hydrostatic atmosphere, $\beta^{-1}$ decreases monotonically with time
as matter continues
to assemble near the NS, reaching $\langle \beta^{-1} \rangle_t < 10^{-2}$ near the NS surface 
at the end of the \texttt{Fiducial} and \texttt{LargeLoop} simulations.
The radial profile then approximately follows a power law $\langle\beta^{-1}\rangle_t \propto r^{3/2}$ 
out to $\approx 10r_\star$ (see Figure~\ref{fig:qmvsr}), which implies that the comoving magnetic field is distributed in the atmosphere roughly as $b \propto r^{-3/2}$. Outside the atmosphere and within the disk ($r \gtrsim 6\, r_\star$), the relative magnetic pressure saturates at $\langle \beta^{-1} \rangle_t \sim 0.2$ in the \texttt{Fiducial} simulation, increasing to $\langle \beta^{-1} \rangle_t \sim 0.5-0.8$
in both \texttt{LargeLoop} simulations.  

\begin{figure}
    \centering
    \includegraphics[width=1\linewidth]{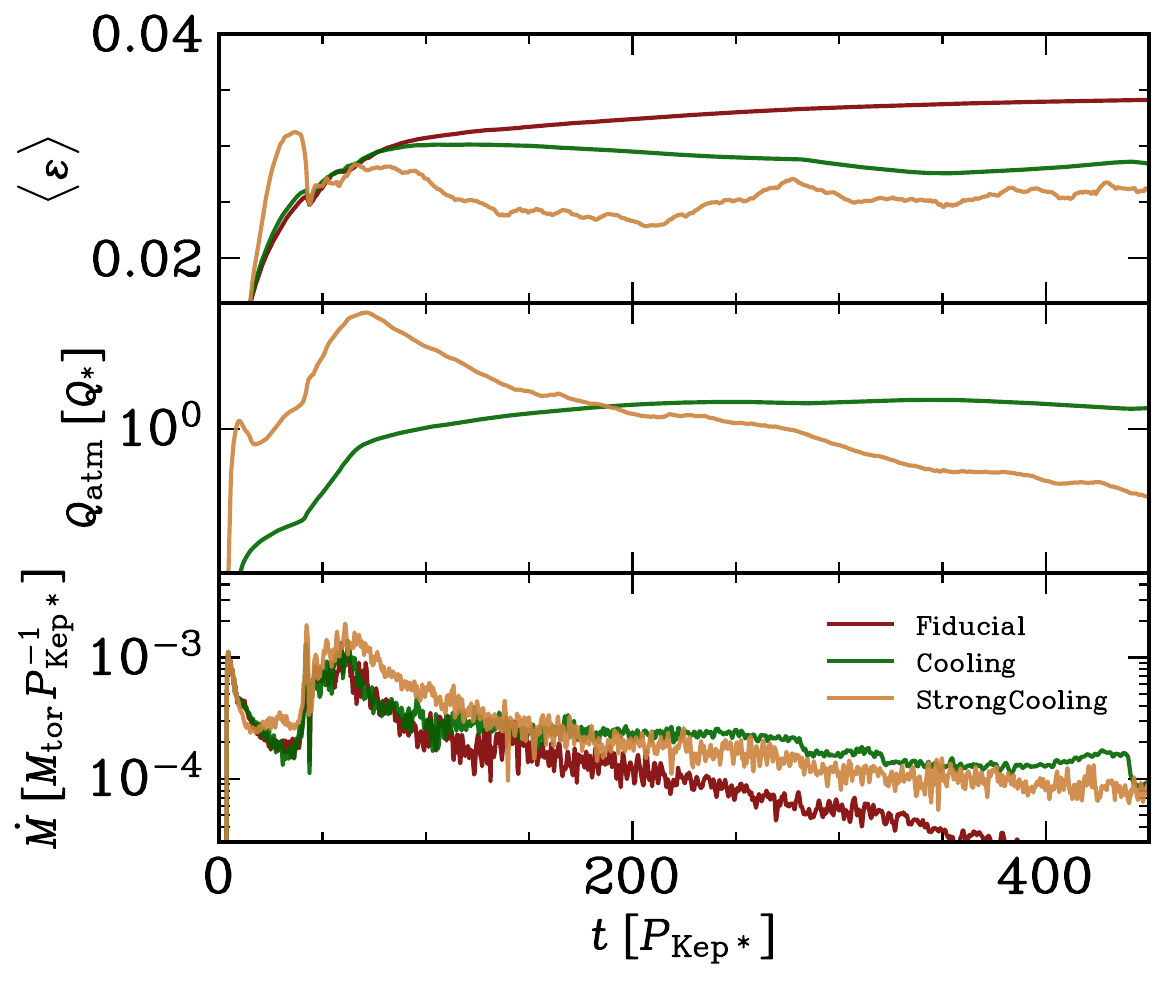}
    \caption{Top: internal energy $\epsilon = u/\rho$ averaged over the spherical volume within the hydrostatic atmosphere $r_\star<r<5\,r_\star$. 
    Center: volume-integrated ($r_\star<r<5\,r_\star$) total neutrino energy losses in units of $Q_\star = G \rho_0 (M_{\rm tor}\,P_{\rm Kep\star}^{-1}) r_\star^{-1}$.
    Bottom: rest-mass accretion rate at radius $2\,r_\star$. Results are shown for simulations with and without neutrino cooling.}
    \label{fig:cooling}
\end{figure}

\subsection{Neutrino-cooled Accretion Flows}
\label{sec:coolingresults}

Neutrino cooling is enhanced by the formation of a hydrostatic density cusp around the NS, as is
seen in our simulations
(Section~\ref{sec:atmosphere} and Figure~\ref{fig:qvsr}).
The flow remains non-radiative outside a thin atmospheric layer adjacent to the NS
(Section~\ref{sec:analytic}).  In the \texttt{Cooling} and \texttt{StrongCooling} simulations, we include an optically thin, isotropic cooling term $Q \propto \rho p^{3/2}$, representing radiative losses by neutrinos due to $e^{\pm}$ captures and pair annihilation. The normalization of the neutrino emissivity
regulates the accretion rate and the size of the atmosphere
(Section~\ref{sec:cooling} and Appendix \ref{app:density_scale}).

In the \texttt{Cooling} model, the pressure and density near the NS are not at first large enough to
generate significant cooling. 
Next, a hydrostatic atmosphere emerges, similar to that seen in the \texttt{Fiducial} simulation;
the atmosphere in the two simulations experiences similar growth out to a time $\approx 100\,P_{\rm Kep\star}$
(Figure~\ref{fig:r_t}, upper panel).  Around this time, the neutrino losses grow
strong enough to feed back on the accretion flow and the net rate of energy loss reaches a steady state 
(Figure~\ref{fig:cooling}, center panel). 

The neutrino emission is concentrated in a thin layer around the NS surface, as a consequence of
the strong temperature dependence of the emissivity (see Section~\ref{sec:analytic}).  Although
the hydrostatic density structure is similar outside this layer in the \texttt{Cooling} and \texttt{Fiducial}
simulations (Figure~\ref{fig:q_r_all}, bottom panel), there are significant differences in the mass flow.
The neutrino losses mediate a steady influx of mass and energy through the 
adiabatic part of the atmosphere ($\dot M$ is
nearly uniform inside radius $r_{\rm atm}$), whereas accretion is eventually suppressed near the star in the adiabatic case ($\dot M \rightarrow 0$ at radius $r_\star$).  See the lower panel of Figure~\ref{fig:cooling}.

The long term growth of the atmosphere is also regulated by neutrino cooling. The specific internal energy of the atmosphere saturates at around $100\,P_{\rm Kep\star}$ in the \texttt{Cooling} simulation, while it keeps increasing in the adiabatic \texttt{Fiducial} run (Figure~\ref{fig:cooling}, upper panel).  
Whereas the atmosphere grows as $r_{\rm atm}(t) \sim t^{0.37}$ over the full duration of the
\texttt{Fiducial} simulation (several hundred $P_{\rm Kep\star}$), we find that $r_{\rm atm}$ plateaus
at $\sim 3\,r_\star$ in the \texttt{Cooling} run, beyond the point where cooling competes with (viscous) heating.
There is little impact of near surface cooling on 
the evolution of the outer RIAF, as indicated by the similar evolution of the inflow-outflow radius 
in both runs (Figure~\ref{fig:r_t}, lower panel). 
The atmosphere in the \texttt{Cooling} simulation remains hydrostatic and  
pressure supported (Figure~\ref{fig:forces}, dashed lines), and maintains a nearly flat 
rotation profile (Figure~\ref{fig:q_r_all}, upper panel).

In the \texttt{StrongCooling} simulation, neutrino cooling is enhanced by three orders of magnitude relative to the \texttt{Cooling} run. Here, radiative losses are dynamically important as soon as the accretion flow reaches the NS surface. As a consequence of the enhanced cooling, a nearly hydrostatic atmosphere
does not form, and the internal energy near the NS is reduced (Figures~\ref{fig:q_r_all} and \ref{fig:rhocool_xz}).   Although rotation is sustained at a nearly Keplerian rate at the NS surface, outflow is suppressed and $\dot M(r)$ flattens out inside $5\,r_\star$ radius (see Figure~\ref{fig:q_r_all}, top and center panels). The interaction of a Keplerian disk with the non-rotating surface of the star resembles a classic boundary layer problem (with the typical thin disk replaced by a RIAF). New simulations at much higher resolutions are needed to resolve the scale-height of the boundary and understand whether the angular velocity reaches a steady state or not (see, e.g., \citealt{belyaev2018inefficient}).

\begin{figure}
    \centering
    \includegraphics[width=1\linewidth]{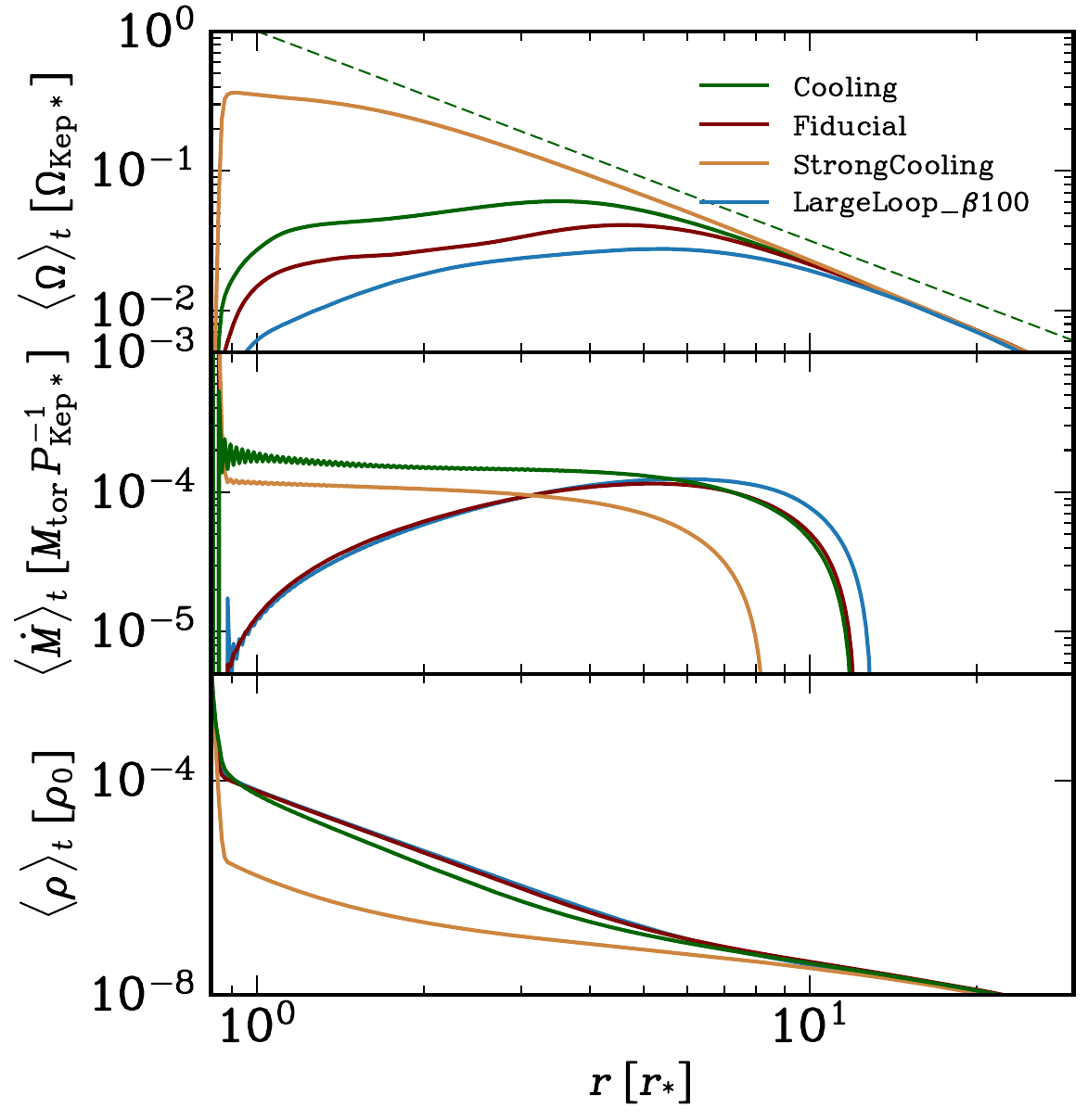}
    \caption{Radial profiles of spherically averaged angular velocity (top), accretion rate in units of $M_{\rm tor}\,P_{\rm Kep\star}^{-1}$ (center), and
    surface rest-mass density (bottom) in the final time interval $\Delta t_{\rm B}$ of Figure~\ref{fig:omegar}. 
    The adiabatic simulations (\texttt{Fiducial} (maroon lines) and \texttt{LargeLoop}\_$\beta100$ (blue lines)) show a pronounced drop in the accretion rate near the star's surface, while the cooling simulations (\texttt{Cooling}, green lines, and \texttt{StrongCooling}, yellow lines) show inflow equilibrium from the surface outward. 
    All simulations show an almost flat $\Omega$ profile inside the atmosphere except \texttt{StrongCooling}, which remains almost Keplerian near the surface. 
    For reference, the green dashed line in the top panel represents the Keplerian angular velocity.
    }
    \label{fig:q_r_all}
\end{figure}

\begin{figure}
    \centering
    \includegraphics[width=1\linewidth]{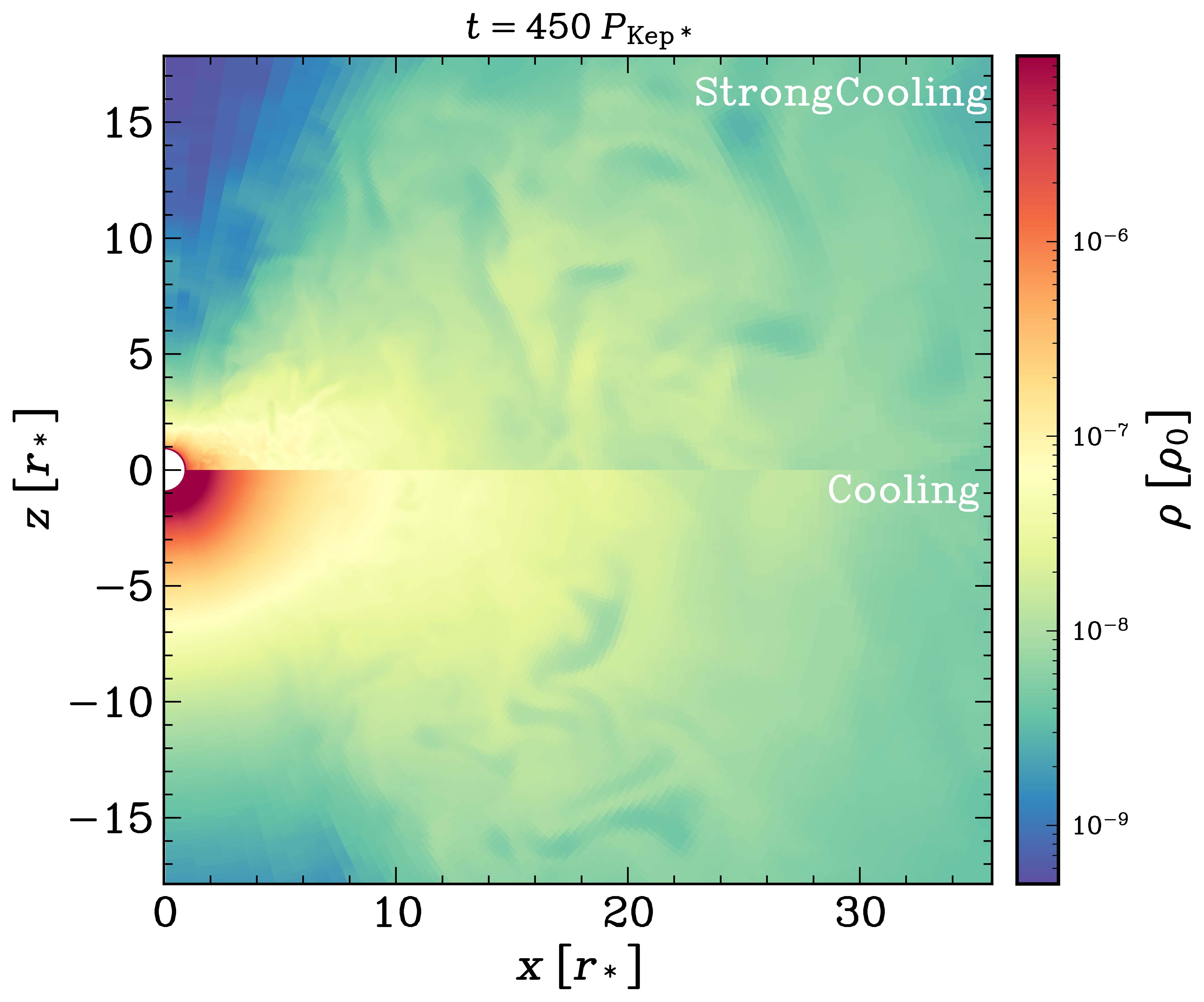}
    \caption{Snapshot of the rest-mass density in the $x$-$z$ plane at $t=450\,P_{\rm Kep\star}$. Upper half-plane: \texttt{StrongCooling} simulation with enhanced neutrino cooling. Lower half-plane: \texttt{Cooling}
    simulation with weak neutrino cooling. 
    A steady, nearly hydrostatic core is formed in the simulation with weak cooling, while a RIAF structure extends to the star's surface in the simulation with strong cooling.}
    \label{fig:rhocool_xz}
\end{figure}

\subsection{Outflows and Energy Transport}
\label{sec:outlfows}

Internal torques in circulating RIAFs transport energy outward, producing unbound flows (i.e.~with positive Bernoulli parameter; \citealt{blandford1999fate}). 
In the case of accreting BHs, outflows near the horizon can take the form of powerful jets or winds.
In our simulations, the energy that is advected into the NS atmosphere is mostly either stored
there or released to neutrinos.
We now consider the radial structure of the energy flow around the NS,
and the appearance of outflows at greater distances from the star.

The quasi-spherical, nearly hydrostatic atmosphere in the \texttt{Fiducial} and \texttt{LargeLoop} simulations is gravitationally bound to the star, with a reduced specific energy ($-hu_t \lesssim  0.995$; see the contours in Figure~\ref{fig:outflowsfield}). This is in contrast to the rest of the disk, which is only marginally bound ($-hu_t \gtrsim 1)$. Gravitationally unbound outflows $(\dot{M}<0)$ emerge outside the atmosphere ($r\gtrsim 1.5 r_{\rm atm}$), with a peak in the mass-loss rate at $r\sim 30 \,r_\star$.  Beyond $r \gtrsim 20 r_*$, the disk is still out of equilibrium by the end of the simulation (Figure~\ref{fig:qvsr}). In the \texttt{Fiducial} run, mass outflows around $10\,r_\star$ (where inflow-outflow equilibrium has been established; Figure~\ref{fig:r_t}) amount to $60\%-80 \%$ of the peak inflow rate;  this is consistent with previous studies of BH RIAFs (e.g. \citealt{white2020, chatterjee2022flux}). 

The net hydrodynamical energy flux, defined as $L_{\rm H} = \lbrace -{{T_{\rm H}}^r}_{t} \rbrace$ (see Appendix \ref{app:tmunu}) closely follows the mass flux. In the simulations without neutrino cooling, we observe polar outflows above the atmosphere, which are absent in the cooling simulations (see lower half-planes of the left panels in Figure~\ref{fig:outflowsfield}). These polar outflows, however, only constitute a small portion of the total outflow from the system, which is dominated by thermally driven winds originating in the equatorial zone at larger radii. The total luminosity carried by unbound material in all simulations reaches similar values, only differing by a few percent among the various runs (Figure~\ref{fig:out_r}).

A large-scale (mostly poloidal) field anchored to the atmosphere will, in the presence of an angular velocity gradient, stimulate the transport of electromagnetic energy. The Poynting luminosity across a sphere is defined as $L_{\rm M} = \lbrace -{{T_{\rm M}}^r}_{t} \rbrace$, where the convention $L_{\rm M}>0$ corresponds to radially outward directed net flux. As illustrated in the top panel of Figure~\ref{fig:out_r}, the time-averaged luminosity $\langle L_{\rm M}\rangle_t$ peaks around radius $r_{\rm atm}$ in all simulations. In the \texttt{Fiducial}, \texttt{LargeLoop}, and \texttt{Cooling} runs, the efficiency of extracting electromagnetic energy from the accretion flow is 
\begin{equation}
    \eta_{\rm EM} \equiv \frac{\langle L_{\rm M}\rangle _t}{c^2\langle\dot{M}\rangle_t} \sim 1\%
\end{equation}
by the end of the respective simulation, where $\langle\dot{M}\rangle_t$ is the corresponding peak 
value of $\langle \dot M(r)\rangle_t$.

The weak variation observed in $\eta_{\rm EM}$ over a wide range of seed magnetic flux can
be related to the near constancy of the magnetic flux trapped in the atmosphere
(Section~\ref{sec:magnetized_accretion} and Figure~\ref{fig:qmvst}).  The outward-directed Poynting flux is mainly launched in a narrow wedge along the sheath of the polar funnel and taps directly this trapped flux (Figure~\ref{fig:outflowsfield}, top half-planes on the left).

\begin{figure}
    \centering
    \includegraphics[width=1\linewidth]{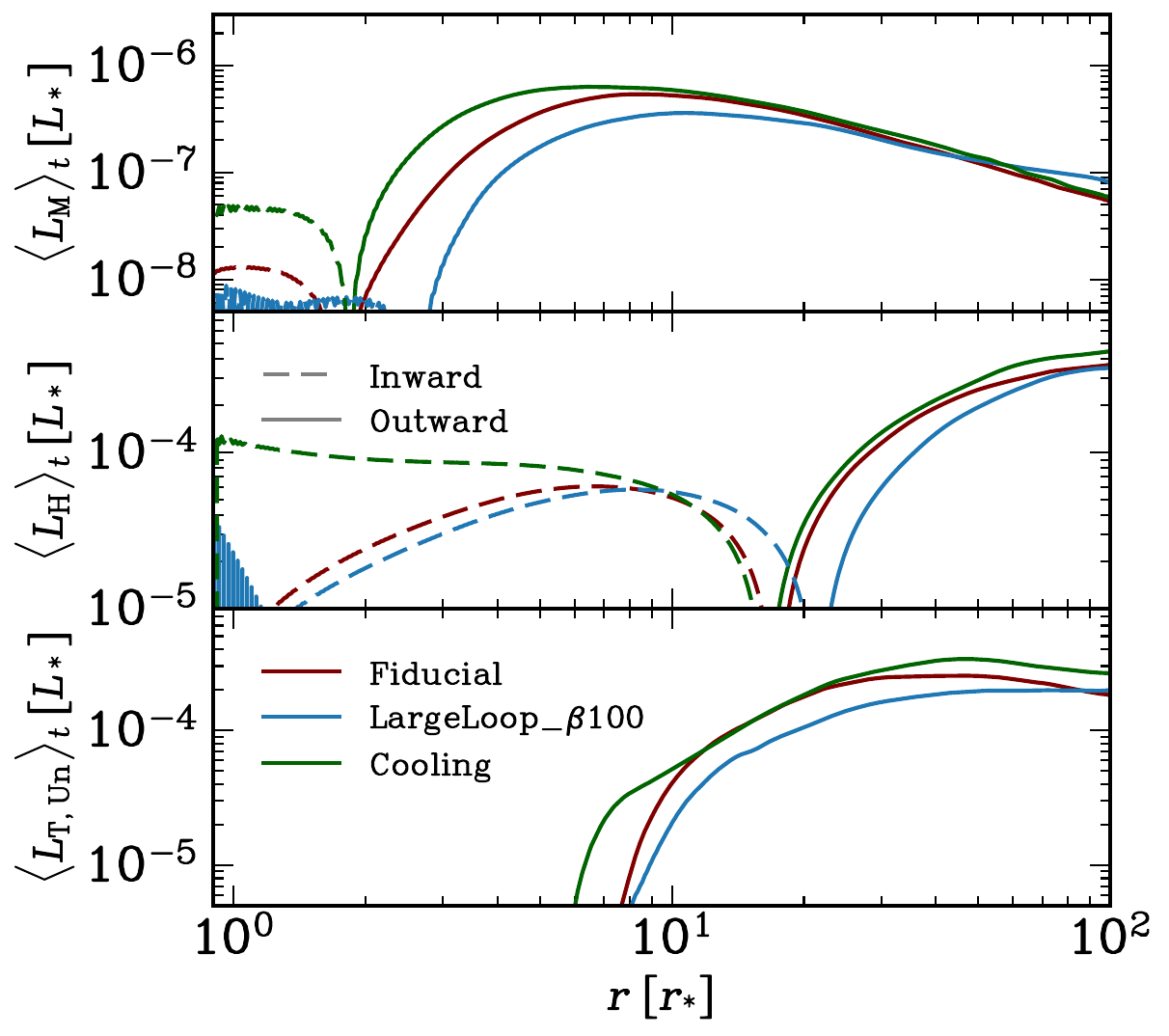}
    \caption{Radial profiles of Poynting luminosity (upper panel), hydrodynamical luminosity, including rest-mass energy (middle panel), and total luminosity {carried by unbound material} (lower panel) in the \texttt{Fiducial}, \texttt{LargeLoop}, and \texttt{Cooling} simulations in units of $L*=c^2 M_{\rm tor}/P_{\rm Kep\star}$ and averaged over $\Delta t_{\rm C}= [449, 750]\,P_{\rm Kep\star}$. Solid curves: outflowing energy.  Dashed curves:  inflowing energy.  
    }
    \label{fig:out_r}
\end{figure}

\begin{figure}
    \centering
    \includegraphics[width=1\linewidth]{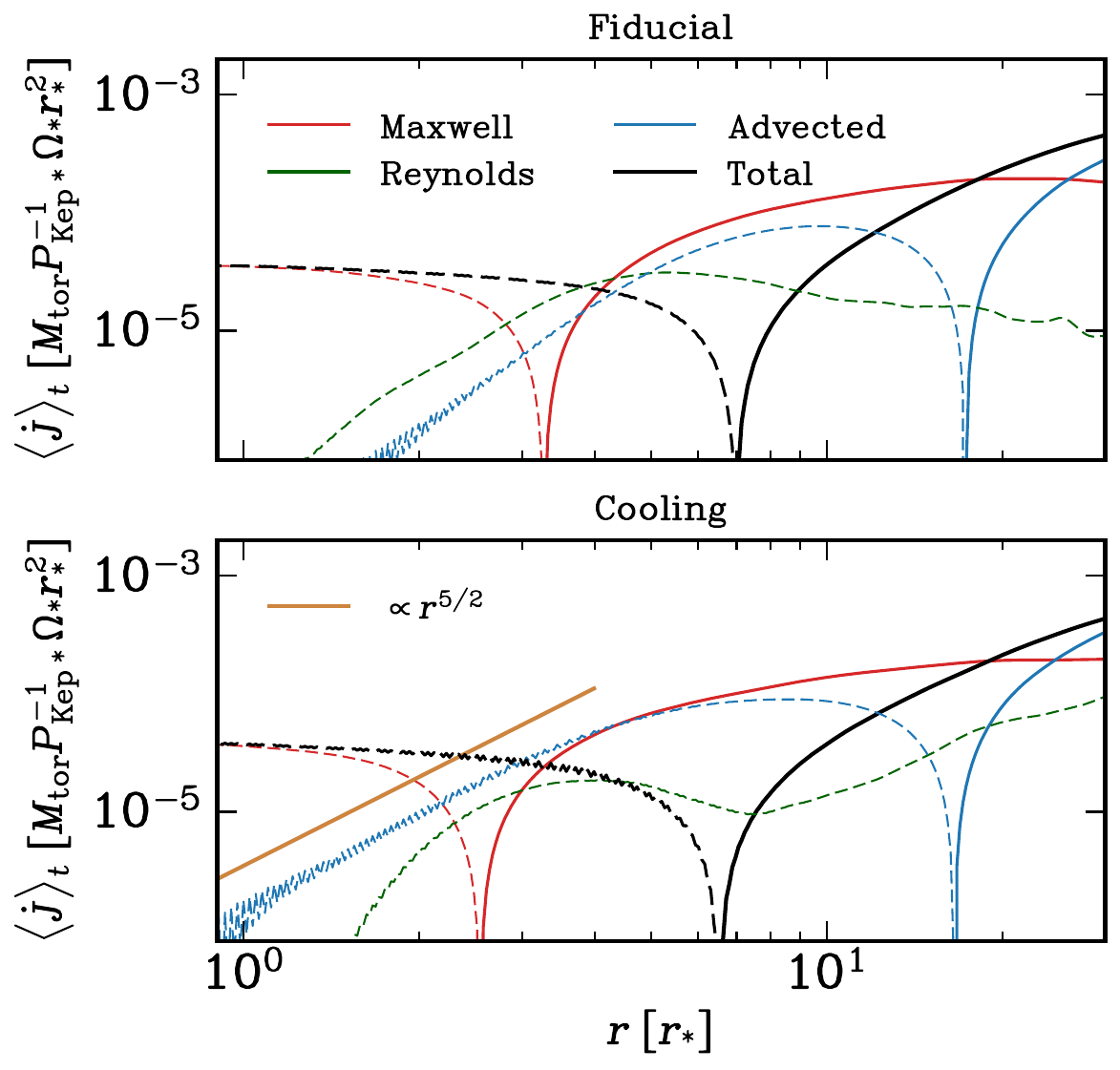}
    \caption{Radial profiles of the spherically averaged angular momentum flux due to Maxwell, Reynolds, and advective stresses (as defined in Appendix~\ref{sec:am}) as well as their total, for the \texttt{Fiducial} (top) and \texttt{Cooling} (bottom) runs, time-averaged over $\Delta t_{\rm C} = [449, 750]\,P_{\rm Kep\star}$. Solid (dashed) lines indicate outward (inward) transport of angular momentum.
    The outer RIAF shows outward transport of angular momentum driven by the MRI, while the inner part shows inward transport of angular momentum driven by magnetic winding.
    }
    \label{fig:angflux}
\end{figure}

\begin{figure}
    \centering  \includegraphics[width=1\linewidth]{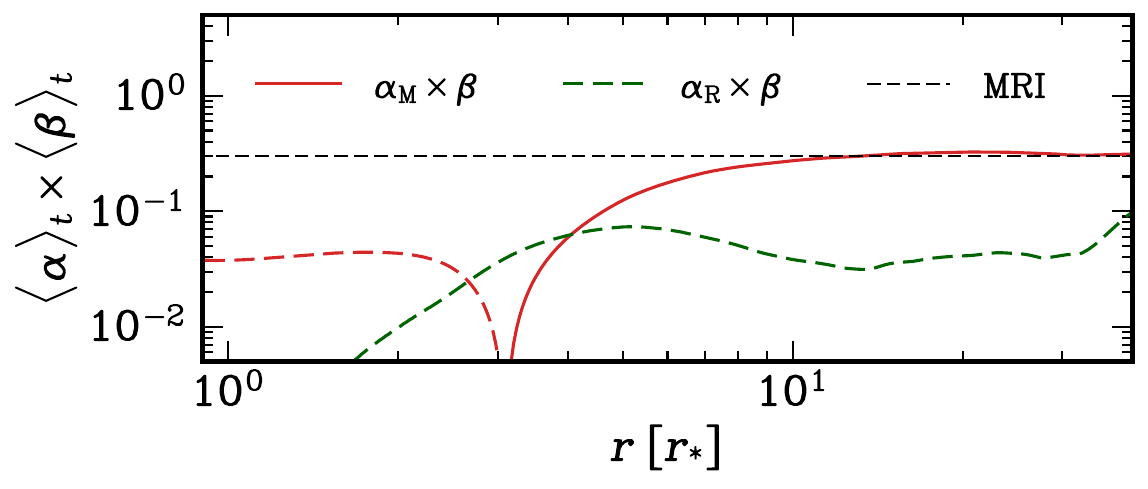}
    \caption{Radial profiles of the spherically averaged effective Shakura-Sunyaev $\alpha$-viscosity parameter multiplied by plasma $\beta$ for the \texttt{Fiducial} run, decomposed into Maxwell and Reynold components,  time-averaged over $\Delta t_{\rm C} = [449, 750]\,P_{\rm Kep\star}$. Solid (dashed) lines correspond to inward (outward) transport of angular momentum. The black dot-dashed line indicates the numerical prediction $\alpha\beta = 0.3$ from local shearing-box simulations of the MRI. 
    }
    \label{fig:alpha_vs_r}
\end{figure}

\subsection{Angular Momentum Transport}\label{sec:amt}

We analyze angular momentum transport in the atmosphere and the surrounding RIAF by decomposing the total angular momentum flux $\dot{J}_{\rm tot}$ into Maxwell ($\dot{J}_{\rm M}$), turbulent Reynolds ($\dot{J}_{\rm R}$), and advected ($\dot{J}_{\rm adv}$) components; see Appendix \ref{sec:am} for definitions. The radial profiles of these components are similar, both qualitatively and quantitatively, 
in the \texttt{Fiducial} and \texttt{Cooling} runs, as shown in Figure~\ref{fig:angflux}. Outward directed fluxes dominate angular momentum transport starting just outside of the nearly hydrostatic atmosphere at $r\gtrsim 7 r_\star$ ($r\gtrsim 6 r_\star$) in the \texttt{Fiducial} (\texttt{Cooling}) simulation (black lines). As expected for magnetized RIAFs, the electromagnetic component of the fluxes dominates the outward transport of angular momentum in the outer region of the accretion flows (red lines), with magnetic stresses dominating over magnetic advection, $\dot{J}_{\rm M} \approx \lbrace -b^r b_{\phi} \rbrace$.

To understand better the origin of magnetic and Reynold stresses in the outer RIAF, we compute the effective magnetic ($\alpha_{\rm M}$) and turbulent ($\alpha_{\rm R}$) Shakura-Sunyaev viscosity parameters, defined as
\begin{equation}
    \alpha_{\rm M} = \langle p\rangle^{-1}\langle-\hat{b}_{{\phi}} \hat{b}_{{r}} \rangle, \quad     \alpha_{\rm R} =  \langle p\rangle^{-1} {\langle \rho h \delta\hat{u}_{{\phi}} \delta\hat{u}_{{r}} \rangle },
\end{equation}
respectively. The hatted quantities are measured in a local orthonormal frame. Local shearing box simulations of the MRI \citep{salvesen2016accretion} as well as global simulations of RIAFs (e.g. 
\citealt{begelman_what_2022}) find an empirical, almost constant value of $\alpha \times \beta \sim 0.3$. We recover this result in the outer RIAF in all simulations, which suggests that magnetic stresses here are sustained by the MRI. As an example, Figure~\ref{fig:alpha_vs_r} presents the case of the \texttt{Fiducial} run. 

\begin{figure}
    \centering
    \includegraphics[width=1\linewidth]{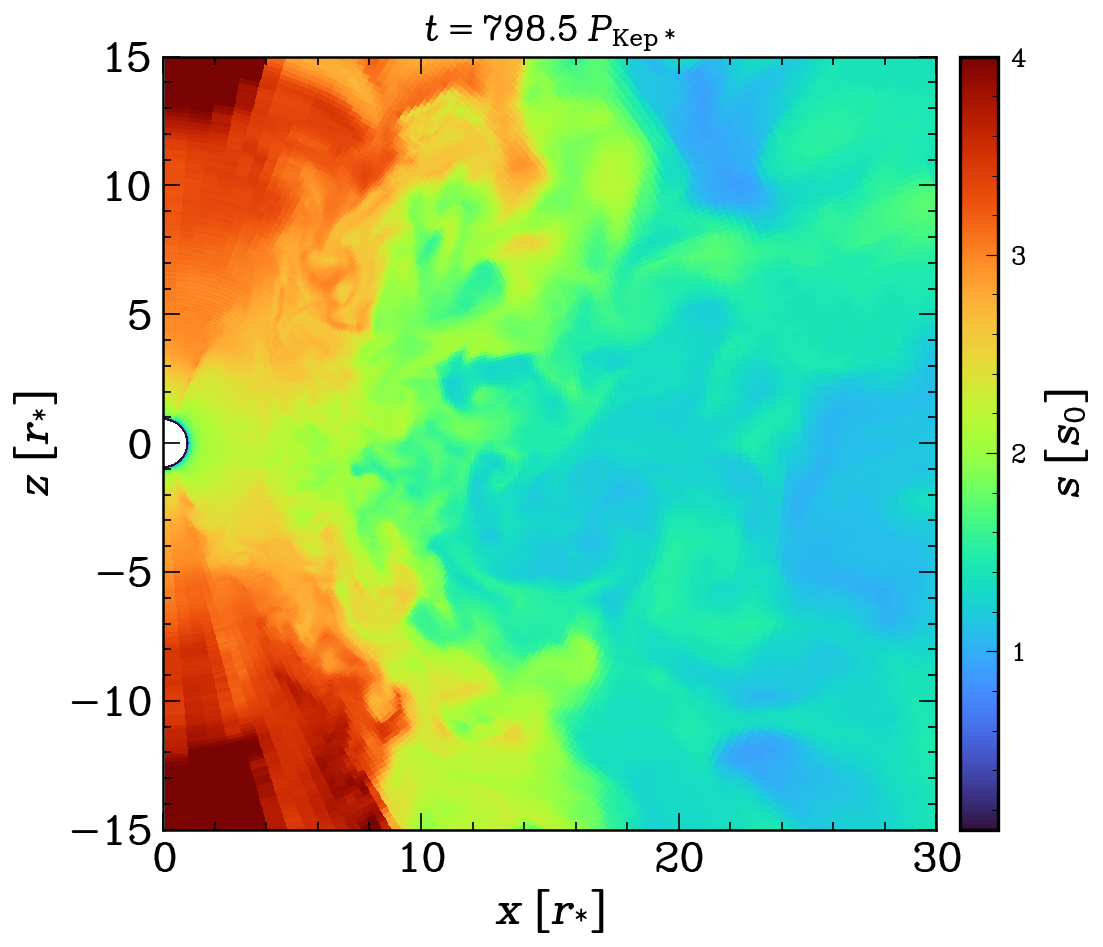}
    \caption{Snapshot of the entropy $s=p/\rho^{\Gamma}$ in the meridional plane at $t=798\,P_{\rm Kep\star}$ for the \texttt{Cooling} simulation normalized to $s_0=\rho^{1-\Gamma}$. A radial entropy gradient is observed between the approximately isentropic atmosphere region and the RIAF onto it. The high entropy zones in the polar region correspond to the magnetic funnel.}
    \label{fig:entropy}
\end{figure}

Turbulent hydrodynamic (Reynolds) stresses in the outer RIAF are subdominant with respect to magnetic stresses by almost an order of magnitude (see Figure \ref{fig:angflux}). Moreover, these Reynolds stresses transport angular momentum inward across the entire radial domain in both the \texttt{Fiducial} and \texttt{Cooling}
simulations. MRI-driven accretion flows, including SANE RIAFs, usually
transport angular momentum outward. 
Inward transport by Reynolds stresses suggests that convective motions are present in the flow; this association has been demonstrated in several numerical and analytical analyses of accretion \citep{igumenshchev2000numerical, ryu1992convective, stone1996angular, narayan2012GRMHD, white2020}. 

It has recently been argued that magnetically arrested RIAFs are convectively unstable and lead to inward transport of angular momentum 
by Reynold stresses \citep{begelman_what_2022}. It is an open question whether the MRI is active in
this situation (see \citealt{chatterjee_flux_2022} for a contrasting view), but this effect cannot operate in our simulations, where no part of the accretion flow reaches a MAD state.
No persistent eruptions are observed in the later evolution (see Section \ref{sec:magnetized_accretion}) and in the RIAF zone, outside the atmosphere, we measure $\beta^{-1}\sim 0.2$ with an angular velocity 
that is close to Keplerian, $\Omega(r) \sim 0.8 \, \Omega_{\rm Kep}(r)$ for $r>r_{\rm atm}$. Convection in our model could be triggered instead by radial entropy gradients (Figure \ref{fig:entropy}) extending from the
nearly isentropic atmosphere into the RIAF zone. We find further evidence for convection in the form of 
turbulent, radially directed eddy motion in the rest-mass density. The contribution from this turbulent hydrodynamical component 
to the Reynolds stress is subdominant except at the outer edge of the atmosphere.

We find that the direction of the angular momentum flux reverses in the atmosphere, pointing inward.
Here, the angular velocity profile is nearly flat and sub-Keplerian, $\Omega\sim r^{0.5}$ (Figure~\ref{fig:omegar}), and the MRI is no longer active.  The hydrodynamic and magnetic stresses
are comparable around the transition to inward transport.
The advected angular momentum flux $\dot J_{\rm adv}$ decreases as a power law in $r$ as $r\rightarrow r_\star$. In the \texttt{Fiducial} run, this power law is particularly steep, as mass accretion is suppressed close to the stellar surface. In the \texttt{Cooling} run, some angular momentum is advected all the way to the NS surface, because the mass accretion reaches a steady state and $d\dot{M}/dr \rightarrow 0$ (Figure~\ref{fig:q_r_all}). The advective flux of angular momentum then scales as
$\dot{J}_{\rm adv} \sim \dot{M}\,\Omega \,r^2 \propto r^{2.5}$, which is in good agreement with 
what we find (Figure~\ref{fig:angflux}). 

The angular momentum flux close to the NS ($r \lesssim 3\,r_\star$) is therefore dominated by 
the magnetic stress 
in the \texttt{Fiducial} and \texttt{Cooling} runs.  
Here, the slowly rotating polar cap of the NS is connected
to the equatorial accretion flow by field lines that become strongly sheared
(as described in Section \ref{sec:magnetized_accretion}).  This shearing transfers angular momentum
from the more rapidly rotating flow to the star.
By contrast, in the \texttt{LargeLoop} simulations, the stronger coherent vertical field does not
connect the star to the flow in this way; $\dot J_M$ is reduced significantly as a result.  
The angular momentum flux is dominated by advection near the star in the \texttt{LargeLoop}\_$\beta100$ run,
decreasing strongly at late times (Figure~\ref{fig:jdot_t}).  Here, the angular velocity profile in the atmosphere is steeper relative to the more weakly magnetized simulations.

\begin{figure}
    \centering
    \includegraphics[width=1\linewidth]{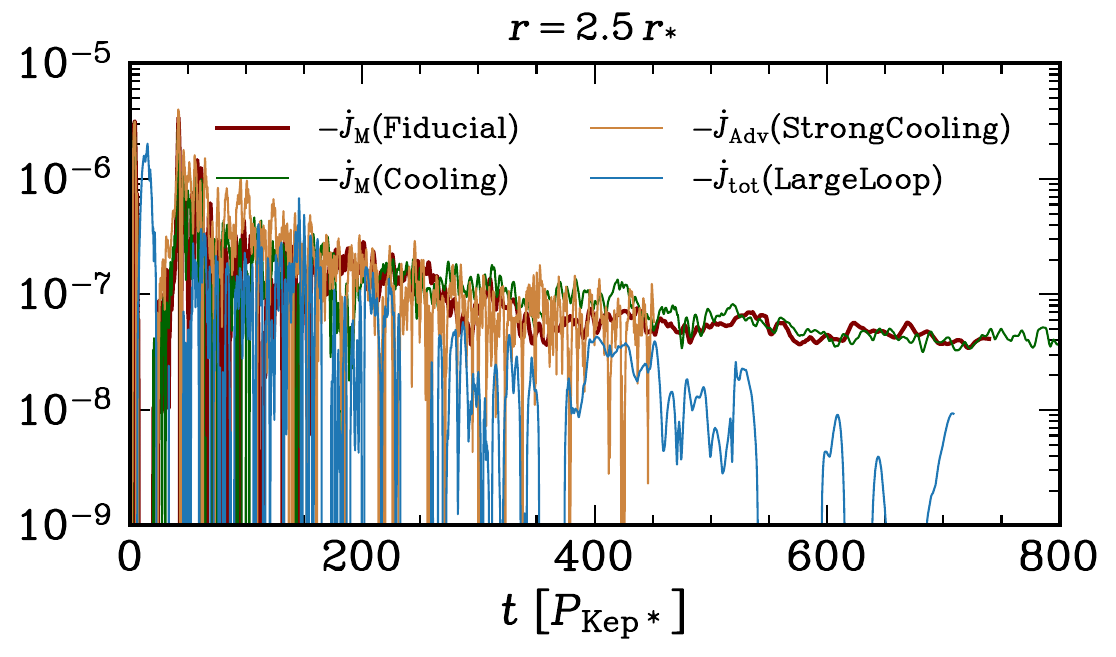}
    \caption{Magnetic fluxes $(\dot{J}_{\rm M})$ for \texttt{Cooling} and \texttt{Fiducial}, advective flux $(\dot{J}_{\rm Adv})$ for \texttt{StrongCooling}, and total flux $(\dot{J}_{\rm tot})$ for $\texttt{LargeLoop}$\_$\beta100$ through a sphere of radius $r=2.5\,r_\star$.
    Magnetic field lines connecting the non-rotating polar cap of the star and the rotating atmosphere transport angular momentum inward in both \texttt{Fiducial} and \texttt{Cooling} simulations at a constant rate. This does not happen in the \texttt{LargeLoop} simulation, given the larger initial field, and the total flux decays. In the \texttt{StrongCooling} case, the total angular momentum is driven predominantly by advection.} 
    \label{fig:jdot_t}
\end{figure}

To summarize, the profile of the angular momentum flux and its various
components (Figure~\ref{fig:angflux}) differs considerably from a classic rotationally supported accretion disk, one that extends all the way to the compact object.  
In this later case, angular momentum is transported outward by turbulence
(generated, e.g., by the MRI) relatively close to the gravitating object;
the specific angular momentum accreted near the stellar surface or BH 
innermost stable circular orbit is (approximately) the Keplerian 
value \citep{noble2010}.  In both our adiabatic and \texttt{Cooling} simulations,
$\dot J_{\rm M}$ is positive outside the atmosphere, but there is a transition to inward transport within the atmosphere, one that 
is dominated by magnetic stresses close to the star.
In the \texttt{StrongCooling} run, more similar to a BH RIAF, the inward angular momentum flux is dominated by advection (Figure~\ref{fig:jdot_t}), whereas
in the \texttt{Fiducial} and \texttt{Cooling} runs $\dot J_{\rm adv}$ and
$\dot J_{\rm M}$ are comparable in magnitude near the atmospheric boundary.
Magnetic fields therefore maintain a key role in transferring angular momentum
from the RIAF to the star in the presence of an atmosphere.

\section{Implications}
\label{sec:implications}

The simulations of magnetized accretion onto a NS presented in Section~\ref{sec:results} are consistent with the expectations outlined in Section~\ref{sec:analytic}. In the absence of strong cooling, the accretion flow indeed develops a central, nearly hydrostatic atmosphere around the NS with roughly uniform angular velocity $\Omega_{\rm atm}(r)$ and a density profile only slightly steeper than expected ($\rho \sim r^{-3}$) for an isentropic, spherical, hydrostatic atmosphere in the absence of rotation. Beyond roughly the predicted radius $r_{\rm atm}$, there is a transition to a flatter density profile characteristic of a RIAF, $\rho(r) \propto r^{-1}$.

{The thermodynamical regimes we considered in this paper apply to astrophysical scenarios where a NS intercepts gas from the envelope of a star, e.g., in a common envelope phase, in late supernovae fallback, or a tidal disruption event in a populated cluster. Current simulations resolving the physics near the Bondi radius do not include the NS explicitly in the domain, missing the feedback to the global flow structure; the accretion rate estimates from these calculations should thus be taken as approximate. 
    
Recent supernova fallback simulations show that accretion rates onto the NS lie between $\dot{M} \sim 10^{-2}-10^{-6}\,M_{\odot}\,{\rm s}^{-1}$ evolving as $\propto t^{-5/3}$ for a duration of $\sim 10^5 \:{\rm s}$  \citep{janka_fb_2022}. For the largest masses and at early times, this suggests that cooling at the NS surface might be strong and thus a RIAF structure can extend close to the NS, similar to our \texttt{StrongCooling} simulation. For most of the evolution, however, a nearly hydrostatic atmosphere is expected to mediate the accretion, as we show in our \texttt{Cooling} simulations. 

Tidal disruption event simulations of a neutron star making a close passage by a main-sequence star predict smaller accretion rates of $\dot{M}\sim 10^{-5}-10^{-6}\,M_{\odot}\,{\rm s}^{-1}$ lasting $\sim 10^5-10^6\: {\rm s}$ \citep{kremer_formation_2022}. Mass transfer to the NS would then be mediated by an atmosphere, with important consequences for angular momentum transfer (see below); the long-term behavior of the magnetic field in this case is, however, crucial to determine the final spin of the NS and remains an open question. 

In a common-envelope phase, as the NS inspirals toward the center of the star companion, the accretion rate might be limited by the asymmetry of the flow resulting in $\dot{M}\sim10^{-9}\,M_{\odot}\,{\rm s}^{-1}$ (smaller than the expected Bondi-Hoyle-Lyttleton rate, \citealt{macleod2014accretion}) for orbital periods of $\sim 10^{7}-10^{8}\:{\rm s}$. Accretion onto the NS would likely be mediated by a large atmosphere as in our \texttt{Cooling} and \texttt{Fiducial} simulations.}  


\subsection{Implications for NS Spin and Magnetism}

The presence of the atmosphere has consequences for the spin acquired by the NS through accretion, and the interaction of the accretion flow with the NS magnetic field. The NS will accrete a net mass $M_{\rm acc} \sim (r_{\rm atm}/r_0)^{1/2}M_0$
(where $M_0$ is the initial bound torus mass and $r_0$ its initial radius).  Given that initially the NS
is weakly magnetized and slowly rotating, the net angular momentum accreted by the star is
$\Delta J_\star \sim (2/3)M_{\rm acc}r_\star^2\Omega_{\rm atm}$, where $\Omega_{\rm atm}
\sim (GM/r_{\rm atm}^3)^{1/2}$.  From Eq.~(\ref{eq:Ratm}), we have
\beq{
  M_{\rm acc} \sim 1\times 10^{-3}\,\alpha_{-1}^{1/8}R_{0,10}^{-3/16}
  \left({M_0\over 10^{-2}\,M_\odot}\right)^{27/32}\; M_\odot,
}
  and
\beq{
  P_\star < {2\pi I_\star\over \Delta J_\star} \sim 180\,\alpha_{-1}^{1/4}R_{0,10}^{9/8}
    \left({M_0\over 10^{-2}\,M_\odot}\right)^{-21/16}\; {\rm s}.
}

Greater angular momentum will be acquired by the star as the accreted magnetic field becomes tied
to it.  For example, in Figures \ref{fig:outflowsfield} and \ref{fig:alpha_vs_r} one observes
in the \texttt{Cooling} simulation that these connecting field lines extend to $r_{\rm mag} \sim 2.5$ 
stellar radii, where the Maxwell stress transmits angular momentum inward.   The accreted angular momentum increases by a factor $\sim (r_{\rm mag}/r_\star)^2$.

A global simulation offers insights about the magnitude of the stresses developing near the NS surface (even if
it cannot adequately resolve the long-term evolution of the magnetic field anchored in the solid crust).
The rotational stress
\beq{
[\rho \Omega^2 r^2]_{r_\star} \simeq \left({r_\star\over r_{\rm atm}}\right)^{-1}[\rho \Omega^2 r^2]_{r_{\rm atm}}
}
is smaller by a factor $\sim r_\star/r_{\rm atm}$ than would be implied by 
a direct extrapolation of the RIAF from $r > r_{\rm atm}$ 
into the NS surface.
On the other hand, the total pressure is enhanced by a factor $\sim (r_{\rm atm}/r_\star)^2$.  The corresponding
equipartition magnetic fields are (for $r_\star = 10$ km and $M = 1.4\,M_\odot$)
\begin{eqnarray}\label{eq:beq1}
  B_{\rm eq} &=& [4\pi \rho\Omega^2r^2]_{r_\star}^{1/2}\nnn
    &=& 2\times 10^{11}\,\alpha_{-1}^{-1/8}R_{0,10}^{-21/16}
    \left({M_0\over 10^{-2}\,M_\odot}\right)^{21/32}\;{\rm G};\nnn
\end{eqnarray}
\begin{eqnarray}\label{eq:beq2}
  B_{\rm eq} &=& [8\pi P(r_\star)]^{1/2} \nnn
  &=& 1\times 10^{14}\,\alpha_{-1}^{1/4}R_{0,10}^{-3/8}\left({M_0\over 10^{-2}\,M_\odot}\right)^{3/16}\;{\rm G}.\nnn
\end{eqnarray}
We observe (Figure \ref{fig:forces}) that the accreted magnetic field is slightly larger than given by Equation (\ref{eq:beq1})
but much smaller than (\ref{eq:beq2}).

A strong toroidal field is observed in the atmosphere (Figure \ref{fig:outflowsfield}), indicating that
a large-scale field mediates the transport of angular momentum, thereby flattening the 
angular velocity profile.  {Consider, in particular, 
the zone in the \texttt{Cooling} simulation extending from 
$\theta \sim 30-70^\circ$ and outward to several stellar radii.  
Here, the poloidal magnetic field is approximately radial,
the flow is generally inward, and the toroidal field is relatively strong and of the opposite sign to $B_r$.}
This leads us to consider the profiles of $B_\phi$ and $\Omega$ in a hydrostatic 
atmosphere through which mass settles slowly toward the NS at a uniform rate $\mathrm{d}\dot M = -\rho v_r \mathrm{d}A$, and
is threaded with magnetic flux $\mathrm{d}\Phi_r = B_r \mathrm{d}A =$ const.
We neglect non-radial forces and focus on the exchange of angular momentum between matter and electromagnetic field
within a column of solid angle $\mathrm{d}\Omega$ and area $\mathrm{d}A = r^2 \mathrm{d}\Omega$.  Balancing the change in angular momentum flux
against the hydromagnetic torque gives
\beq{
{\partial\over\partial r}\left(\Omega r^2\sin^2\theta\cdot \rho v_r \mathrm{d}A\right) = 
{\partial\over\partial r}\left({B_rB_\phi\over 4\pi}r\sin\theta\, \mathrm{d}A\right).  
}
Integrating this equation gives (taking $\Omega_{\rm atm} \simeq$ const),
\beq{
B_\phi r 
\;\simeq\; (B_\phi r)_{r_{\rm atm}} + {4\pi \Omega(r_{\rm atm})r_{\rm atm}^2\over \mathrm{d}\Phi_r/\mathrm{d}\dot M}\sin\theta.
}
The angular velocity profile is likewise readily estimated in a quasi-steady state,
\beq{
{\partial B_\phi\over\partial t} = {1\over r}{\partial\over\partial r}\left[r(-v_rB_\phi + \Omega r\sin\theta B_r)\right],
}
giving
\begin{eqnarray}
{\partial\Omega\over\partial r} &=& 
-{1\over 2\pi\sin\theta (\mathrm{d}\Phi_r/\mathrm{d}\dot M)}{\partial\over \partial r}\left({B_\phi\over\rho r}\right)\nnn
&\simeq&  -{B_\phi\over 2\pi \rho r\sin\theta (\mathrm{d}\Phi_r/\mathrm{d}\dot M)}.
\end{eqnarray}
{One sees that opposing signs of $B_r$
and $B_\phi$, as we described in Figure \ref{fig:outflowsfield}, are consistent with inflow through a hydrostatic atmosphere with positive $\partial\Omega/\partial r$.

\subsection{NS Propeller?}

A final check on the consistency of our calculation involves 
evaluating the possibility of a propeller-driven outflow during the
earlier stages of accretion, before a hydrostatic atmosphere is 
established.  Let us compare
the initial stress that is applied to the NS boundary by a RIAF
with the magnetic stress that might be applied back to the flow by
a rotating magnetosphere.  As the initial torus of mass $M_0$
and radius $r_0$ spreads inward toward the NS, we take a
density profile $\rho(r) \sim M_0/2\pi r_0^2r$ and
rotation profile $\Omega^2 \sim \Omega_{\rm Kep}^2/2$ for the flow
at radius $r < r_0$.  We also allow the initial NS to have a magnetic
moment $\mu_\star$ and spin rate $\Omega_\star$.

A magnetosphere forms when $\mu_\star$ is large enough.
The magnetospheric boundary sits approximately at the radius $r_A$ where
$(\mu_\star/r_A^3)^2/4\pi \sim \rho(r_A)\Omega^2(r_A)r_A^2$,
or equivalently,
\begin{equation}
    r_A = {\mu_\star^{1/2}r_0^{1/2}\over (GMM_0)^{1/4}}.
\end{equation}
Consistency here requires a torus column $M_0/r_0^2$ small enough
that $r_A > r_\star$, or equivalently
\begin{equation}
    \mu_\star > 6\times 10^{30}\,{M_{1.4}^{1/2}\over r_{0,10}}\left({M_0\over 10^{-2}M_\odot}\right)^{1/2}\;
    {\rm G\,cm^3}.
\end{equation}
Entrainment in this rotating magnetic field can push plasma 
outward when $\Omega_\star^2 \gtrsim \Omega_{\rm Kep}^2(r_A)/2$.
Taking a fixed magnetic moment $\mu_\star = 10^{30}\,\mu_{\star,30}$ G cm$^3$
and spin period $P_\star = 1\,P_{\star,0}$ s,
this requires a torus mass smaller than
\begin{equation}
M_0 \;<\; 9\times 10^{-11}\,{\mu_{\star,30}^2 r_{0,10}^2
\over M_{1.4}^{7/4}P_{\star,0}^{8/3}}\;M_\odot.
\end{equation}
Unless the neutron star is strongly magnetized compared with 
ordinary radio pulsars ($\mu_\star \gg 10^{30}$ G cm$^3$) 
and/or still rapidly rotating (with $P_\star \ll 1$ s), 
this column is too small for our calculation to be internally
consistent:  photons will not remain trapped in the flow
(see Equation (\ref{eq:trad})).  Otherwise, we may neglect the
possibility that the neutrino-cooled flow is suppressed by
propeller heating.}

\section{Summary}
\label{sec:conclusion}

We perform novel GRMHD simulations of a hypercritical plasma flow around and onto a non-rotating and initially unmagnetized NS.  The initial conditions represent a magnetized, radiation-pressure dominated torus.  The photon field 
is effectively locked into the flow;  radiative losses to neutrinos may be triggered only after the flow reaches the NS surface.
The outer layers of the NS are represented as a fixed, hydrostatic, TOV star solution.  We compare and analyze simulations with strong, mild, and negligible optically thin neutrino cooling corresponding to different plasma densities. 
We also probe the effects of small and large initial magnetic flux reservoirs in the accretion disk. 

The main results can be summarized as follows:
\begin{enumerate}

    \item In all simulations with mild and negligible neutrino cooling, we find the self-consistent development of a nearly hydrostatic atmosphere around the NS.  This atmosphere has a steep density profile, $\rho \propto r^{-3.5}$, which transitions smoothly into a RIAF with significant centrifugal support and 
    a softer density profile, $\rho \propto r^{-1}$.
    
    \item In the adiabatic simulations (\texttt{Fiducial} and \texttt{LargeLoop}), no inflow equilibrium is found. The accretion rate drops near the surface of the NS as an atmosphere forms.  The size of the inner hydrostatic zone
    increases $\propto t^{0.37}$ as additional plasma is brought in from the RIAF.
    In the simulation with weak neutrino cooling (\texttt{Cooling}), a balance between accretion power and radiative losses is reached, thereby stalling the growth of the atmosphere and establishing an equilibrium between inflow from
    the RIAF and settling of the atmosphere onto the NS.  
    The amplitude of neutrino emission is high enough in the \texttt{StrongCooling} simulation to prevent the formation
    of a hydrostatic atmosphere; the RIAF structure extends all the way to the NS surface.
    
    \item When an atmosphere form, the angular velocity profile 
    flattens at $r < r_{\rm atm}$.
    The rotation rate peaks at $\Omega_{\rm atm} \sim \Omega_{\rm Kep} (r_{\rm atm})$ and maintains a small positive slope, $\Omega_{\rm atm}\propto r^{1/2}$. The atmosphere is almost entirely pressure-supported, with centrifugal forces being negligible except within an outer transition layer.  
    In the \texttt{StrongCooling} runs, an atmosphere is absent and the Keplerian angular velocity profile extends almost up to the NS surface.

    \item {The slowing of
    rotation in the inner atmosphere limits the angular momentum that can be
    accreted by the NS, and the strength of the velocity shear near the stellar surface.  For example, this effect will limit the rate of accretion of 
    angular momentum from a small amount ($\lesssim 10^{-2}\,M_\odot$) of
    fallback material in the later stages of a core collapse event.  Existing
    simulations compute the angular momentum assembled near the Bondi radius
    10 or more seconds after collapse, but do not fully resolve the structure of
    the accretion flow near the NS surface \citep{janka_fb_2022}.}
    
    \item Simulations with 
    a large-scale seed magnetic field give rise to flux eruptions during the initial part of the simulation (until $t\approx 100\,P_{\rm Kep\star}$), similar to those found in MADs around BHs. We find that these eruptions cease because the growing matter-dominated atmosphere efficiently traps the accreted magnetic flux and shields it from dynamic interaction with the RIAF, in stark contrast with accretion around BHs in the MAD state. For strong cooling and/or different boundary conditions, this conclusion might change.

    \item In regions where magnetic field lines anchor to the atmosphere, a strong large-scale toroidal field forms. Inside the nearly hydrostatic atmosphere, a small positive angular velocity gradient amplifies the toroidal field by magnetic winding. Electromagnetic energy can be extracted from the accreting NS system with an efficiency of $\langle L_{\rm M}\rangle _t / \langle\dot{M}\rangle_t \sim 1\%$ relative to the accretion power into the atmosphere, largely independent of the configurations investigated here.
    
    \item Magnetic stresses mediate the outward transport of angular momentum in the RIAF zone (thick red lines in Figure~\ref{fig:angflux}).  
    The measured $\alpha$ parameter is consistent
    with transport by the MRI (Figure~\ref{fig:alpha_vs_r}). 
    On the other hand, $d\Omega/dr$ is positive in the atmosphere,
    meaning that the
    MRI is suppressed there.  We also find that the transport of angular momentum is 
    inward through the atmosphere, being mediated mainly by magnetic stresses from winding. Turbulent hydrodynamic stresses are subdominant; they transport angular momentum inward across almost the entire domain.
    Finally, the advective transport of angular momentum reaches all the way to the NS in the \texttt{Cooling} run, whereas
    it is strongly suppressed at small $r$ in the adiabatic simulations.    

\end{enumerate}

L.C.~thanks Claire Ye, Alex Dittmann, Geoff Ryan, and Brian Metzger for useful discussions, and Scott Noble for developing and providing the {\sc HARM3D} code used in this work. This research was enabled in part by support provided by SciNet (www.scinethpc.ca) and Compute Canada (www.computecanada.ca) and by the VSC (Flemish Supercomputer Center), funded by the Research Foundation Flanders (FWO) and the Flemish Government---department EWI. The authors gratefully acknowledge the computing time made available to them on the high-performance computer ``Lise'' at the NHR Center NHR@ZIB. This center is jointly supported by the German Federal Ministry of Education and Research and the state governments participating in the NHR (www.nhr-verein.de/unsere-partner). L.C.~is a CITA National fellow and  acknowledges the support by the Natural Sciences and Engineering Research Council of Canada (NSERC), funding reference DIS-2022-568580. C.T.~acknowledges the support of NSERC, funding reference number RGPIN-2023-04612.  D.M.S.~acknowledges the support of NSERC, funding reference number RGPIN-2019-04684. D.M.S.~acknowledges a Visiting Fellow position at Perimeter Institute.  A.P. acknowledges support by an Alfred P.~Sloan Research Fellowship and a Packard Foundation Fellowship in Science and Engineering.  This research was supported in part by Perimeter Institute for Theoretical Physics. Research at Perimeter Institute is supported in part by the Government of Canada through the Department of Innovation, Science and Economic Development Canada and by the Province of Ontario through the Ministry of Colleges and Universities. C.T., A.P.~and B.R.~are supported by a grant from the Simons Foundation (MP-SCMPS-00001470).
B.R.~is supported by NSERC, and the Canadian Space Agency (23JWGO2A01). B.R.~acknowledges a guest researcher position at the Flatiron Institute, supported by the Simons Foundation. 


\software{\texttt{Matplotlib} \citep{hunter2007Matplotlib}, \texttt{NumPy} \citep{harris2020Array}, \texttt{SciPy} \citep{virtanen2020SciPy}, and \texttt{hdf5} \citep{hdf5}.}

\appendix

\section{Density Scale and Neutrino Emission}
\label{app:density_scale}

Although we use a specific TOV solution to define fixed boundary conditions at the surface of the NS, the disk mass is orders of magnitude smaller than the total system mass;  this means that the maximum density of the initial torus $\rho_{\rm max}$ is effectively an independent parameter, at least when neutrino losses may be neglected.
An absolute density scale is introduced by the choice of the neutrino cooling function.
In our simulations, there are two potentially relevant emission processes: $e^{\pm}$ annihilation and $e^{\pm}$ captures on free neutrons and protons. These processes dominate over each other at different accretion rates, and they can be described by the same functional form $Q \propto p^{3/2} \rho$ in a nearly adiabatic hydrostatic atmosphere with $p \propto T^4$ and $\rho \propto T^3$. 

The code assumes geometric units and $M=1$. After we recover $G$ and $c$ in physical units and specify the total mass of the spacetime $M$, the length and time scales are 
\begin{equation}
    r_{\rm cgs} = r_{\rm cu} r_g =  r_{\rm cu} \frac{GM}{c^2}; \quad t_{\rm cgs} = t_{\rm cu} \frac{GM}{c^3},
\end{equation}
where ``cu'' denotes code units.  
The density and pressure are converted to physical units as
\begin{equation}
    \rho_{\rm cgs} = \rho_{\rm cu} {M\over r_g^3};\quad p_{\rm cgs} = p_{\rm cu}{Mc^2\over r_g^3};
\end{equation}
The normalization factor $A_Q$ in the cooling function \eqref{eq:Qfunc} fixes the density scale in the following way. Let us consider the net neutrino emission rate \citep{fernandez2009stability},
\begin{eqnarray}\label{eq:qnu}
    Q &=& Q^{\rm cap} + Q^{\rm ann} = Q^{\rm cap}\left(1+{m_p s\over 60k_{\rm B}}\right)\nnn
    &=& 1.4 \times 10^{36} \rho_{12} T_{\rm MeV}^6 \left(1+{m_p s\over 60k_{\rm B}}\right) \: \textrm{MeV} \: \textrm{cm}^{-3} \: \textrm{s}^{-1} =  A_{Q,\rm cgs} \rho_{\rm cgs} \: p_{\rm cgs}^{3/2}.
\end{eqnarray}
Here, the pressure is dominated by radiation and relativistic pairs, $p = (11a/12) T^4$, 
and $A_{Q,\rm cgs} = 1.6\times 10^{-21}(1+m_p s/60k_{\rm B})~\rm{erg~cm^{-3}~s^{-1}}$.   
The plasma enthalpy is radiated to neutrinos over a characteristic time $t_{\rm cool} = 4p/Q$.  
Assuming a hydrostatic atmosphere with $p = \rho c^2 (r_g/4r)$, and 
cancelling factors of $r_{\rm cu}$ and $t_{\rm cool,cu} = ct_{\rm cool,cgs}/r_g$, one finds
\begin{equation}
    \rho_{\rm cgs} = \rho_{\rm cu}
    \left(\frac{A_{Q,\rm cu}}{r_g A_{Q,\rm cgs}}\right)^{2/3}.
\end{equation}

Using $M=1.4\:M_{\odot}$ and $A_{Q,\rm cu} =10^{11},10^8$ for \texttt{StrongCooling} and \texttt{Cooling}, respectively, we obtain that the {time-averaged} accretion rates measured at the end of the simulations are, in physical units, $\langle \dot{M} \rangle_t \sim 6.1 \times 10^{-8} M_{\odot}\:s^{-1}$ for \texttt{Cooling} and $\langle \dot{M} \rangle_t = 6.1 \times 10^{-5} M_{\odot}\:s^{-1}$  for \texttt{StrongCooling}. 

\section{Settling Flow Mediated by Near-Surface Neutrino Emission}\label{sec:cooling_layer}

Consider a spherical atmosphere around a NS that is supported by radiation pressure.  Close
to the stellar surface, the temperature is high enough for neutrino cooling to remove energy;
the temperature is therefore high enough near the surface for $e^\pm$ pairs to contribute significantly
to the enthalpy.  Combining the equations of hydrostatic equilibrium and energy conservation,
and assuming solid rotation with angular velocity $\Omega_{\rm atm}$ in the atmosphere, we have
\begin{eqnarray}\label{eq:basic}
    &&-{1\over\rho}{dp\over dr} - {GM\over r^2} + \Omega_{\rm atm}^2 r = 0;\nnn
    &&\rho v_r T{ds\over dr} = -{\dot M\over 4\pi r^2}
    \left({dh\over dr} - {1\over\rho}{dp\over dr}\right) = -Q.
\end{eqnarray}
Outside the surface neutrino cooling layer, the hydrostatic profile is 
$h = T s_{\rm atm} =  GM/r + {1\over 2}\Omega_{\rm atm}^2 r^2$.  As the atmosphere settles 
toward the NS with negligible radiative losses or heat conduction, the entropy $s_{\rm atm}$ per
unit baryonic mass may be assumed
constant.   The density near the surface is then $\rho(r) \simeq {11\over 3}a(GM/r)^3s_{\rm atm}^{-4}$.

Adopting dimensionless variables $T = m_ec^2\widetilde T$, $\rho = (a/c^2)(m_ec^2/k_{\rm B})^4\widetilde\rho$ 
and $r = r_g\widetilde r$, Equations (\ref{eq:basic}) may be combined to give (near the NS surface)
\begin{eqnarray}
    {d\over d\widetilde r}\left({11\over 12}\widetilde T^4\right) &=& 
    -{\widetilde\rho\over\widetilde r};\nnn
    {11\over 3}{\widetilde T^4\over\widetilde\rho^2}{d\widetilde\rho\over d\widetilde r} 
    + {3\over\widetilde r} &=& - 2.5\times 10^{-12} \widetilde r^2 \widetilde\rho \widetilde T^6
    \left(1+{m_p s_{\rm atm}\over 60k_{\rm B}}\right)\left({\dot M\over 10^{-5}\,M_\odot~{\rm s}^{-1}}\right)^{-1}
    M_{1.4}^3.
\end{eqnarray}
In the cooling layer, the temperature increases monotonically inward, but not as rapidly as the $e^\pm$ Fermi energy.  These equations are integrated inward from the outer adiabatic solution until the $e^\pm$ gas 
becomes degenerate.  This gives a direct
relation between $s_{\rm atm}$ and $\dot M$, as reported in Equation (\ref{eq:svsmdot}) and the
corresponding relation before Equation (\ref{eq:Ratm}).

\section{Energy momentum tensor decomposition and stresses}\label{app:tmunu}

The energy-momentum tensor in ideal MHD is defined as ${T^{a}}_{b} = (T_{\rm M})^{a}_{\: b} + (T_{\rm H
})^{a}_{\: b}$, where
\begin{equation}
{T_{\rm M}}^{a}_{\: b} = b^2 u^a u_b + \frac{b^2}{2} {\delta^a}_b - b^a b_b
\end{equation}
is the electromagnetic part and 
\begin{equation}
{T_{\rm H}}^{a}_{\: b} = h \rho u^{a} u_{b} + p \, \delta^{a}_{\:b}
\end{equation}
is the hydrodynamic part. The hydrodynamical contribution can be further split into Reynold (turbulent) stresses ${R^a}_b$ and advected angular momentum ${A^a}_b$: ${T_{\rm H}}^{a}_{\: b} = {R^a}_b + {A^a}_b$. The Reynolds stress is defined with respect to the comoving mean flow: 
\beq{
{R^a}_b = \rho h \, \delta u^a \,  \delta u_b
\,. \label{reynolds-stress}
}
We denote perturbations with respect to the mean flow by `$\delta$', and define these as deviations from the energy-weighted, spherically averaged flow. For example, for the 4-velocity, we have
\beq{
\delta u^a \equiv u^a - \langle u^a \rangle_{\rho h}
\,. \label{velocity-perturbation}
}
Taking the spherical average of the Reynold stress and using Eq.~\eqref{velocity-perturbation}, we find that the advection part is given by the uncorrelated spherical average of each component:
\begin{equation}
    \langle {A^{a}}_{b} \rangle = \langle \rho h \rangle \langle u^a \rangle_{\rho h} \langle u_b \rangle_{\rho h},
\end{equation}
and thus
\begin{equation}
    \langle {R^a}_{b} \rangle = \langle \rho h u^a u_b\rangle - \langle \rho h \rangle \langle u^a \rangle_{\rho h} \langle u_b \rangle_{\rho h}.
\end{equation}

The energy-momentum tensor can also be split as
\begin{equation}
    {T^a}_b = \rho (h+\sigma) u^a u_b  + p(1+\beta){\delta^{a}}_b + \left(\frac{b^2}{2} u^a u_b - b^a b_b\right). \label{eq:decomposition_energy-momentum_tensor}
\end{equation}
The first term represents the total energy, with $\sigma = b^2/2\rho$ the magnetization, the second term the total isotropic pressure, and the third term represents the magnetic stresses.

\section{Forces from conservation of radial momentum fluxes} \label{sec:force}

We compare the different forces acting on the fluid by decomposing the conservation equation of the radial momentum $\nabla_{a} T^{a}_r = 0$. Because the vector field $\partial_r$ is not associated with any symmetry of the spacetime, the radial momentum flux equation can be written as a conservation equation with an additional source term:
\begin{equation}
    \partial_{a} ({\sqrt{-g}} T^a_{r}) = \sqrt{-g} \, \Gamma^b_{cr} T^c_b.
\end{equation}

Integrating this equation over a 3D volume, one finds that the change of radial momentum over this volume, i.e. the total force, depends on a non-local term:
\begin{equation}
    F^r =\frac{d P^r}{d(ct)} = -\int T^r_{r} \sqrt{-g} \,\mathrm{d}A + \int \Gamma^b_{cr} T^c_b \sqrt{-g} \, \mathrm{d}V.
\end{equation}

In spherically symmetric spacetimes, it is more convenient to work with the radial force density per area to avoid integrating over volume, i.e. we work with the total force per unit volume, where the volume is a spherical shell:
\begin{equation}
    \frac{\partial_r F^r}{A}  = \frac{dF^r}{dV_{\rm shell}} =  -\frac{\partial_r \lbrace T^r_r \rbrace}{A} + \frac{\lbrace  \Gamma^b_{cr} T^c_b \rbrace}{A}. \label{eq:force_per_volume}
\end{equation}
In areal (Schwarzschild-like) coordinates, one has $A= 4\pi r^2$. The terms in this equation have units of force per unit volume. 

We now want to study the different contributions to Eq.~\eqref{eq:force_per_volume}. For a Schwarzschild background spacetime, the second term on the right-hand side takes the form
\begin{equation}
     \Gamma^b_{cr} T^c_b = -\frac{1}{r} (T^{\theta}_{\theta} + T^{\phi}_{\phi}) + \frac{M}{r^2(1-2(GM/c^2)/r)} (T^r_{r}-T^t_t).
\end{equation}
Decomposing the first term into different parts and re-arranging the quantities we have
\begin{equation}
\frac{dF^r}{dV_{\rm shell}} = \textrm{(kinetic)} + \textrm{(centrifugal)} +\textrm{(pressure)} + \textrm{(gravitational)} + \textrm{(magnetic)}, 
\end{equation}
where
\begin{equation}
    \textrm{(kinetic)} \equiv 
    \Big \langle \rho (h+\sigma) u^r u_r \frac{2\alpha r +M}{(\alpha r)^2} \Big\rangle
    +\Big \langle \partial_r ( \rho (h+\sigma) u^r u_r) \Big \rangle,
\end{equation}
\begin{equation}
    \textrm{(centrifugal)} \equiv - \frac{1}{r} \Big \langle T^{\theta}_{\theta} + T^{\phi}_{\phi}-2p(1+\beta) \Big \rangle,
\end{equation}
\begin{equation}
    \textrm{(gravitational)} \equiv -\Big \langle {\rho (h+\sigma) u^tu_t} \frac{GM}{(\alpha r)^2} \Big \rangle,
\end{equation}
\begin{equation}
    \textrm{(thermal)} \equiv \Big \langle \partial_r ( p(1+\beta) \Big \rangle,
\end{equation}
and 
\begin{equation}
    \textrm{(magnetic)} \approx 
    \Big \langle   ({M^r}_r-{M^t}_t) \frac{GM}{(\alpha r)^2} \Big\rangle+
    \Big \langle \partial_r {M^r}_r \Big \rangle + \Big \langle  \frac{2}{r}{M^r}_r \Big \rangle.
\end{equation}
Here, we have used that $\partial_r \lbrace Q \rbrace/4\pi r^2 = \langle \partial_r Q \rangle + \langle 2 Q/r \rangle $, which is valid for a spherically symmetric spacetime, and we have defined ${M^a}_b= b^a b_b- \frac{b^2}{2} u^a u_b$ as the magnetic stress tensor (see Eq.~\eqref{eq:decomposition_energy-momentum_tensor}). We can further decompose the kinetic term, defining a turbulent pressure as $ \rho h \delta u^r \delta u_r$, where $\delta u^r = u^r-\langle u^r \rangle_{\rho}$. We then obtain $\lbrace \rho h \delta u^r \delta u_r \rbrace \approx \lbrace \rho h u^r u_r \rbrace - \lbrace \rho h u^r \rbrace^2/\lbrace \rho h \rbrace $, where the second term represents advection. We find that the turbulent component dominates the kinetic term.

If there is no motion in the $\theta$-direction, the hydrodynamic part of the centrifugal pressure gradient reduces to (centrifugal) $\sim-((T^{\theta}_{\theta} + T^{\phi}_{\phi})-2p(1+\beta))/r \sim  -\rho h u^\phi u_{\phi}/r = -\rho h (\gamma \alpha \Omega)^2 r$. In the Newtonian limit, this expression takes the well-known form (centrifugal)$\sim -\rho \Omega^2 r$, or, in terms of pressure, $-\rho (\Omega r)^2/2$. Similarly, the Newtonian limit of the gravitational term is (gravitational)$\approx -\rho(GM/r^2)$.

\section{Angular momentum flux balance}\label{sec:am}

Since our spacetime is axisymmetric and stationary, there are two Killing vector fields $t^a$ and $\phi^a$. These are aligned with the time and azimuthal directions in the simulations, respectively. It is easy to show that the current $j^a = T^{ab} \phi_{b}$ is conserved, $\nabla_{a} j^a =0$. Using Gauss's theorem, we can derive a balance law for the total angular momentum, defined as
\begin{equation}
J\equiv \int T^{ab} t_{a} \phi_{b} \sqrt{-g}\,\mathrm{d}^3x,
\end{equation}
which is given by
\begin{equation}
    \dot{J}_{\rm tot} \equiv \frac{\mathrm{d}J}{\mathrm{d}(ct)} = -\int T^r_{\phi} \sqrt{-g}\, \mathrm{d}\phi \mathrm{d}\theta.
\end{equation}

One can show from equations~(\ref{reynolds-stress}) and (\ref{velocity-perturbation}) that 
\beq{
\dot{J}_{\rm R} \equiv \left\{ {R^r}_\phi \right\} 
\ = \ \left\{ \rho h \, \delta u^r \, \delta u_\phi  \right\}
\ \simeq \  \dot{J}_{\rm H} - \dot{J}_{\rm Adv},
\label{reynolds-derivation-11}
}
where $\dot{J}_{\rm Adv}$ is calculated as the energy-averaged uncorrelated component originating in $T^r_\phi$:
\beq{ 
\dot{J}_{\rm Adv} \simeq 4\pi r^2 \times \langle \rho h \rangle  \langle u_{\phi} \rangle_{\rho h} \langle u^r \rangle_{\rho h} \simeq  \frac{\left\{ \rho h u_{\phi} \right\}  \left\{ \rho h u^r \right\} }{\left\{ \rho h \right\}}
\,. \label{advected-flux}
}
We define analogously the Maxwell component of the angular momentum flux, $\dot{J}_{\rm M} = \lbrace (T_{\rm EM})^{r}_{\phi}\rbrace$, in which we include both advection of magnetic energy and the shear term. The total angular momentum is then given by
\begin{equation}
    \dot{J}_{\rm tot} = \dot{J}_{R} + \dot{J}_{\rm Adv} + \dot{J}_{\rm M}.
\end{equation}

\section{Convergence and quality factor for the magneto-rotational instability} \label{sec:convergence}

 The quality of our resolution to resolve the MRI can be determined by monitoring the number of grid cells available to resolve the wavelength of the fastest growing unstable mode of the instability, defined for a given direction as
\begin{equation}
    \lambda^{(i)}_{\rm MRI} \equiv \textrm{(rotational period)} \times \textrm{(Alfv\'{e}n speed)} = \frac{2\pi}{\Omega} \frac{b^{(i)}}{\sqrt{\rho h +b^2}},
\end{equation}
where $i = r,\theta, \phi$. If this linear mode is under-resolved, the magnetic field amplification by the MRI is suppressed and, in extreme cases of very poor resolution, would lead to the gradual decay due to numerical dissipation, thus forcing the accretion rate to decrease in concert. We define the quality factor of the MRI as
\begin{eqnarray}
    Q^i_{\rm MRI}\equiv \frac{\lambda^{(i)}_{\rm MRI}}{\Delta x^{(i)}},
\end{eqnarray}
where $\Delta x^{(i)}$ is the grid cell size in a given coordinate direction. We find in all our simulations with a standard resolution of $n_{r}\times n_{\theta} \times n_{\phi} = 400 \times 256 \times 160$ that the quality factor satisfies, on average, $Q^i_{\rm MRI} \gg 10$ in all directions, as shown in Figure~\ref{fig:qmri}. 

\begin{figure}
    \centering
    
    \includegraphics[width=0.49\linewidth]{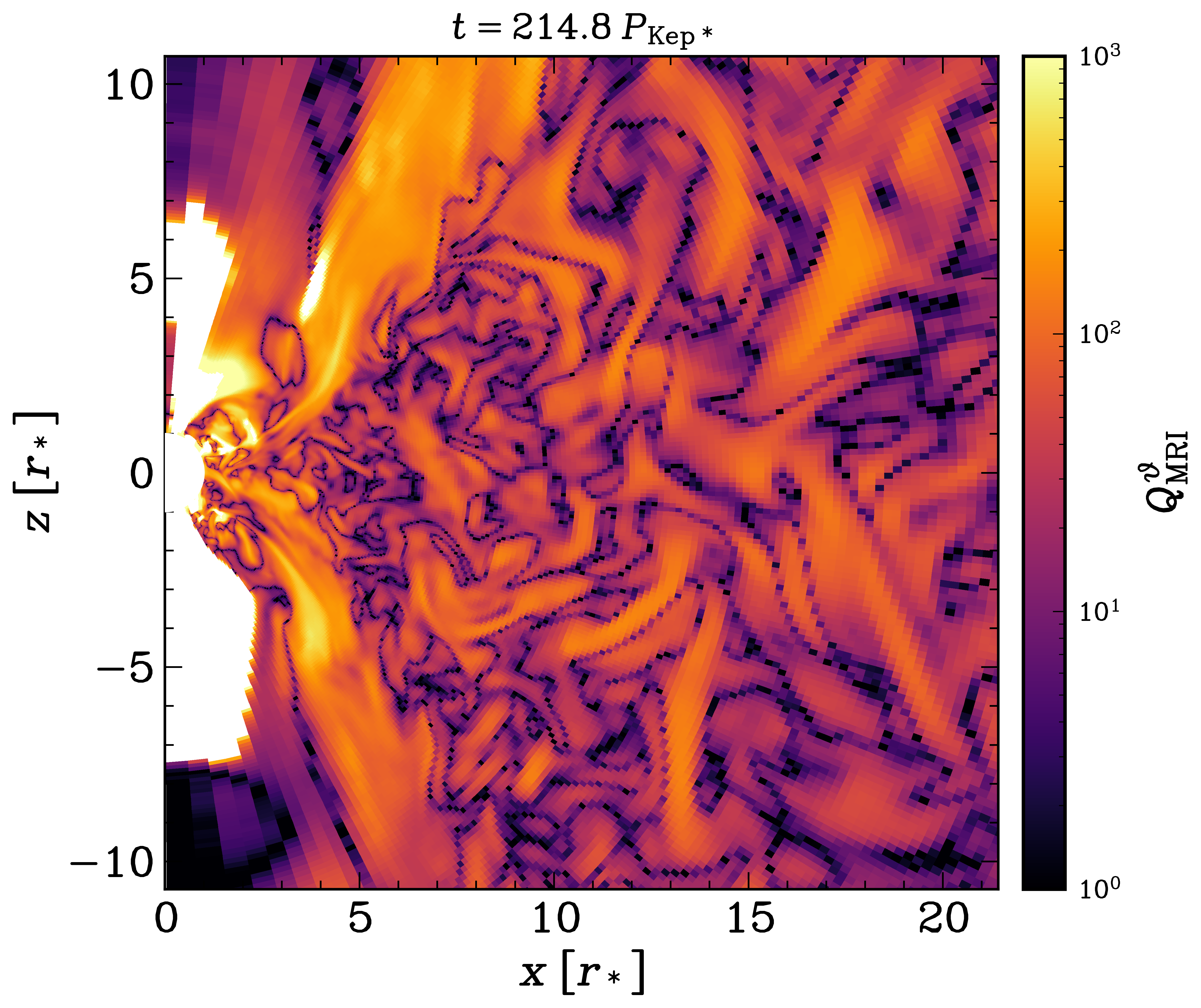}
    \includegraphics[width=0.49\linewidth]{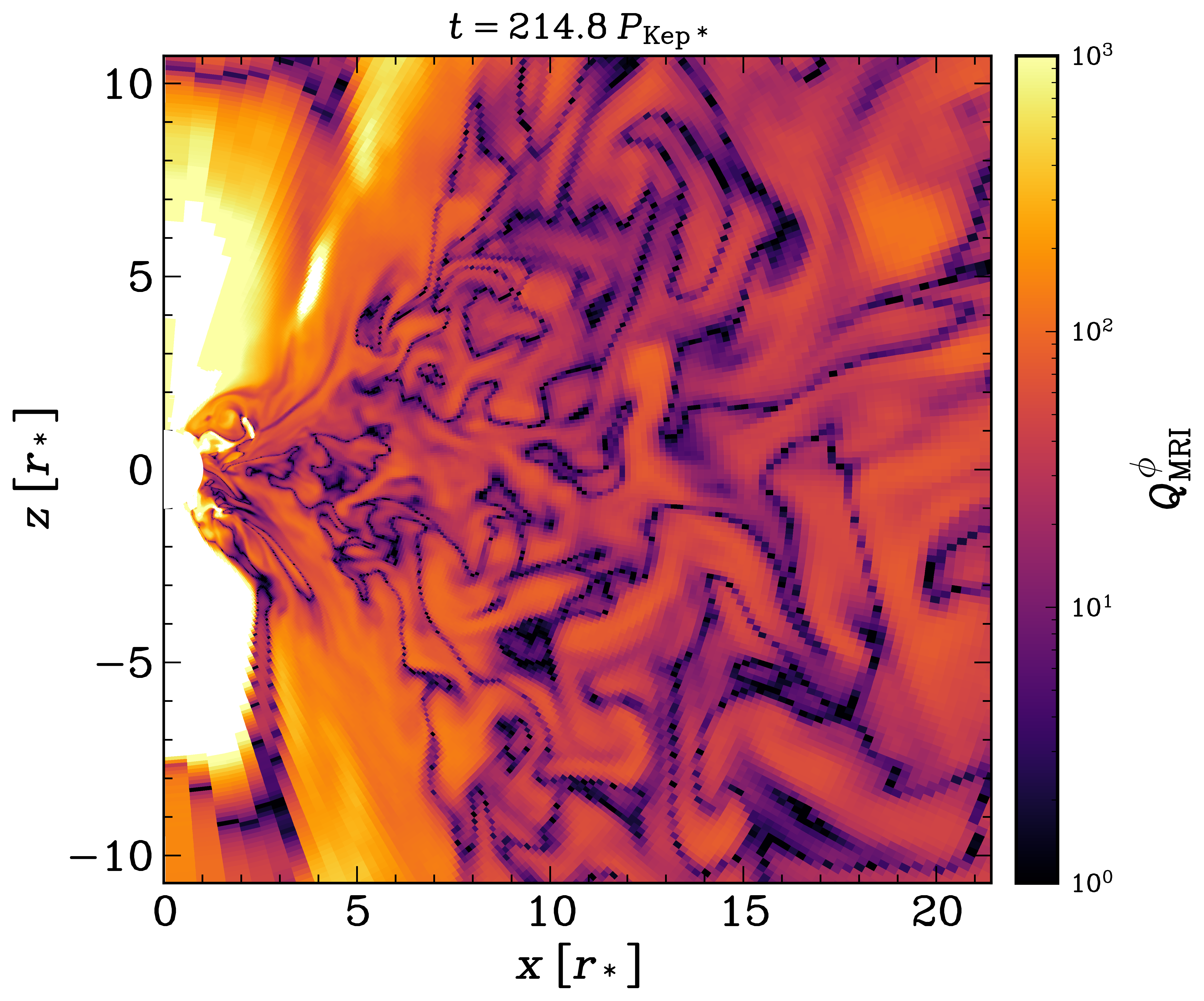}
    \includegraphics[width=0.49\linewidth]{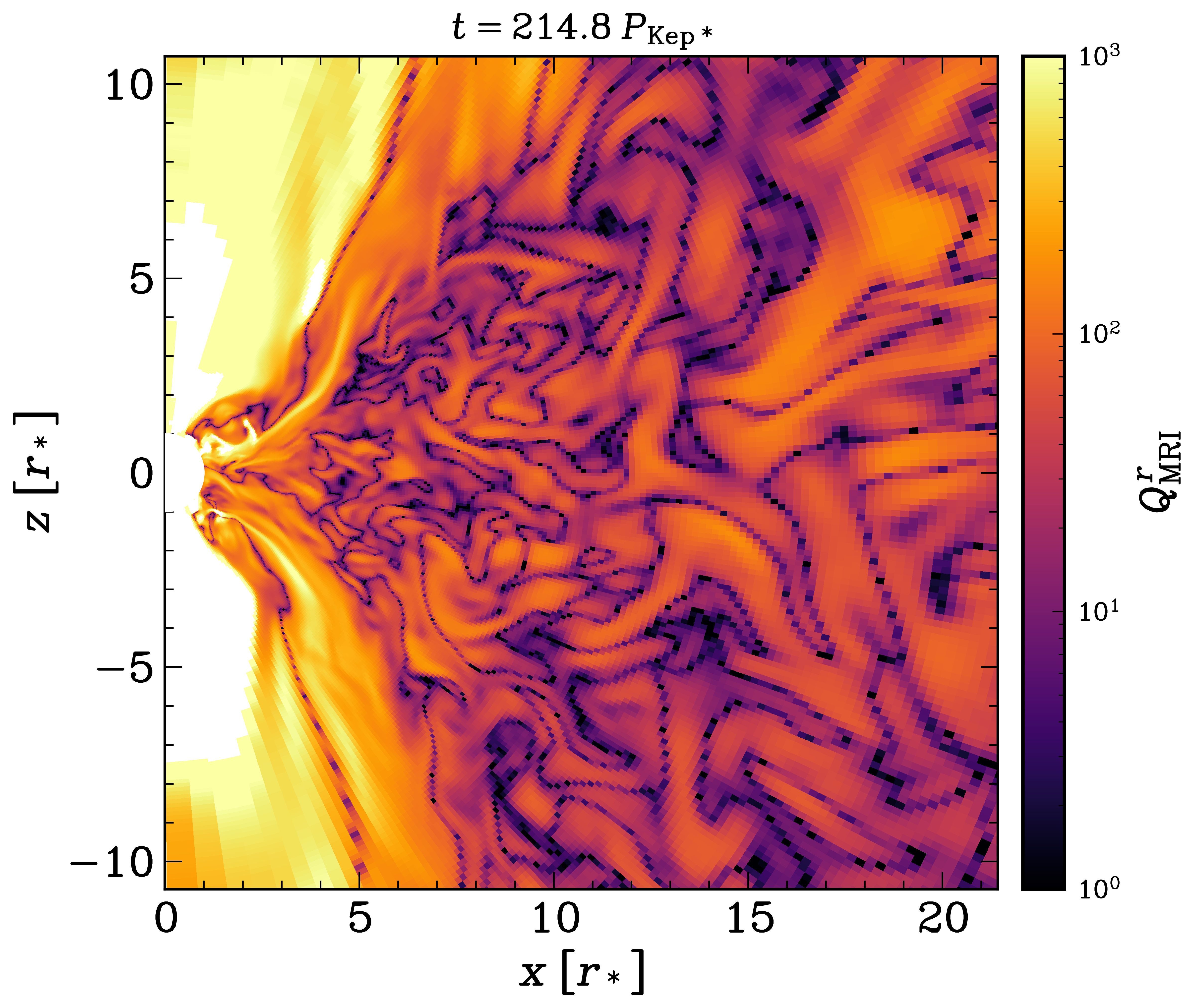}    
    \caption{Representative snapshots of the quality factor $Q_{\rm MRI}$ for the MRI in the \texttt{Fiducial} simulation at around $t=215\,P_{\rm Kep\star}$. The polar ($\theta$; top left), azimuthal ($\phi$; top right), and radial ($r$; bottom) quality factors indicate a well-resolved fastest growing unstable mode ($Q^i_{\rm MRI}>10$) throughout the turbulent accretion flow.
    }
    \label{fig:qmri}
\end{figure}

We also performed an additional shorter run at a higher azimuthal resolution (HR) of $n_{r}\times n_{\theta} \times n_{\phi} = 400 \times 256 \times 320$ for the \texttt{LargeLoop\_}$\beta100$  simulation to understand whether the dynamics of the atmosphere is susceptible to resolution. Figure~\ref{fig:conv} shows different time-averaged (semi-global) properties of the accretion flow for both standard and HR resolution runs during the last $100\,P_{\rm Kep\star}$ of evolution of the HR run. The radial profiles of average density, accretion rate, and angular velocity of both runs agree very well. For the purposes of the present study, we consider that the time-averaged quantities of interest to our study are converged with our standard resolution.

\begin{figure}
    \centering
    \includegraphics[width=0.49\linewidth]{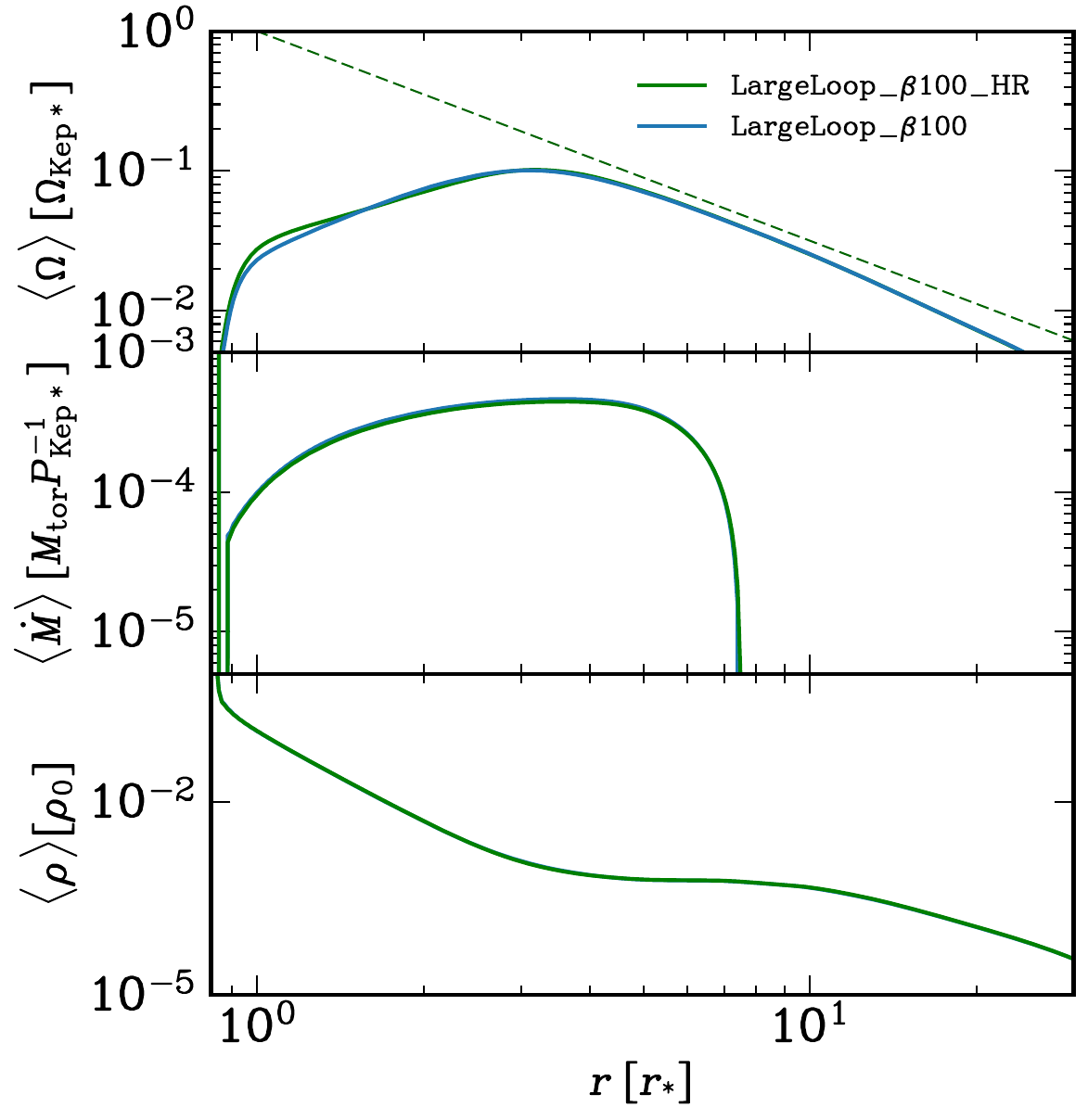}    
    \caption{Time-averaged angular velocity (top), accretion rate (center), and density (bottom) for the \texttt{LargeLoop\_}$\beta100$ simulations with standard and high azimuthal resolution in the last $100\,P_{\rm Kep\star}$ of evolution.}
    \label{fig:conv}
\end{figure}

\bibliography{astrograv,gr-astro}
\bibliographystyle{aasjournal}



\end{document}